\documentclass[12pt,preprint,epsf,eqsecnum]{aastex}
\usepackage{graphicx}
\newcommand{\bra}{\langle}
\newcommand{\ket}{\rangle}
\newcommand{\sn}{\mathit{sn}}
\newcommand{\cn}{\mathit{cn}}
\newcommand{\dn}{\mathit{dn}}
\newcommand{\be}{\begin{equation}}
\newcommand{\ee}{\end{equation}}
\newcommand{\ba}{\begin{eqnarray}}
\newcommand{\ea}{\end{eqnarray}}
\newcommand{\robold}{\mbox{\boldmath$\vec{\rho}\,$}}
\newcommand{\taubold}{\mbox{\boldmath$\vec{\tau}\,$}}

\newcommand{\Rv}{{\bf \vec R}}
\newcommand{\vv}{{\bf \vec v}}
\newcommand{\Lv}{{\bf \vec L}}
\slugcomment{\underline{Invited Review, ~to be published in}\\ 
\underline{``Recent Research Developments in Astrophysics,''}\\
\underline{by Research Signpost, India 2002}}
\shorttitle{Precession Relaxation of Unsupported Rotators}
\shortauthors{Efroimsky, Sidorenko, Lazarian}
\begin{document}
\title{\Large Complex rotation with internal dissipation. \\
Applications to
cosmic-dust alignment and to wobbling\\ comets and asteroids.\\
~\\}

\author{\bf Michael Efroimsky}
\affil{Institute for Mathematics and its
Applications, University of  Minnesota,\\ 
207 Church St. SE, Suite 400, Minneapolis MN 55455,
USA}
\email{efroimsk@ima.umn.edu\\
~}

\author{\bf Alex Lazarian}
\affil{Departments of Astronomy, 
University of Wisconsin, 475 North Charter Street,\\ 
Madison WI 53706, USA}
\email{lazarian@astro.wisc.edu\\
~}

\author{\bf Vladislav Sidorenko}
\affil{Keldysh Institute of Applied Mathematics,
Miusskaja Pl., Dom 4\\ 
Moscow 125047 Russia}
\email{sidorenk@spp.keldysh.ru\\
~}

\author{~\\}
\affil{~\\}
\email{~\\}

\author{~\\}
\affil{~\\}
\email{~\\}

 \author{~\\}
 \affil{~\\}


\begin{abstract}

  Neutron stars, asteroids, comets, cosmic-dust granules,  spacecraft,
  as well as whatever other freely spinning body dissipate energy when
  they  rotate  about  any  axis  different from principal. We discuss
  the internal-dissipation-caused  relaxation  of  a freely precessing
  rotator  towards  its minimal-energy mode  (mode that corresponds to
  the spin about the maximal-inertia axis).  We show  that this simple
  system contains in itself  some quite unexpected physics.  While the
  body nutates at some rate,  the internal stresses and strains
  within the body oscillate at frequencies  both  higher and  (what is
  especially surprising)  lower than  this rate.  The internal
  dissipation  takes place  not so much  at the frequency  of nutation
  but rather at the second and higher harmonics.  In other words, this
  mechanical system provides  an example  of an  extreme non-linerity.
  Issues  like chaos and separatrix  also come into play.  The earlier
  estimates,  that ignored non-linearity,  considerably underestimated
  the efficiency  of the internal relaxation of wobbling asteroids and
  comets.  At the same time, owing to the non-linearlity  of inelastic
  relaxation,  small-angle  nutations  can  persist for very long time
  spans.  The latter circumstance is important  for  the analysis  and
  interpretation  of NEAR's data on  Eros'  rotation state.  Regarding
  the comets,  estimates  show  that  the currently available  angular
  resolution  of  spacecraft-based instruments  makes it  possible  to
  observe wobble damping  within year- or  maybe even month-long spans
  of time. Our review also covers pertinent topics from the cosmic-dust
  astrophysics;  in particular,  the role played by precession damping 
  in the dust alignment. We show that this damping provides coupling of
  the grain's rotational and vibrational degrees of freedom; this
  entails occasional flipping of dust grains due to thermal fluctuations.
  During such a flip, grain preserves its angular momentum, but the
  direction of torques arising from H$_2$ formation reverses. As a
  result, flipping grain will not rotate fast in spite of the action of
  uncompensated H$_2$ formation torques. The grains get ``thermally
  trapped,'' and their alignment is marginal. Inelastic relaxation
  competes with the nuclear and Barnett relaxations, so we define
  the range of sizes for which the inelastic relaxation dominates.

\end{abstract}

\maketitle

\pagebreak

\section{\underline{INTRODUCTION}}

\subsection{Motivation}

On the 14-th of February 1967 the Soviet Union
launched artificial spacecraft Kosmos 142, to carry out
some ionospheric research. The sputnik had the shape of a cross
constituted by four 15-meter-long rods. A separate container, shaped as a 
cylinder with hemispheres on its ends, was attached in elastic manner to the 
cross, in a position orthogonal to its plane. This block had dimensions of 
about 1.6 $m$ $\times$ 0.8 $m$, and carried in itself all the scientific 
equipment. It was connected to the cross frame by a joint, and it turned out 
that
this perpendicular position of the container was not secured with a sufficient 
strength.
The mission planners wanted the satellite to rotate in the plane of the cross
at a rate of 2 revolutions per second. At a certain point, when the spacecraft 
was yet gaining rotation, deformation started. The cylindrical container 
overpowered the locking device in the joint, and bent towards the plane of the 
cross-shaped frame. This phenomenon was addressed by 
Vasil'ev \& Kovtunenko (1969) who pointed out that the intensity of the effect 
depends, among other things, upon the angular velocity of rotation of the 
cross frame.
In 22 months after that event, on the 14-th of December 1968, 
a similar sputnik, Kosmos 259, was launched. Its rotation
rate was not so swift: less that one revolution per second. This time
no deformations of the spacecraft was observed, and the mission 
succeeded.

The misadventure of Kosmos 142 resulted from the first principles of mechanics:
a freely rotating top must end up in the spin state that minimises the kinetic 
rotational energy, for a fixed angular momentum. This spin mode can be achieved
by one or both of the following means: adjustment of shape or/and alteration of
the rotation axis. Since the Russian spececraft was easily deformable, it 
``preferred'' the first option. Things would go differently if the satellite's 
construction were more rigid. The latter effect was observed back in 1958, when
the team operating the first American artificial satellite was surprised by 
some unexpected maneuvres that the spacecraft suddenly began to carry out. The 
satellite, called Explorer I, was a very elongated body with four flexible 
antennas on it. After launching and getting to the orbit, it was set to perform
steady rotation about its longest dimension. However, the flight operators 
never managed to keep the spacecraft in the designed spin state: Explorer 
persistently deviated from the simple rotation and went into a wobble, 
exhibiting slowly changing complex spin. Naturally, the rotation state was 
evolving toward that of minimal kinetic energy (the angular momentum being 
fixed). We would remind that the state of rotation about the maximal-inertia 
axis is the one minimising the kinetic energy, while spin about the 
least-inertia axis corresponds to the maximal energy. Hence, one should expect 
that the body will (through some dissipative processes) get rid of the 
excessive energy and will change the spin axis. 

Another example of unsupported rotator subject to internal dissipation is a
cosmic-dust granule. Due to various spin-up mechanisms (the main of which is
catalytic formation of $H_2$ moleculae on the granule surface (Purcell 1979)),
these particles spend most part of their life in rotation. This circumstance 
gives birth to a whole sequence of subtle effects, which determine alignment of
the dust relative to the interstellar magnetic field. This alignment can be 
indirectly observed through measuring the polarisation degree of the starlight
passing through the dust cloud (Lazarian 2000, 1994). It turns out that 
theoretical description of alignment in based on one's knowledge of the 
granules' typical rotation state: it is important whether the dust particles
are, predominantly, in their principal spin states or not (Lazarian \& 
Efroimsky 1999; Efroimsky 2002). 

Similar to spacecraft and interstellar grains, a comet or an asteroid in a 
non-principal rotation mode will dissipate energy and will, accordingly, return
to the stable spin (Prendergast 1958, Burns \& Safronov 1973, Efroimsky \& 
Lazarian 1999). Nevertheless, several objects were recently found in excited 
states of rotation. These are asteroid 4179 Toutatis (Ostro et al. 1993, Harris
1994, Ostro et al. 1995, Hudson and Ostro 1995, Scheeres et al. 1998, Ostro et 
al. 1999) and comet P/Halley (Jewitt 1997; Peale \& Lissauer 1989; Sagdeev et 
al. 1989; Peale 1991; Wilhelm 1987). Quite possibly, tumbling are also comet 
46P/Wirtanen (Samarasinha, Mueller \& Belton 1996; Rickman \& Jorda 1998), 
comet 29P/Schwachmann-Wachmann 1 (Meech et al 1993). The existing observational
data on asteroid 1620 Geographos may, too, be interpreted in favour of wobble 
(Prokof'eva et al. 1997; Prokof'eva et al. 1996; Ryabova 2002).

The dynamics of a freely rotating body is determined, on the one hand, by the 
initial conditions of the object's formation and by the external factors 
forcing the body out of its principal spin state. On the other hand, it is 
influenced by the internal dissipation of the excessive kinetic energy 
associated with wobble. Two mechanisms of internal dissipation are known. The 
so-called Barnett dissipation, caused by the periodic remagnetisation, is 
relevant only in the case of cosmic-dust-granule alignment (Lazarian \& Draine 
1997). The other mechanism, called inelastic relaxation, is, too, relevant for 
mesoscopic grains, and plays a primary role in the case of macroscopic bodies.
Inelastic relaxation results from alternating stresses generated inside a 
wobbling body by the transversal and centripetal acceleration of its parts. The
stresses deform the body, and inelastic effects cause energy dissipation. 

The external factors capable of driving a rotator into an excited state are 
impacts and tidal interactions, the latter being of a special relevance for 
planet-crossers. In the case of comets, wobble is largely impelled by jetting. 
Even gradual outgassing may contribute to the effect because a spinning body 
will start tumbling if it changes its principal axes through a partial loss or 
redistribution thereof. Sometimes the entire asteroid or comet may be a 
wobbling fragment of a progenitor disrupted by a collision (Asphaug \& Scheeres
1999, Giblin \& Farinella 1997, Giblin et al. 1998) or by tidal forces. All 
these factors, that excite rotators, compete with the inelastic dissipation 
that always tends to return the rotator to the minimal-energy state.

Study of comets' and asteroids' rotation states may provide much information 
about their recent history and internal structure. However, theoretical 
interpretation of the observational data will become possible only after we 
understand quantitatively how inelastic dissipation affects rotation. The 
kinetic energy of rotation will decrease at a rate equal to that of energy 
losses in the material. Thus, one should first calculate the elastic energy 
stored in a tumbling body, and then calculate the energy-dissipation rate, 
using the material quality factor $\, Q $. This empirical factor is introduced 
for a phenomenological description of the overall effect of the various 
attenuation mechanisms (Nowick \& Berry 1972; Burns 1986, 1977; Knopoff 1963; 
Goldreich \& Soter 1965). A comprehensive discussion of the $Q$-factor of 
asteroids and of its frequency- and temperature-dependence is presented in 
Efroimsky \& Lazarian (2000).

\subsection{Complex Rotation of a Rigid Body}

Our review addresses unsupported rotation of rigid and not-entirely-rigid 
objects. Stated differently, we intend to describe behaviour of unsupported 
rotators of two sorts: ideal (i.e., those that are exempt from internal 
dissipation) and realistic (i.e., those subject to dissipation). While the role
of dissipative phenomena in the rotating top has become an issue only less than
half a century ago, complex rotation of an ideal (absolutely rigid) top has 
been on the scientific agenda since, at least, the mid of XVIII-th century. 
This problem generated some of the major mathematical advances carried out by
Jacobi, Poinsot and other eminent scholars. However, the founding father of 
this line of study was Euler whose first notes on the topic date back to 
1750's.

Leonhard Euler, the most prolific scientist of all times, will forever retain 
an aura of mistery in the eyes of historians. Very few researchers, if any, 
shared his power of insight and his almost superhuman working ability. His life
in science consisted of three major periods: the first Russian period
(which began in 1730, when young Euler retired from the Russian navy for the
sake of academic career), the Berlin period (that started in 1741, when
Euler assumed a high administrative position at the Berlin Academy), and
the second Russian period (which began in 1765, when major disagreements with
King Frederich the Second moved Euler to accept an invitation from
Empress Catherine the Great, to return to St.Petersburg). Each of these three
periods in Euler's life was marked by numerous scientific achievements in 
all areas of mathematics known at that epoch. 

One of the fields, that grossly benefitted from Euler's attention during his 
tenure in Berlin, was mechanics of an unsupported top. Euler wrote ca 1760 
his pivotal result on the topic (Euler 1765), his famous equations:
\ba
\frac{d}{dt} \left(I_i \; {{\Omega}}_i\right) \; - \; \left(I_j \; - \; I_k \protect\right) \; 
\Omega_j \; \Omega_k \; = \; \tau_i \; \; \; ,
\label{1.1}
\ea
where $\;I_{1,2,3}\;$ are the eigenvalues of inertia tensor of the body. The
tensor is defined through
\ba
I_{ij}\;\equiv\; \int dm\,\left\{{\robold}^2 \delta_{ij}\;-\;\rho_i \rho_j 
\right\}\;\;\;,
\label{1.2}
\ea
$\;\robold\;$ being the position of mass element $\;dm\;$ relative to the
centre of mass of the body. Equations (\ref{1.1}) are merely a reformulation 
of the 
simple fact that the torque equals the rate of change of the angular 
momentum. They express this law in {\bf a} body frame. Among the body frames,
there exists one (called principal) wherein the inertia tensor is diagonal. 
In (\ref{1.1}), $\;\Omega_{1,2,3}\;$ are the angular-velocity components 
as measured in that, principal, coordinate system. Quantities $ \; \tau_i \;$
are principal-axes-related components of the total torque acting on the body.
As ever, we shall assume without loss of generality that $\;I_3\,\geq\,I_2\,
\geq\,I_1\;$. Hence the third axis will always be that of major inertia.

In the body frame, the period of angular-velocity precession about the principal axis $\;3\;$ is: 
$\;\tau\;=\;2\,\pi/\omega\,.$ Evidently, 
\begin{eqnarray}
{\dot{\Omega}}_i/{\Omega}_i \; \approx \; {\tau}^{-1} \; \; \; , 
\; \; \; \; \; 
{\dot{I}}_i/{I}_i \; \approx \; {\tau}^{-1} \, \epsilon \; \; \;,\;\;\;
\label{22.3}
\end{eqnarray}
$\epsilon\;$ being a typical value of the relative strain that is several 
orders less than unity. These estimates lead to the inequality  $\; \dot{I_i}
\,
\Omega_i\;\ll \;I_i\, \dot{\Omega_i}\;$, thereby justifying the commonly used
approximation to Euler's equations\footnote{~Rigorously speaking, in the case
when the approximation (\ref{22.3}) is not satisfied, not only (\ref{22.4}) 
fail but even equations (\ref{1.1}) must be 
somewhat amended. The problem arises from the ambiguity in the choice of the 
body frame. For example, if we prefer to choose the coordinate system wherein 
the inertia tensor always remains diagonal, then the angular momentum will be
different from zero in the body frame (and will be of order $\epsilon$). If,
though, we choose the coordinates in which the angular momentum vanishes, then
the inertia matrix will no longer be diagonal. With this choice of the body 
frame, (\ref{1.1}) should rather be written down not in terms of $I_i$ but in
terms of all $I_{ij}$. We shall not elaborate on this issue in our review.}:
\be
I_i \; {\dot{\Omega}}_i \; - \; \left(I_j \; - \; I_k \protect\right) \; 
\Omega_j \; \Omega_k \; = \tau_i \; \; \; \; .
\label{22.4}
\ee
Naturally, this elegant system of equations has carried since its birth the 
name of its author. It took scientists some more  years to 
understand that
the system deserves its given name for one more reason: formulae (1.1) 
are exactly the Euler-Lagrange equations for the Lagrangian of an
unsupported rigid body. Proof of this fact demands a certain effort.
A brief (but still not trivial) derivation offered in 1901 by Poincare can be
found in the textbook by Marsden (2000).

The Euler equations simplify considerably when two of three moments
of inertia $\;I_i\;$ are equal. This is called dynamic symmetry, to 
distinguish it from the full geometric symmetry. Further on, whenever we 
refer to symmetric top, we shall imply only the dynamic symmetry, not the 
geometric one. This case was addressed by Euler (1765) himself,
and later by Lagrange (1813) and Poisson (1813). For prolate
symmetric rotators (i.e., when $\;I_3\,=\,I_2\,>\,I_1\;$), in the absence of
external torques, the solution is simple:
\be
{\Omega}_1 \; \; = \; \; const\;\;,\;\;\;
{\Omega}_2 \; \; = \; \; {\Omega}_{\perp} \sin {\omega}t~~,~~~
{\Omega}_3 \; \; = \; \; {\Omega}_{\perp} \cos {\omega}t~~,~~~
\label{1.3}
\ee
where $\omega =(I_1/I_3  - 1) \Omega_1 $. We see that, from the viewpoint of
an observer placed on the rotating body, the vector of inertial 
angular velocity $\;\bf \vec \Omega\;$ describes a circular cone about the 
minor-inertia axis (1) of the body. So does the angular-momentum vector 
$\,\bf \vec J\,$. Both $\;\bf \vec \Omega\;$ and $\,\bf \vec J\,$ precess
about the least-inertia axis at the same rate $\omega =(I_1/I_3  - 1) 
\Omega_1 $, though at different angles from the axis.

In an inertial coordinate system, the angular momentum $\,\bf \vec J\,$
will not precess, because it must keep constant for a free rotator.  Instead,
it is the least-inertia axis (1) and the angular velocity $\;\bf \vec 
\Omega\;$ that precess about $\,\bf \vec J\,$, in an inertial observer's
opinion. (For brevity, we denote each vector by one letter, though we, of
course, imply that every vector transforms appropriately whenever the
coordinate system is changed.)

In the case of oblate (dynamic) symmetry ($\;I_3\,>\,I_2\,=\,I_1\;$) 
free precession will be expressed, in the body frame, by solution
\be
{\Omega}_1 \; \; = \; \; {\Omega}_{\perp} \cos {\omega}t~~,~~~
{\Omega}_2 \; \; = \; \; {\Omega}_{\perp} \sin {\omega}t~~,~~~
{\Omega}_3 \; \; = \; \; const
\label{1.4}
\ee
where $\,\omega = (I_3/I_1 - 1) \Omega_3\;$ is the mutual rate of circular
precession of $\;\bf \vec \Omega\;$ and $\;\bf \vec J\;$ about the major inertia 
axis (3).

The general case of $I_3>I_2 \geq I_1$ is quite involved and demands numerics (see
Mitchell \& Richardson (2001); Richardson \& Mitchell (1999), and references 
therein). Still, in the absence of external torques the problem can be solved 
analytically, and Euler coped with it (Euler 1765), though to that end he had to 
introduce functions similar to what we now call elliptic integrals. The solution 
much simplifies when expressed through the elliptic functions of Jacobi {\textit 
{sn, cn, dn}}. These were defined and studied by Karl 
Jacobi (1829) and used by him (Jacobi 1849, 1882) to 
describe free rotation. Jacobi's functions are generalisations of the 
trigonometric ones, in the following sense: while for symmetric prolate 
and oblate rotators the circular precession is expressed by (\ref{1.3}) 
and (\ref{1.4}) 
correspondingly, in the general case $I_3 \geq I_2 \geq I_1$ precession
is expressed by very similar formulae that contain Jacobi's finctions 
instead of {$\;\sin\;$} and {$\;\cos\;$}: 
\begin{eqnarray}
\Omega_1\;=\;\gamma\;\,{\it{dn}}\left(\omega t , \; k^2 \protect\right)\;\;,
\;\;\;\;\Omega_2 \; = \; \beta \, \; sn\left(\omega t , \; k^2 \protect\right)
\;\;,\;\;\;\;\Omega_3 \; = \; \alpha\;\,{\it{cn}}\left(\omega t,\;k^2
\protect\right)\;\;,\;\;\;
\label{1.5}
\end{eqnarray}
for $\;{\bf{J}}^2 \; < \; 2\;I_2 \; T_{\small{kin}} \; $, and  
\begin{eqnarray}
\Omega_1\;=\;{\tilde \gamma}\;\,{\it{cn}}\left({\tilde \omega} t,\;{\tilde k}^2
\protect\right)\;\;,\;\;\;\;\Omega_2 \; = \;{\tilde \beta}\,\;{\it sn}\left(
{\tilde \omega} t ,\;{\tilde k}^2 \protect\right)\;\;,\;\;\;\;\Omega_3 \; = \; 
{ \alpha}\;\,{\it{dn}}\left({\tilde \omega} t,\;{\tilde k}^2
\protect\right)
\label{1.6}
\end{eqnarray}
for $\;\;{\bf{\vec J}}^2\;>\;2\;I_2\;T_{\small{kin}}\;$. Here the precession 
rate 
$\;\it \omega\;$ and the parameters $\;\alpha ,\;\beta ,\;{\tilde \beta}, \;
\gamma ,\;{\tilde \gamma}, \;{\tilde \omega},\;k\;$ and $\;{\tilde k}\;$ are
certain combinations of $\;I_{1,2,3}, \;T_{\small {kin}}\;$ and 
$\;{\bf \vec J}^2\;$. We see that 
(\ref{1.5}) is a generalisation of (\ref{1.3}), while (\ref{1.6}) is that of
 (\ref{1.4}). Solution (\ref{1.5}) approaches (\ref{1.3}) in the 
limit of prolate symmetry, $\;(I_3\,-\,I_2)/I_1\,\rightarrow\,0\;$, while 
solution (\ref{1.6}) approaches (\ref{1.4}) in the limit of oblate symmetry, 
$\;(I_2\,-\,I_1)/I_1\,\rightarrow\,0\;$. This situation is illustrated by 
Figure 1. To understand this 
\begin{figure}
\plotone{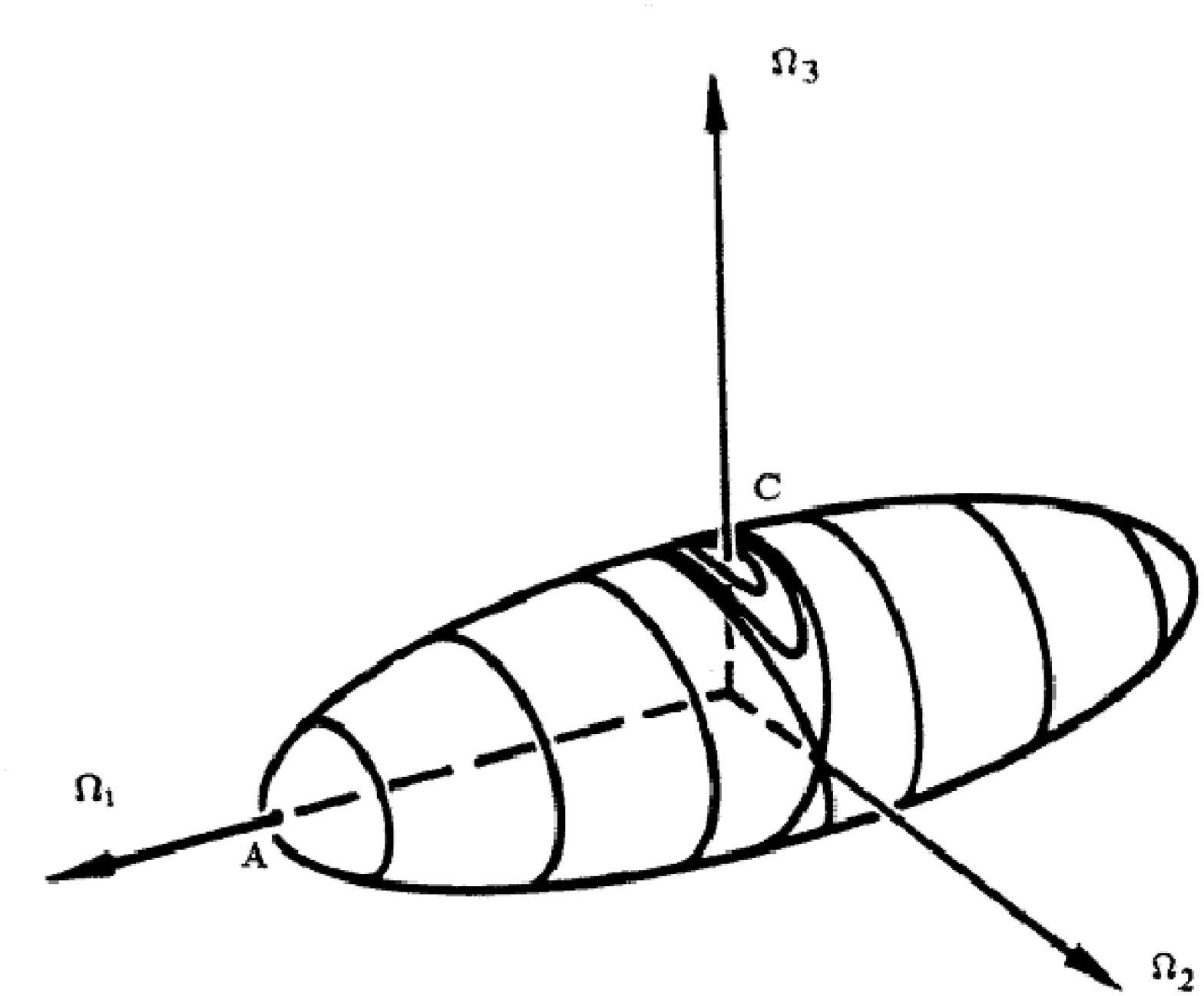}
\caption{The constant-angular-momentum ellipsoid, 
in the angular-velocity space. The 
lines on its surface are its intersections with the kinetic-energy ellipsoids
corresponding to different values of the rotational energy. The quasi-stable 
pole A is the maximal-energy configuration, i.e., the state wherein the body 
spins about its minimal-inertia axis. The stable pole C symbolises the 
minimal-energy state, i.e., rotation about the maximal-inertia axis. The 
angular-velocity vector describes the constant-energy lines, and at the same 
time slowly shifts from one line to another, approaching pole C. The picture 
illustrates the case of an elongated body: $I_3 \stackrel{>}{\sim}I_2>I_1$. 
The trajectories are circular near A and remain (in the case of an elongated 
body) virtually circular almost up to the separatrix. The trajectories 
will regain a circular shape only in the closemost proximity of C.}
\end{figure}
picture, one should keep in mind that two 
quantities are conserved for a freely-spinning body: the angular momentum 
\be
{\bf{\vec J}}^2 \;=\;I_1^2\,{\Omega}_1^2\;+\;I_2^2\,{\Omega}_2^2\;+\;I_3^2\,
{\Omega}_3^2 \;\;,\;\;\;
\label{1.7}
\ee
(which is conserved exactly) and the kinetic energy
\be
T_{\small{kin}}\;=\;\frac{1}{2}\;\left\{I_1\,{\Omega}_1^2\;+\;I_2\,{\Omega}_2^2
\;+\;I_3\,{\Omega}_3^2 \right\}\;\;.\;\;\;
\label{1.8}
\ee
(which is conserved only approximately because of the internal dissipation).
Expressions (\ref{1.7}) and (\ref{1.8}) define ellipsoids in the 
angular-velocity space $\;(\Omega_1,\,\Omega_2,\,\Omega_3)\;$. Intersection 
of these gives the trajectory described by the tip of vector $\;\bf \vec \Omega\;$
in the angular-velocity space. On the picture we see the angular-momentum 
ellipsoid (\ref{1.7}) with lines marked on its surface. These lines are its 
intersection with ellipsoids (\ref{1.8}) appropriate to several different 
values of energy $\,T_{\small{kin}}\;$. For a fixed value of ${\bf{\vec  J}}^2$, i.e., for a fixed angular-momentum surface
(\ref{1.7}), there exists an infinite family of kinetic-energy surfaces 
(\ref{1.8}) intersecting with it. The largest surface of kinetic energy 
(corresponding to the maximal value of $\;T_{kin}\;$) will be an ellipsoid 
that
fully encloses our angular-momentum ellipsoid and only touches it in point A 
and its opposite. Similarly, the smallest surface of kinetic energy 
(corresponding to minimal $\;T_{kin}\;$) will be an ellipsoid escribed by our 
angular-momentum ellipsoid and only touching it from inside, at point C and its
opposite. For a fixed $\;\bf \vec J\;$, the maximal and minimal possible values
of the kinetic energy are achieved during rotations about the minimal-inertia 
and maximal-inertia axes, appropriately. In the case of a non-dissipative 
torque-free rotation, the tip of vector $\;{\bf{\vec \Omega}}\;$ will be 
describing, on Fig. 1, a curve along which the angular-momentum and energy 
ellipsoids intersect (Lamy \& Burns 1972).
Solution (\ref{1.5}) is valid for higher energies, i.e., from point A 
through the separatrix. In astronomy such rotations are called LAM 
(~=~ Long-Axis Modes). Solution (\ref{1.6}) works for lower energies, 
i.e., from the separatrix through point C. Such rotations are called SAM 
(~=~ Short-Axis Modes). Wherever the trajectories on Fig.1, i.e, in the space 
($\,\Omega_1\,$, $\,\Omega_2\,$, $\,\Omega_3\,$), are 
almost circular\footnote{Be mindful that the trajectory in the space ($\,\Omega_1\,$, $\,\Omega_2\,$, $\,\Omega_3\,$) being almost circular does not 
necessarily mean that the precession cone of the major-inertia axis about $\bf
 \vec J$ is circular or almost circular.}, the solutions (\ref{1.5}) and (\ref{1.6}) may be approximated by (\ref{1.3}) and (\ref{1.4}), correspondingly. In 
the limit of an oblate rotator, the applicability domain of (\ref{1.5}) will
shrink into a point (or, to be more exact, into two points: A and its 
opposite). Similarly, in the limit of a prolate body, the applicability region 
of (\ref{1.6}) will shrink into two points: C and its opposite.

\subsection{Realistic Rotators}

The formalism developed by Euler and refined by Jacobi might be a perfect 
tool
for description of rotation of asteroids, comets, cosmic-dust granules, 
spacecraft and whatever other unsupported rigid rotators, if not for 
one circumstance, inner dissipation. Because of this circumstance, the 
Euler-Jacobi theory of precession works only for time spans
short enough to neglect kinetic-energy losses. 

The necessity of internal 
dissipation follows from the basic principles of mechanics. 
A freely spinning body of a fixed angular momentum has kinetic energy whose 
values are constrained to lie within a certain bounded range. Hence, from the
physical viewpoint, it is very natural for this body to be seeking ways of 
relaxation. In other words, the body must ``do its best'' to get rid of the 
excessive kinetic energy, in order to approach the minimal-energy 
configuration. Thence the necessity of some dissipation mechanism. 

Two such mechanisms are known. One is relevant only for mesoscopic 
rotators, like interstellar-dust grains, and therefore plays a certain role in 
the cosmic-dust alignment. This is the Barnett dissipation, a phenomenon called
into being by periodic remagnetisation of a precessing paramagnetic body 
(Lazarian \& Draine 1997).

The second mechanism, inelastic dissipation, is, too, relevant for mesoscopic
grains (Lazarian \& Efroimsky 1999), and it plays the decisive role in the
macroscopic bodies' relaxation. The effect results from the alternating 
stresses produced inside a wobbling body by the time-dependent acceleration of 
its parts. The stresses deform the body, and the inelastic effects cause 
dissipation of the rotational energy.

Dissipation entails relaxation of the precession: the major-inertia axis of the 
body and its angular-velocity vector $\;\bf \vec \Omega \;$ tend to align along
the angular momentum $\bf \vec J$. In other words, the precession cone described 
by $\;\bf \vec \Omega \;$ about $\bf \vec J$ will be narrowing until $\;\bf  \vec 
\Omega \;$ aligns along $\bf \vec J$ completely. A simple calculation (Efroimsky 
2001, Efroimsky 2000, Efroimsky \& Lazarian 2000, Lazarian \& Efroimsky 1999)
shows that in this case the major-inertia axis of the body will align in the 
same direction, so that, from the body-frame viewpoint, $\;\bf \vec \Omega \;$ 
will eventually be pointing along this axis. This configuration will correspond 
to the minimal kinetic energy, the angular momentum being fixed. 

An inertial observer will thus see the unsupported body miraculously changing 
its rotation axis. This is exactly what happened in 1958 when, to mission 
experts' surprise, rotating satellite Explorer I changed its rotation axis
and went into wobble (Thomson 1961).

This was, probably, the first example of a practical need for a further 
development of the Eulerian theory of a free top, a development that would 
address an unsupported top with dissipation. However, Chandrasekhar realised 
this already in  mid-50s, after having been alerted by Kuiper, and asked a 
postdoc, Kevin Prendergast, to look into that\footnote{~The authors are 
grateful to Tom Gehrels for providing these historical facts.}. The most 
general question was
(and still is): how many asteroids in the Solar System can be in non-principal 
(i.e., nutating) spin states, and how can this evidence of the impact frequency
in the main belt? Prendergast in his paper (1958) implied that it is 
collisions\footnote{~The collisions within the main belt became a popular topic
much later, in 90-s. See, for example, (dell'Oro, Paolicchi, Cellino, Zappala, Tanga, \& Michel 2001).} that  
drive asteroids out of the principal state and make them wobble. An 
important point made by Prendergast (1958) was generation of the second harmonic in a 
symmetrical oblate rotator: if a body is precessing at an angular rate $\;\omega\;
$, then the dissipation is taking place not only at this frequency but also at 
double thereof. Prendergast considered only the deceptively simple case of 
symmetrical rotator, and therefore failed to notice the emergence of harmonics
higher than the second. Besides, the mathematical treatment of the problem, 
offered in his paper, was erroneous in several other aspects. Nevertheless, the 
fact that he noticed the second harmonic was by itself an important contribution 
for which Prendergast should be credited. His paper was published much ahead of time and, 
therefore, was forgotten. Independently from Prendergast, Lazarian and Efroimsky 
(1999) came across the second harmonic some 40 years later. 
Generation of the higher harmonics was pointed out only in (Efroimsky 2000).
The reason why the important work by Prendergast was not fully appreciated by
his contemporaries is that back in 50-s the observational astronomy lacked any 
reliable data on wobbling asteroids. So, Prendergast's paper went almost
unnoticed, and his successors had to start up from scratch.

The interest in the asteroidal precession re-emerged in 70-s, after the 
publication of the milestone work by Burns \& Safronov (1973) that suggested 
estimates for the relaxation time, based on the decomposition of the 
deformation pattern into bulge flexing and bending, and also on the conjecture 
that ``the centrifugal bulge and its associated strains wobble back and forth 
relative to the body as the rotation axis {\bf $\; \bf \omega\;$} moves through
the body during a wobble period.'' As turned out later, the latter conjecture 
was a too strong statement, because the inelastic dissipation, for the most 
part of it, is taking place not near the surface but in the depth of the body, 
i.e., not right under the bulge but deep beneath it. Thus, the bulge is much 
like an iceberg 
tip. This became clear when the distribution of precession-caused stresses was 
calculated, with improved boundary conditions (Efroimsky \& Lazarian 2000), 
(Lazarian \& Efroimsky 1999)\footnote{~This topic will be discussed in section 
4. Our treatment of stresses, demonstrated there, is not mathematically 
rigorous either. It is a polynomial approximation which satisfies the boundary 
conditions only approximately. From the physical viewpoint, it is not worth
further refining that treatment, because the slight mishandling of the boundary
conditions ``spoils'' the solution much less than the irregularity and 
inhomogeneity of a the realistic body.}. Burns \& Safronov's 
treatment neglected the nonlinearity, i.e., generation of frequencies higher 
and lower than the nutation rate. The nonlinearity, in fact, is essential. Its 
neglect leads to a large underestimation of the damping rate, because in many 
spin states a considerable input comes from the harmonics
(Efroimsky \& Lazarian 2000), (Efroimsky 2000). The neglect of nonlinearity 
leads to up to a two-order underestimate of the precession-damping rate.

In the same year, Peale published an article dealing with inelastic relaxation
of nearly spherical bodies (Peale 1973), and there he did take the second 
harmonic into account.

In 1979 Purcell addressed a similar problem of interstellar-grain precession  
damping. He ignored the harmonics and mishandled the
boundary conditions upon stresses: in (Purcell 1979) the normal stresses had 
their maximal values on the free surfaces and vanished in the centre of the 
body (instead of being maximal in the centre and vanishing on the surfaces). 
These oversights lead to a several-order underevaluation of the dissipation 
effectiveness and, thereby, of the relaxation rate.

\subsection{~~Precession ~damping}

The dynamics of precession relaxation is described by the angular rate of
alignment of the maximal-inertia axis (3) along the angular momentum $\bf 
\vec J$, 
i.e., by the decrease in angle $\;\theta\;$ between these. In the case of 
oblate symmetry (when $\;I_3\;>\;I_2\;=\;I_1\;$), this angle remains 
adiabatically 
unchanged over the precession period, which makes $d{\theta}/dt$ a perfect 
measure of the damping rate (Efroimsky \& Lazarian 2000). However, in the 
general case of a triaxial body angle $\;\theta\;$ evolves periodically through 
the precession cycle. To be more exact, it evolves $\it{almost}$ periodically, 
and its value at the end of the cycle is only slightly different from that in 
the beginning of the cycle. The relaxation is taking place through accumulation
of these slight variations over many periods. This is called adiabatic 
regime, i.e., regime with two different time scales: we have a ``fast'' process
(precession) and a ``slow'' process (relaxation). Under the adiabaticity 
assumption, one may average  $\;\theta\;$, or some function thereof, over the 
precession cycle. Then the damping rate will be described by the evolution of 
this average. Technically, it is convenient to use the average of its squared 
sine  (Efroimsky 2000). One can write for a triaxial rotator:
\be
\frac{d\;<\sin^2 \theta >}{dt}\;=\;\frac{d\;<\sin^2\theta>}{dT_{kin}}\;\;
\frac{dT_{kin}}{dt}\;\;\;,\;\;\;\;
\label{2.1}
\ee
while for an oblate one the expression will look simpler:
\be
\left(\frac{d\, \theta }{dt}\right)_{(oblate)}\;=\;\left(\frac{d\,\theta}{
dT_{kin}}\right)_{(oblate)}\;\;\frac{dT_{kin}}{dt}\;\;\;.\;\;\;\;
\label{2.2}
\ee
The derivatives $\;d\;<\sin^2\theta>/dT_{kin}\;$ and 
$\;\left(d\,\theta/dT_{kin}\right)_{(oblate)}\;$ appearing in (\ref{2.1}) and 
(\ref{2.2}) indicate how the rotational-energy dissipation affects the value of
$\;<\sin^2 \theta>\;$ (or simply of $\;\theta\;$, in the oblate case). These 
derivatives can be calculated from the equations of motion (see Efroimsky \& 
Lazarian (2000) and Efroimsky (2000)). The kinetic-energy decrease, $\;
dT_{kin}/dt\;$, is caused by the inelastic dissipation:  
\be
d{T}_{kin}/dt \; = \; < d{W}/dt > \;\; \;,\;\;\;
\label{2.3}
\ee
$W\;$ being the energy of the alternating stresses, and $\;<...>\;$ 
denoting an average over a precession cycle. (This averaging is justified 
within the adiabatic approach. For details see section III below.) Finally, in the
general case of a triaxial top, the alignment rate will read:
\be
\frac{d\,<\sin^2\theta>}{dt}\;=\;\frac{d\,<\sin^2\theta>}{dT_{kin}}\;\;
\frac{d\,<W>}{dt}\;\;\;,\;\;\;
\label{2.4}
\ee
and for a symmetrical oblate top:
\be
\left(\frac{d\, \theta }{dt}\right)_{(oblate)}\;=\;\left(\frac{d\,\theta}{
dT_{kin}}\right)_{(oblate)}\;\;\frac{d\,<W>}{dt}\;\;\;.\;\;\;\;
\label{2.5}
\ee
Now we are prepared to set out the strategy of our further work. While 
calculation of $\;d {\small \left. \langle \right.} \sin^2 \theta {\small 
{ \left. \ket \right.}} / 
dT_{kin}\;$ and $\;\left(d\theta /dT_{kin}\right)_{oblate}\;$ is an easy 
exercise\footnote{See 
formula (\ref{5.18}) below and also
formulae (A12 - A13) in Efroimsky 2000.}, our main goal will be to find the 
dissipation rate $\;d\,<W>/dt\;$. This quantity will consist of inputs from the 
dissipation rates at all the frequencies involved in the process, i.e., from the 
harmonics at which stresses oscillate in a body precessing at a given rate $\,
\omega\,$. The stress is a tensorial extension of the notion of a pressure or 
force. Stresses naturally emerge in a spinning body due to the centripetal and 
transversal accelerations of its parts. Due to the precession, these stresses 
contain time-dependent components. If we find a solution to the boundary-value 
problem for alternating stresses, it will enable us to write down explicitly the 
time-dependent part of the elastic energy stored in the wobbling body, and to 
separate contributions from different harmonics:
\be
<W>\;=\;\sum_{n}\;\,<W(\omega_n)>\;\;\;\;\;.\;\;
\label{4.9}
\ee
$W(\omega_n)\;$ being the elastic energy of stresses alternating at frequency 
$\,\omega_n$. One should know each contribution $W(\omega_n)$, for these will 
determine the dissipation rate at the appropriate frequency, through the 
frequency-dependent empirical quality factors. The knowledge of these factors, 
along with the averages $\,<W(\omega_n)>\,$, will enable us to find the 
dissipation rates at each harmonic. Sum of those will give the entire dissipation 
rate due to the alternating stresses emerging in a precessing body.

\subsection{Inelastic dissipation caused by complex rotation}

Equation (\ref{4.9}) implements the most important observation upon which all our
study rests: generation of harmonics in the stresses inside a precessing rigid 
body. The harmonics emerge because the acceleration of a point inside a precessin 
body contains centrifugal terms that are quadratic in the angular velocity $\bf{
\vec \Omega}$. In the simpliest case of a symmetrical oblate body, for example, 
the body-frame-related components of the angular velocity are given in terms of $
\,\sin \omega t\,$ and $\,\cos \omega t\,$ (see formulae (\ref{1.3}) - 
(\ref{1.4})). Evidently, squaring of $\bf{\vec \Omega}$ will yield terms both with
$\,\sin \omega t\,$ or $\,\cos \omega t\,$ and with $\,\sin 2 \omega t\,$ or $\,
\cos 2 \omega t\,$. The stresses produced by this acceleration will, too, contain 
terms with frequency $\,\omega t\,$ as well as those with the harmonic $\,2\omega 
t$. In the further sections we shall explain that a triaxial body precessing at 
rate $\,\omega\,$ is subject, in distinction from a symmetrical oblate body, to a 
superposition of stresses oscillating at frequencies $\;\omega_n\,=\,n\,\omega_1\,
$, the ''base frequency'' $\,\omega_1\,$ being lower than the precession rate $\,
\omega$. The basic idea is that in the general, non-oblate case, the time 
dependence of the acceleration and stresses will be expressed not by trigonometric
but by elliptic functions whose expansions over the trigonometric functions 
will generate an infinite number of harmonics. In subsection 4.3 we 
shall explain this in more detail.

The total dissipation rate will be a sum of the particular rates (Stacey 1992) 
to be calculated empirically. The empirical description of attenuation is
based on the quality factor $\,Q(\omega)\,$ and on the assumption of 
attenuation rates at different harmonics being independent from one 
another:
\be
\dot{W}\;=\;\sum_{n} \; \dot{W}{({\omega_n})}\;=\;-\;\sum_{n}\;
\frac{\omega_n\;W_0({\omega}_n)}{Q({\omega_n})}\;=\;-\;2\;\sum_{n}\;
\frac{\omega_n\;\,<W({\omega}_n)>}{Q({\omega_n})}\;\;\;\;\\
\label{4.10}
\ee
$\;Q(\omega)\;$ being the quality factor of the material, and  
$\;W_0({\omega}_n)\;$ and $\;\,<W({\omega}_n)>\,\;$ being the maximal and 
the average values of the appropriate-to-$\omega_n\;$ fraction of elastic 
energy stored in the body. This expression will become more general if we put 
the quality factor under the integral, implying its possible coordinate 
dependence\footnote{In strongly inhomogeneous nutating bodies 
attenuation may depend on location.}:
\be
\dot{W}\;=\;-\;2\;\sum_{\omega_n}\;\int\;dV\;\left\{\frac{\omega_n}{
Q({\omega_n})}\;\,\frac{d\,<W(\omega_n)>}{dV}\;\right\}\;\;\;,\;\;
\label{4.11}
\ee
The above assumption of attenuation rates at different harmonics being mutually 
independent is justified by the extreme smallness of strains (typically, much less 
than $\;10^{-6}$) and by the frequencies being extremely low ($10^{-5}\,-\,10^{-3}\;
Hz$). One, thus, may say that the problem is highly nonlinear, in that we shall take 
into account the higher harmonics in the expression for stresses. At the same time, 
the problem remains linear in the sense that we shall neglect any nonlinearity stemming 
from the material properties (in other words, we shall assume that the strains are linear 
function of stresses). We would emphasize, though, that the nonlinearity is most 
essential, i.e., that the harmonics $\;\omega_n\;$ come to life unavoidably: no matter 
what the properties of the material are, the harmonics do emerge in the expressions for 
stresses. Moreover, as we shall see, the harmonics interfere with one another due to $\;
W\;$ being quadratic in stresses. Generally, all the infinite amount of multiples of $\;
\omega_1\;$ will emerge. The oblate case, where only $\;\omega_1\;$ and $\;2\omega_1\;$ 
show themselves, is an exception. Another exception is the narrow-cone precession of a
triaxial rotator studied in Efroimsky (2000): in the narrow-cone case, only the first
and second modes are relevant (and $\,\omega_1\,\approx \,\omega$). 

Often the overall dissipation rate, and therefore the relaxation rate is determined 
mostly by harmonics rather than by the principal frequency. This fact was discovered 
only recently (Efroimsky \& Lazarian 2000, Efroimsky 2000, Lazarian \& Efroimsky 1999), and 
it led to a considerable re-evaluation of the effectiveness of the 
inelastic-dissipation mechanism. In some of the preceding publications, its 
effectiveness had been underestimated by several orders of magnitude, and the main 
reason for this underestimation was neglection of the second and higher harmonics. As
for the choice of values of the quality factor $\;Q\,$, Prendergast (1958) and
Burns \& Safronov (1973) borrowed the terrestial seismological data for $Q$. 
In Efroimsky \& Lazarian (2000), we argue that these data may be inapplicable to 
asteroids.

To calculate the afore mentioned average energies $\,<W(\omega_n)>\,$, we use 
such entities as stress and strain. As already mentioned above, the stress is a tensorial 
generalisation of the notion of pressure. The strain tensor is analogous to the stretching 
of a spring (rendered in dimensionless fashion by relating the displacement to the base 
length). Each tensor component of the stress consists of two inputs, elastic and plastic. 
The former is related to the strain through the elasticity constants of the material; the 
latter is related to the time-derivative of the strain, through the viscosity coefficients.
As our analysis is aimed at extremely small deformations of cold bodies, the 
viscosity may well be neglected, and the stress tensor will be approximated, to a 
high accuracy, by its elastic part. Thence, according to Landau \& Lifshitz (1976), the 
components of the elastic stress tensor $\sigma_{\it{ij}}$ are interconnected 
with those of the strain tensor $\epsilon_{\it{ij}}$ like:
\be
\epsilon_{ij} \; \; = \; \; \delta_{ij} \; \; \frac{Tr \; 
\sigma}{9 \; K} \; \; + \; \; 
\left( \; \sigma_{\it{ij}} \; \; - \; \; \frac{1}{3} \; \; \delta_{ij} 
\; \; Tr \; \sigma \right) \; \frac{1}{2 \; \mu} \; \; \; ~~~,
\label{4.6}
\ee
$\mu$ and $K$ being the {\it{adiabatic}} shear and bulk moduli, and $Tr$ 
standing for the trace of a tensor. 

To simplify the derivation of the stress tensor, the body will be modelled  by 
a rectangular prism of dimensions $\,2\,a\,\times\,2\,b\,\times\,2\,c\,$ where 
$\,a\,\ge\,b\,\ge\,c$. The tensor is symmetrical and is defined by 
\be
\partial_{i}\sigma_{ij}\;=\;\rho\;a_j\;\;,\;\;
\label{4.12}
\ee
$a_j$ being the time-dependent parts of the acceleration components, and 
$\,\rho\,a_j$ being the time-dependent parts of the components of the 
force acting on a unit volume\footnote{Needless to say, these acceleration 
components $\,a_j\,$ are not to be mixed with $\,a\,$ which is the longest 
dimension of the prism.}. Besides, the tensor $\,\sigma_{ij}$ must obey the 
boundary conditions: its product by normal unit vector, $\,\sigma_{ij}n_j\,$, 
must vanish on the boundaries of the body (this condition was 
not fulfilled in Purcell (1979)). 

Solution to the boundary-value problem provides such a distribution of the 
stresses and strains over the body volume that an overwhelming share of 
dissipation is taking place not near the surface but in the depth of the body. 
For this reason, the prism model gives a good approximation to realistic 
bodies. Still, in further studies it will be good to generalise our solution to
ellipsoidal shapes. The first step in this direction has been made by Molina,
Moreno \& Martinez-L{\'o}pez (2002).

Equation (\ref{4.12}) has a simple scalar analogue\footnote{This example was 
kindly offered to us by William Newman.}. 
Consider a non-rotating homogeneous liquid planet of radius $\,R\,$ and density
$\,\rho\,$. Let $\;g(r)\;$ and $\,P(r)\,$ be the free-fall acceleration and the
self-gravitational pressure at the distance $\,r\,\le \,R\,$ from the centre. 
(Evidently, $\,g(r)\,=\,(4/3) \, \pi \, G \, \rho \, r\,$.) Then the analogue 
to (\ref{4.12}) will read:
\be
\rho \; g(r) \;=\;-\;\frac{\partial P(r)}{ \partial r} \;\;\;\;\;\;,\;\;\;\;
\label{444}
\ee
the expression $\;\rho \, g(r)\;$ standing for the gravity force acting upon a 
unit volume, and the boundary condition being $\;P(R)\,=\,0$. Solving equation 
(\ref{444}) reveals that the pressure has a maximum at the centre of the 
planet, although the force is greatest at the surface. Evidently, the maximal 
deformations (strains) also will be experienced by the material near the centre
of the planet. 

In our case, the acceleration $\,\bf \vec a\,$ of a point inside the precessing
body will be given not by the free-fall acceleration $\,g({\bf{ \vec r}})\,$ 
but will be a sum of the centripetal and transversal accelerations: $\,{\bf{ 
\vec \Omega}}\,\times\,( {\bf{ \vec \Omega}}\,\times\,{\bf{ \vec r}} )\,+\,{\bf
{\dot{ \vec \Omega}}}\,\times\,{\bf{ \vec r}}\;$, the Coriolis term being 
negligibly small. Thereby, the absolute value of $\,{\bf{ \vec a}}\,$ will be 
proportional to that of $\,\bf  \vec r\,$, much like in the above example. In 
distinction from the example, though, the acceleration of a point inside a 
wobbling top will have both a constant and a periodic component, the latter 
emerging due to the precession. For example, in the case of a symmetrical 
oblate rotator, the precessing components of the angular velocity $\,\bf \vec 
\Omega\,$ will be proportional to $\,\sin \omega t\,$ and $\,\cos \omega t\,$, 
whence the transversal acceleration will contain frequency $\,\omega\,$ while 
the centripetal one will contain $\,2 \omega$. The stresses obtained through 
(\ref{4.12}) will oscillate at the same frequencies, and so will the strains. 
As we already mentioned, in the case of a non-symmetrical top an infinite 
amount of harmonics will emerge, though these will be obertones not of the 
precession rate $\,\omega\,$ but of some different ''base frequency'' $\,
\omega_1\,$ that is less than $\,\omega.$

Here follows the expression for the (averaged over a precession period) elastic
energy stored in a unit volume of the body:
\ba
\frac{d\;\,<W>}{dV}\;=\;\frac{1}{2}\;\,<\epsilon_{ij}\;\sigma_{ij}>\;=
\;\frac{1}{4 \mu}\; 
\left\{ \left(\frac{2 \; \mu}{9\; K} \; - \; \frac{1}{3} \protect\right) \;
\,<\,\left(Tr\;\sigma \protect\right)^2\,>\;+\;<\sigma_{ij}\,\sigma_{ij}> 
\protect\right\} \; = 
\nonumber \\
\nonumber \\
\frac{1}{4\mu}\;\left\{\,-\,\frac{1}{1\,+\,\nu^{-1}}\;\,<\left(Tr\;\sigma
\protect\right)^2>\,+\,<\sigma_{xx}^2>\,+\,<\sigma_{yy}^2>\,+\,<\sigma_{zz}^2
> \, + \, 2 \,<\,\sigma_{xy}^2 \, + \, \sigma_{yz}^2 \, + \, 
\sigma_{zx}^2 \,> \protect\right\}\;\;
\label{4.7}
\ea 
where $2\mu/(9K)-1/3= -\nu/(1+\nu) \approx -1/5$, $\,\nu$ being
Poisson's ratio (for most solids $\,\nu \approx 1/4$). Naturally\footnote{Very naturally indeed, 
because, for example, $\,\sigma_{xx} \epsilon_{xx} dV\,=\,(\sigma_{xx}\,dy\,dz)(\epsilon_{xx}\,dx)\,$
is a product of the $\,x$-directed pressure upon the $\,x$-directed elongation of the elementary
volume $dV$.}, the total averaged 
elastic energy is given by the integral over the body's volume: 
\be
<W>\;=\;\frac{1}{2}\;\int\;dV\;\sigma_{ij}\;\epsilon_{ij} \;\;\;\;,\;\;\;
\label{4.8}
\ee
and it must be expanded into the sum (\ref{4.9}) of inputs from oscillations of 
stresses at different frequencies. Each term $\,\bra W(\omega_n) \ket\,$ emerging 
in that sum will then be plugged into the expression (\ref{4.10}), together with the 
value of $\,Q\,$ appropriate to the overtone $\,\omega_n$.

\section{\underline{NONLINEARITY, CHAOS, SEPARATRIX}}
\subsection{~~The ~Origin ~of ~the ~Nonlinearity}

When (\ref{4.11}) is inserted into (\ref{2.4}) and (\ref{2.5}), one can 
explicitly see the contributions to the
entire effect, coming from the principal frequency $\,\omega_1\,$ and from the 
harmonics $\,\omega_n\,\equiv\,n\,\omega_1\,$. When vector $\bf{ \vec \Omega}$ 
describes approximately circular trajectories on Figure 1, the principal 
frequency
$\,\omega_1\,$ virtually coincides with the precession rate $\,\omega\,$. 
This doesn't hold, though, when $\bf{ \vec \Omega}$ get closer to the separatrix: there $\,\omega_1
\,$ becomes {\it{lower}} than the precession rate. The analysis of 
the stress and strain distributions, and the resulting expressions for $\,d\,
<W>/dt\,$ written down in (Efroimsky \& Lazarian 2000) and (Efroimsky 
2000) shows that the nonlinearity is essential, in that the
generation of harmonics is not a high-order effect but a phenomenon playing a 
key role in the relaxation process. In other words, dissipation associated with
the harmonics is often of the same order as that at the principal 
frequency. Near the separatrix it may be even higher.

The nonlinearity emerges due to the simple fact that the acceleration of a 
point within a wobbling object contains centrifugal terms that are quadratic in
the angular velocity $\;{\bf{\Omega}}\;$. In neglect of small terms caused by 
the body deformation, the acceleration will read:
\be
{\bf{\vec a}}\;\;=\;\;{\bf{{\dot{\vec \Omega}}}}\;\times\;{\bf{\vec r}}\;+ \; 
{\bf{\vec \Omega}} \; \times \; ( {\bf{\vec \Omega}} \times {\bf{\vec r}}) \; \; \; \; .
\label{2.9}
\ee
${\bf{a}} \,$ being the acceleration in the inertial frame, and  $\, {\bf{\vec r}}
\,$ being the position of a point. In the simpliest case of oblate symmetry, 
the 
body-frame-related components of the angular velocity are expressed by 
(\ref{1.3}) plugging whereof into (\ref{2.9}) produces terms containing 
$\,\sin \omega t \,$ and  $\,\cos  \omega t \,$, as well as those containing 
$\,\sin 2  \omega t \,$ and  $\,\cos 2  \omega t \,$. The alternating stresses 
and strains caused by this acceleration are linear functions of $\,\bf a\,$ and, thus, will also contain the second harmonic, along with the principal frequency. 
Calculation of the stresses, strains, and of the appropriate elastic energy $\,W\,
$ is then only a matter of some elaborate technique. This technique (presented in 
(Efroimsky \& Lazarian 2000) and (Efroimsky 2001)) leads to an expression for $\,W
\,$, with contributions from $\,\omega\,$ and $\,2 \omega\,$ explicitly separated.
The nonlinearity is essential: in many rotation states the $\,2 \omega$ input in  
(\ref{2.4}), (\ref{2.5}). is of order and even exceeds that coming from the 
principal frequency $\,\omega$. To explain in brief the reason why the 
nonlinearity is strong, we would mention that while the acceleration and the 
stresses and strains are quadratic in the (precessing) angular velocity $\,\bf 
\vec \Omega\,$, the elastic energy is proportional to the product of stress and 
strain tensors. Hence the elastic energy is proportional to the fourth power of $
\,\bf \vec \Omega\,$.
                   
\subsection{~~The ~Near-Separatrix ~Slowing-Down ~of ~the ~Precession\\
(Lingering ~Effect)}                 
                                                                              
In the general case of a triaxial rotator, precession 
is described by (\ref{1.5}) or (\ref{1.6}). The acceleration of a point inside 
the body (and, therefore, the stresses and strains in the material) will,     
according to (\ref{2.9}), contain terms quadratic in the Jacobi functions.    
These functions can be decomposed in converging series (the so-called nome    
expansions) over $\sin$'es and $\cos$'ines of $\,n \nu$, $n\,$ being odd      
integers for $\;{\it{sn( \omega t,\,k^2)}}\;$ and $\;{\it{cn( \omega t,\,k^2)}}
\;$  and even integers for $\;{\it{dn( \omega t,\,k^2)}}\;$. Here $\;\nu\;$ is 
a frequency {\it{lower}} than the precession rate $\;\omega\,$:              
\be                                                                           
\nu\;=\;\omega\;\frac{2\,\pi}{4\,K(k^2)} \;\;\;\;\;,\;\;\;                    
\label{3.2}                                                                   
\ee                                                                           
$4K(k^2)$ being the mutual period of ${\it{sn}}$ and $                        
{\it{cn}}$. Near the poles $\nu \rightarrow \omega $,                         
while on approach to the separatrix $ \nu \rightarrow 0$. When two such      
expansions get multiplied by one another, they produce a series               
containing all sorts of products like $(\sin\,$m$\nu t\;\sin\,$n$\nu t)\;$, $\; 
(\cos\,$m$\nu t\;\cos\,$n$\nu t)$, and cross terms. Hence the                 
acceleration, stress and strain contain the entire multitude of overtones. 
Even though the further averaging of $W$ over the precession cycle weeds
out much of these terms, we are eventually left with all the harmonics on our hands. 

As explained in (Efroimsky 2000), higher-than-second harmonics will bring
only high-order contributions to the precession-relaxation process when the 
rotation state is described by a point close to poles A or C. Put differently, 
it is sufficient to take into account only the frequencies $\nu \approx \omega
$ and $2 \nu \approx 2 \omega $, insofar as the trajectories on 
Figure 1 are approximately circular (i.e., when (\ref{1.4}) and (\ref{1.5}) are 
well approximated by (\ref{1.2}) and (\ref{1.3})). Near the separatrix the 
situation is drastically 
different, in that all the harmonics become important. We 
thus transit from the domain of essential nonlinearity into the regime of 
extreme nonlinearity, regime where the higher harmonics bring more in the 
process than $\nu$ or $2 \nu $. We are reminded, however, that en route
to the separatrix we not just get all the multiples of the principal frequency,
but we face a change of the principal frequency itself: according to 
(\ref{3.2}), the principal frequency $ \nu $ will be lower than the 
precession rate $ \omega $! This regime may be called 
 ``exotic nonlinearity''. 

Without getting bogged down in involved mathematics (to be attended to 
in subsection 4.3 below), we would just mention here that in the limit of 
$\bf{ \vec \Omega}$ approaching the separatrix the dissipation rate will vanish,
in the adiabatic approximation. This may be guessed even from the fact that in 
the said limit $\nu \rightarrow 0$. We thus come to an important conclusion 
that the relaxation rate, being very high at a distance from the separatrix, 
decreases in its closemost vicinity. Can we, though, trust that the relaxation
rate completely vanishes on the separatrix? No, because in the limit of $\bf{ \vec
\Omega}$ approaching the separatrix the adiabatic approximation will fail. 
In other words, it will not be legitimate to average the energy 
dissipation over the precession cycle, because near the separatrix the 
precession rate will not necessarily be faster than the relaxation rate. A 
direct calculation shows that even on the separatrix itself the acceleration
of a point within the body will remain finite (but will, of course, vanish at 
the unstable middle-inertia pole). The same can be said about stress,
strain and the relaxation rate. So what we eventually get is not a 
near-separatrix trap but just lingering: one should expect relaxing tops to
considerably linger near the separatrix. As for Explorer, it is now 
understandable why it easily went wobbling but did not rush to the 
minimal-energy spin state: it couldn't cross the separatrix so quickly. We would call 
it "lingering effect". There is nothing mysterious in it. The capability of 
near-intermediate-axis spin states to mimic simple rotation was pointed out by 
Samarasinha, Mueller \& Belton (1999) with regard to comet Hale-Bopp. A similar setting 
was considered by Chernous'ko (1968) who studied free precession of a tank filled with 
viscous liquid and proved that in that case the separatrix is crossed within a finite 
time interval\footnote{Such problems have been long known also to mathematicians 
studying the motion with a non-Hamiltonian perturbation: the perturbation wants the 
system to cross the separatrix, but is not guaranteed to succeed in it, because
some trajectories converge towards the unstable pole (Neishtadt 1980)} .

\section{\underline{THE APPLICABILITY REALM OF THE ADIABATIC APPROACH}}

In the beginning of subsection 1.2 we explained that
Euler's equations (\ref{1.1}) may be written down in their approximate
form (\ref{22.4}) in case the nutation-caused deformations of the body are
negligibly small. Thus it turns out that in our treatment the same phenomenon is neglected in one
context and accounted for in another: on the one hand, the very process of the 
inelastic dissipation stems from the precession-inflicted small deformations;  
on the other hand, we neglect these deformations in order to write down
(\ref{22.4}). This approximation (also discussed in Lambeck 1988) may be called
adiabatic, and it remains acceptable insofar as the relaxation is slow against 
rotation and precession. To cast the adiabatic approximation into its exact 
form, one should first come up with a measure of the relaxation rate. Clearly,
this should be the time derivative of the angle $\;\theta\;$ made by the 
major-inertia axis $\;3\;$ and the angular momentum $\;\bf \vec J\;$. The axis 
aligns towards $\;\bf \vec J\;$, so $\;\theta\;$ must eventually decrease. Be 
mindful, though, that even in the absence of dissipation, $\;\theta\;$ does 
evolve in time, as can be shown from the equations of motion. Fortunately, this
evolution is periodic, so one may deal with a time derivative of the angle 
averaged over the precession period. In practice, it turns out to be more 
convenient to deal with the squared sine of $\theta$ (Efroimsky 2000) and 
to write the adiabaticity assertion as:
\be
-\;\frac{d\,\bra \sin^2 \theta \ket }{dt}\,\ll\,\omega\;\;,
\label{22.5}
\ee
$\omega$ being the precession rate and $\,<...>\,$ being the average  
over the precession period. The case of an oblate symmetrical 
top\footnote{Hereafter oblate symmetry will imply not a 
geometrical symmetry but only the so-called dynamical symmetry: $\,I_1\,=\,
I_2$.} is exceptional, in that $\,\theta \,$ remains, when dissipation is 
neglected, constant over a precession cycle. No averaging is needed, and 
the adiabaticity condition simplifies to:
\be
-\;\left(\frac{d\,\theta}{dt}\right)_{(oblate)}\,\ll\,\omega\;\;.
\label{22.6}
\ee
We would emphasise once again that the distinction between the oblate and triaxial
cases, distinction resulting in
the different forms of the adiabaticity condition, stems from the difference in
the evolution of $\,\theta \,$ in the weak-dissipation limit. The equations of 
motion of an oblate rotator show that, in the said limit, $\,\theta \,$ stays 
virtually unchanged through a precession cycle (see section IV below). So the 
slow decrease of $\,\theta \,$, accumulated over many periods, becomes an  
adequate measure for the relaxation rate. The rate remains slow, compared to 
the rotation and precession, insofar as (\ref{22.6}) holds. In 
the general case of a triaxial top the equations of motion show that, even in 
the absence of dissipation, angle $\,\theta \,$ periodically evolves, though 
its average over a cycle stays unchanged (virtually unchanged, when 
dissipation is present but weak)\footnote{See formulae (A1) - (A4) in the 
Appendix to Efroimsky 2000.}. In this case we should measure the
relaxation rate by the accumulated, over many cycles, change in the average of
$\,\theta\,$ (or of $\,\sin^2 \theta\,$). Then our assumption about the 
relaxation being slow yields (\ref{2.5})

The above conditions (\ref{22.5}) - (\ref{22.6}) foreshadow the applicability 
domain of our further analysis. For example, of the two quantities,
\be
I_1^2 \, {\Omega}_1^2 \; + \; I_2^2 \, {\Omega}_2^2 \; + \; I_3^2 \, 
{\Omega}_3^2 \; = \; {\bf{\vec J}}^2 \; \; \; , \;\;
\label{22.7}
\ee
\be
I_1 \, {\Omega}_1^2 \; + \; I_2 \, {\Omega}_2^2 \; + \; I_3 \, 
{\Omega}_3^2 \; = \; 2 \; T_{\small{kin}}\;\;\;,\;\; 
\label{22.8}
\ee
only the former will conserve exactly, while the latter will remain virtually 
unchanged through one cycle and will be gradually changing through many cycles 
(just like $\;\;\bra \sin^2 \theta \ket\;\,$).

\section{\underline{SYMMETRIC AND ASYMMETRIC ROTATORS}}

\subsection{{Precession of an Oblate Body.}}

An oblate body has moments of inertia that relate as:
\be
I_3\;>\;I_2\;=\;I_1\;\equiv\;I\;\;.\;\;\;\;\;\;
\label{5.1}
\ee
We shall be interested in $\dot{\theta}$, the rate of the maximum-inertia axis'
approach to the direction of angular momentum $\bf{\vec J}$. To achieve this goal,
we shall have to know the rate of energy losses caused by the periodic 
deformation. To calculate this deformation, it will be necessary to find the 
acceleration experienced by a particle located inside the body at a point ($x$,
$y$, $z$). Note that we address the inertial acceleration, i.e., the one with 
respect to the inertial frame $(X,Y,Z)$, but we express it in terms of 
coordinates $x$, $y$ and $z$ of the body frame $(1,2,3)$ because eventually we
shall have to compute the elastic energy stored in the entire body (through 
integration of the elastic-energy density over the body volume). 

The fast motions (revolution and precession) obey, in the adiabatical 
approximation,  the simplified Euler equations (\ref{2.4}). Their solution,  
with neglect of the slow relaxation, looks (Fowles and Cassiday 1986, Landau and Lifshitz 1976), in the oblate case (\ref{5.1}):
\be
{\Omega}_1 \; \; = \; \; {\Omega}_{\perp} \cos {\omega}t~~,~~~
{\Omega}_2 \; \; = \; \; {\Omega}_{\perp} \sin {\omega}t~~,~~~
{\Omega}_3 \; \; = \; \; const
\label{5.2}
\ee
where
\be
{{\Omega}_{\perp}} \; \; \equiv \; \; {\Omega} \; \; \sin \; {\alpha}~~, ~~~~~
{{\Omega}_{3}} \; \; \equiv \; \; {\Omega} \; \; \cos \; {\alpha}\;\;\;,
\label{5.3}  
\ee
$\;\alpha\;$ being the angle made by the major-inertia axis 3 with $\,\bf{ \vec 
\Omega}\,$. Expressions (\ref{5.2}) show that in the body frame the angular 
velocity ${\bf{ \vec \Omega}}$ describes a circular cone about the principal axis
$3$ at a constant rate 
\be
\omega \; = \; (h - 1) \Omega_3, 
\; \; \; \; \; \; \; \;  h\equiv {I_3}/I~~~. 
\label{5.4}
\ee
So angle $\;\alpha\;$ remains virtually unchanged through a cycle (though in the
presence of dissipation it still may change gradually over many cycles). 
The precession rate $\;\omega \;$ is of the same order as $|{\bf{\Omega}}|$, 
except in the case of $\;h\,\rightarrow\,1\;$ or in a very special case of 
${\bf{\Omega}}$ and ${\bf{J}}$ being orthogonal or almost orthogonal to the 
maximal-inertia axis $3$. Hence one may call not only the rotation but also the
precession ``fast motions'' (implying that the relaxation process is a slow 
one). Now, let $\;\theta\;$ be the angle between the principal axis $3$ and the
angular-momentum $\;\bf  \vec J\;$, so that $\;J_3\,=\,J\,\cos \theta\;$ and
\be
{\Omega}_3 \; \; \equiv \; \; \frac{J_3}{I_3} \; \; = \; \;\frac{J}{I_3} \; \; 
\cos \; \theta~~ ~~~~
\label{5.5}
\ee
wherefrom
\be
\omega\;=\;(h\;-\;1)\;\frac{J}{I_3}\;\cos \theta\;\;\;.
\label{5.6}
\ee
Since, for an oblate object,
\be 
{\bf  \vec J}\,=\,I_1\,\Omega_1\,{\bf e}_1\,+\,I_2\,\Omega_2\,{\bf e}_2\,+\,I_3\,
\Omega_3\,{\bf e}_3\,=\,I\,(\Omega_1\,{\bf e}_1\,+\,\Omega_2\,{\bf e}_2)\,+\,
I_3\,\Omega_3\,{\bf e}_3\,\;,\;
\label{5.7}
\ee
the quantity $\;\Omega_{\perp}\,\equiv\,\sqrt{\Omega_1^2+\Omega_2^2}\;$ is 
connected with the absolute value of $\bf J$ like:
\be
{\Omega}_{\perp} \; \; = \;\;\frac{J}{I}\;\;\sin\;\theta\;\;=\; \; \frac{J}{I_3} \; h \; \; \sin \; \theta\;\;\;,\;\;\;\;\;\;h\;\equiv\;I_3/I\;\;.\;\;
\label{5.8}
\ee
It ensues from (\ref{5.3}) that $\;{{\Omega}_{\perp}}/{{\Omega}_{3}}\;=\; 
\tan \; \alpha \;$. On the other hand, (\ref{5.5}) and (\ref{5.8}) entail: $\;
{{\Omega}_{\perp}}/{{\Omega}_{3}}\,=\,h\,\tan \, \theta\;$. Hence,
\be
\tan \alpha\;=\;h\;\tan \theta\;\;\;.\;\;\;
\label{5.9}
\ee
We see that angle $\;\theta\;$ is almost constant too (though it gradually 
changes through many cycles). We also see 
from (\ref{5.7}) that in the body frame the angular-momentum vector $\;\bf J\;$
describes a circular cone about axis $3$ with the same rate $\;\omega\;$ as 
$\;\Omega\;$. An inertial observer, though, will insist that it is rather axis 
$3$, as well as the angular velocity $\;\Omega\;$, that is describing circular
cones around $\bf{\vec J}$. It follows trivially from (\ref{5.4}) and (\ref{5.7}) 
that
\be
I\;{\bf  \vec \Omega}\;=\;{\bf  \vec J}\;-\;I\;\omega\;{\bf  \vec e}_3 \;\;\;,\;\;\;\;
\label{5.10}
\ee
whence it is obvious that, in the inertial frame, both $\bf  \vec \Omega$ and 
axis $\;3\;$ are precessing about $\bf  \vec J$ at rate $J/I$. (The angular 
velocity of this precession is ${\dot{\bf  \vec e}}_3\,=\,{\bf  \vec \Omega} \,\times\,{
\bf  \vec e}_3\,=\,({\bf  \vec J}/I\,-\,\omega\,{\bf  \vec e}_3)\,\times\,{\bf  \vec e}_3\,=\,({\bf 
J}/I)\,\times\,{\bf  \vec e}_3$.) Interestingly, the rate $\omega\,=\,(h\,
-\,1)\,\Omega_3$, at which $\bf  \vec \Omega$ and $\bf  \vec J$ are precessing 
about axis $\;3\;$ in the body frame, differs considerably from the rate $
J/I$ at which $\bf  \vec \Omega$ and axis $\;3\;$ are precessing around $\bf  \vec J$ in 
the inertial frame. (In the case of the Earth, $J/I\,\approx\,400\,
\omega\;$ because $\,h\,$ is close to unity.) Remarkably, the 
inertial-frame-related precession rate is energy-independent and, thus, stays 
unchanged through the relaxation process. This is not the case for the 
body-frame-related rate $\,\omega\,$ which, according to (\ref{5.6}), gradually
changes because so does $\,\theta$. 

As is explained above, we shall 
be interested in the body-frame-related components $\Omega_{1,2,3}$ precessing 
at rate $\omega$ about the principal axis $\;3$. Acceleration of an arbitrary 
point of the body can be expressed in terms of these components through  
formula
\be
{\bf{ \vec a}} \; \; = \; \; {\bf{ \vec a}}' \; + \;{\bf{{\dot{ \vec \Omega}}}} \; \times \;
{\bf{ \vec r}}' \; + \; 2 \; {\bf{ \vec \Omega}} \; \times \; {\bf{ \vec v}}' \; + \; 
{\bf{ \vec \Omega}} \; \times \; ( {\bf{ \vec \Omega}} \times {\bf{ \vec r}}' ) \; \; \; \; , 
\label{5.11}
\ee
where $\; {\bf{ \vec r}}, \; {\bf{ \vec v}}, \; {\bf{ \vec a}} \;$ are the position, 
velocity and acceleration in the inertial frame, and  $\; {\bf{ \vec r}}', \; 
{\bf{ \vec v}}' \;$ and $\; {\bf{ \vec a}}' \;$ are those in the body frame. Here $\;
{\bf  \vec r}\,=\,{\bf  \vec r}'\;$. Mind though that $\;{\bf{ \vec v}}'\;$ and $\;{\bf{ \vec a}}'\;$ 
do not vanish in the body frame. They may be neglected on the same grounds as  
term $\;{\dot{I}}_i\Omega_i\;$ in (\ref{2.1}): precession of a body of 
dimensions $\;\sim\,{\it l}\;$, with period $\;\tau\;$, leads to 
deformation-inflicted velocities $\;v'\,\approx\;\epsilon\,\it{l}/\tau\;$ 
and accelerations $\;a'\,\approx\;\epsilon\,\it{l}/\tau^{2}\;$, $\;
\epsilon \;$ being the typical order of strains arising in the material. 
Clearly, for very small $\;\epsilon\;$, quantities $\; v' \;$ and $\; a'\;$ are
much less than the velocities and accelerations of the body as a whole (that 
are about $\;{\it{l}}/\tau\;$ and $\;{\it{l}}/{\tau}^2\;$, correspondingly). 
Neglecting these, we get, from (\ref{5.11}) and (\ref{5.2}), for the acceleration at 
point $\;(x,\,y,\,z)\;$:
\ba
\nonumber
{\bf{ \vec a}} \; \; = \; \; {{\bf{ \vec e}}_1} \; \left\{ \;
{\frac{1}{2}} \; {\Omega}_{\perp}^2 \; x \; \; \cos \; 2{\omega}t \; \; + \;\; 
{\frac{1}{2}} \; {\Omega}_{\perp}^2 \; y \; \; \sin \; 2{\omega}t \; \; + \; \;
z \; {\Omega}_{\perp} \; {\Omega}_3 \; h \; \; \cos \; {\omega}t  
\; \right\}\;+\\
\nonumber\\
\nonumber
+ \; \; {{\bf{ \vec e}}_2} \; \left\{
{\frac{1}{2}} \; {\Omega}_{\perp}^2 \; x \; \; \sin \; 2{\omega}t \; \; - \; \;
{\frac{1}{2}} \; {\Omega}_{\perp}^2 \; y \; \; \cos \; 2{\omega}t \; \; + \; \;
z \; {\Omega}_{\perp} \; {\Omega}_3 \; h \; \; \sin \; {\omega}t \; 
\; \right\}\;+\\  
\nonumber  \\
+ \; \; {{\bf{ \vec e}}_3} \; \left\{
{\Omega}_{\perp} \; {\Omega}_3 \; (2 \; - \; h) \; (x \; \; \cos \; {\omega}t 
\; \; + \; \; y \; \; \sin \; {\omega}t \; )
\; \right\}~~~.
\label{5.12}
\ea
Substitution thereof into (\ref{4.12}), with the proper boundary conditions 
imposed, 
yields, for an oblate prism of dimensions $\,2a\,\times\,2a\,\times\,2c\;$, 
$\;a\,>\,c\,$, to the following {\bf approximate} expressions:
\be
\sigma_{xx} = \frac{\rho \Omega_\perp^2}{4}  
(x^2 - a^2) \; \cos 2 {\omega} t  \; \; , \; \; \; \;  
\sigma_{yy} = - \frac{\rho \Omega_\perp^2}{4} (y^2 - a^2) \;  
\cos 2 {\omega} t \; \; , \; \; \; \;  
\sigma_{zz} = 0 
\label{5.13}
\ee
\be
\sigma_{xy} \; \; = \; \; \frac{\rho}{4} \; \Omega_\perp^2 \; 
(x^2 \; \; + \; \; y^2 \; \; - \; \; 2a^2)
\; \; \sin \; 2 {\omega} t \; \; \; ,
\label{5.14}
\ee
\be
\sigma_{xz} \; \; = \; \; \frac{\rho}{2} \; {\Omega_\perp} \; {\Omega_3} \; 
\left[ \; h \; (z^2 \; - \; c^2)
\; \; + \; \; (2 \; - \; h) \; (x^2 \; - \; a^2) \; \right] \; \; 
\cos \; {\omega} t \; \; \; ,
\label{5.15}
\ee
\be
\sigma_{yz} \; \; = \; \; \frac{\rho}{2} \; {\Omega_\perp} \; {\Omega_3} \; 
\left[ \; h \; (z^2 \; - \; c^2)
 \; \; + \; \; (2 \; - \; h) \; (y^2 \; - \; a^2) \; \right] \; \; 
\sin \; {\omega} t \; \; \; .
\label{5.16}
\ee
In (\ref{5.12}) - (\ref{5.16}) we kept only time-dependent parts, because 
time-independent parts of the acceleration, stresses and strains are irrelevant
in the context of dissipation. A detailed derivation of (\ref{5.12}) - 
(\ref{5.16}) is presented in (Lazarian \& Efroimsky 1999). 

Formulae (\ref{5.13}) - (\ref{5.16})
implement the polynomial approximation to the stress tensor. This approximation
keeps the symmetry and obeys (\ref{4.12}) with (\ref{5.12}) plugged into it. The 
boundary condition are satisfied 
exactly for the diagonal components and only approximately for 
the off-diagonal components. The approximation considerably simplifies
calculations and entails only minor errors in the numerical factors in 
(\ref{5.21}).

The second overtone 
emerges, along with the principal frequency $\,\omega\,$, in the expressions for stresses 
since the centripetal part of the acceleration is quadratic in $\; \bf{\Omega}
\;$. The kinetic energy of an oblate spinning body reads, according to
(\ref{1.8}), (\ref{5.3}), and (\ref{5.9}):
\begin{eqnarray}
T_{kin}\;=\;\frac{1}{2}\;\left[I\;\Omega_{\perp}^2\;+\;I_3\;{\Omega_3}^2\right]
\;=\;\frac{1}{2}\;\;\left[\;\frac{1}{I}\;\;\sin^2\;\theta\;\;+\;\; 
\frac{1}{I_3} \; \; \cos^2 \; \theta \; \right] \; J^2\;\;,\;\;\;
\label{5.17}
\end{eqnarray}
wherefrom
\be
\frac{dT_{kin}}{d\theta} \; \; = \; \; 
\frac{J^2}{I_3} \; (h \; - \; 1) \; \; \sin \; \theta \; \; \cos \; \theta 
\; \; =  \; \; \omega \; J \; \; \sin \; \theta \; \; \; .
\label{5.18}
\ee
The latter expression, together with (\ref{2.5}) and (\ref{4.7}), leads to:
\be
\frac{d\theta}{dt} \; = \; \left(\frac{dT_{kin}}{d\theta}\right)^{-1} 
\frac{dT_{kin}}{dt} \; = \; \left( \omega \; J \; \; \sin \; \theta 
\right)^{-1} \; \dot{W}\;\;,
\label{5.19}
\ee
where
\ba
\dot{W}\;= \; \dot{W}^{({\omega})}\; + \; \dot{W}^{(2{\omega})} \; = \;-
\; \omega \; \frac{W_0^{({\omega})}}{Q^{({\omega})}} \; - \; 2 \; \omega \; 
\frac{W_0^{({2\omega})}}{Q^{({2\omega})}}\;\approx 
\\
\nonumber\\
\approx
\; -\;\frac{2\,\omega}{Q} \; 
\left\{ <W^{({\omega})}> \; + \; 2\;< W^{({2\omega})}> \protect\right\}\;, 
\label{5.20}
\ea
the quality factor assumed to depend upon the frequency very 
weakly\footnote{The $\omega$-dependence of $\;Q\;$ should be taken into account
within frequency spans of several orders, but is irrelevant for frequencies 
differing by a factor of two.}. In the above formula, $\;W_0^{\omega}\;$ and 
$\;W_0^{2\omega}\;$ are amplitudes of elastic energies corresponding to the 
principal mode and the second harmonic. Quantities $<W^{\omega}>\,=\,W_0^{
\omega}/2\,$ and $\,<W^{2\omega}>\,=\,W_0^{2\omega}/2\,$ are the appropriate 
averages. Substitution of (\ref{5.13}) - (\ref{5.16}) into (\ref{4.7}), with 
further integration over the volume and plugging the result into (\ref{2.5}), 
will give us the final expression for the alignment rate:
\be
d \theta/dt \; = \; - \; \frac{3}{2^4} \; \sin^3 \theta \; \; \frac{63 \; 
(c/a)^4 \; \cot^2 \theta \; + \; 20}{[1+(c/a)^2]^4} \; \; \frac{a^2 \; 
\Omega^3_0 \; \rho}{\mu \; Q}
\label{5.21}
\ee
where 
\be
\Omega_0\;\equiv\;\frac{J}{I_3} \;\;\;\;
\label{5.22}
\ee
is a typical angular velocity. Deriving (\ref{5.21}), we took into account 
that, for an oblate $\;2a\,\times\,2a\,\times\,2c\;$ prism (where $\;a\,>\,c\;
$), the moment of inertia $\;I_3\;$ and the parameter $\;h\;$ read:
\be
I_3 \;  = \; \frac{16}{3} \; \rho \; a^4 \; c\;\;\;\;\;,\;\;\;\;\;\;\;\;
h \; \; \equiv \; \; \frac{I_3}{I} \; \; = \; \; \frac{2}{1 \; + \; (c/a)^2}  
\; \; \; \; \; . \; \; \; 
\label{5.23}
\ee
Details of derivation of (\ref{5.21}) are presented in (Lazarian \& Efroimsky 
1999)\footnote{Our expression (\ref{5.21}) presented here differs from the 
appropriate formula in Lazarian \& Efroimsky (1999) by a factor of 2, because 
in Lazarian \& Efroimsky (1999) we missed the coefficient 2 connecting 
$W_0^{(...)}$ with $W^{(...)}$.}. 

Formula (\ref{5.21}) shows that the major-inertia axis slows down its alignment
at small residual angles. For $\theta\rightarrow 0 $, the derivative ${\dot{\theta
}}\,$ becomes proportional to $\theta$, and thus, $\theta$ decreases 
exponentially slowly: $\theta = A \exp (-\zeta t)$, where $A$ and $\zeta$ are some 
positive numbers\footnote{This resembles the behaviour of a pendulum: if the pendulum 
is initially given exactly the amount of kinetic energy sufficient for the pendulum to 
move up and to point upwards at the end of its motion, then formally it takes an infinite 
time for the pendulum to stand on end.}. This feature, ''exponentially slow finish'', 
(which was also mentioned, with regard to the Chandler wobble, in Peale (1973), formula (55))
is natural for a relaxation process, and does not lead to an infinite relaxation time 
if one takes into account the finite resolution of the equipment. 
Below we shall discuss this topic at length. 

Another feature one might expect of (\ref{5.21}) would be a ``slow start'': 
it would be good if $\;d\theta/dt\;$ could vanish for $\;\theta\,\rightarrow\,\pi/2\;$. 
If this were so, it would mean that at $\;\theta\,=\,\pi/2\;$ (i.e., when the 
major-inertia axis is exactly perpendicular to the angular-momentum vector) the body 
``hesitates'' whether to start aligning its maximal-inertia axis along or opposite to 
the angular momentum, and the preferred direction is eventually determined by some stochastic 
influence from the outside, like (say) a collision with a small meteorite. This behaviour 
is the simplest example of the famous spontaneous symmetry breaking, and in this setting 
it is desirable simply for symmetry reasons: $\;\theta\,=\,\pi/2\;$ must be a position of an unstable 
equilibrium\footnote{Imagine a knife freely rotating about its longest dimension, and let the rotation
axis be vertical. This rotation mode is unstable, and the knife must eventually come to rotation about 
its shortest dimension, the blade being in the horizontal plane. One cannot say, though, which of the 
two faces of the blade will look upward and which downward. This situation is also illustrated by 
the pendulum mentioned 
in the previous footnote: when put upside down on its end, the pendulum ''hesitates'' in what 
direction to start falling, and the choice of direction will be dictated by some infinitesimally 
weak exterior interaction (like a sound, or trembling of the pivot, or an evanescent flow of air).}. Contrary to these 
expectations, though, (\ref{5.21}) leaves $\; d \theta /dt \;$ nonvanishing for $\;
\theta\,\rightarrow\,\pi/2\;$, bringing an illusion that the major axis  
leaves the position $\; \theta = \pi/2 \;$ at a finite rate. This failure of 
our formula (\ref{5.21}) comes from the inapplicability of our analysis in 
the closemost vicinity of $\;\theta\,=\,\pi/2\;$.  This vicinity simply falls
out of the adiabaticity domain of our treatment, because $\;\omega\;$ given
by (\ref{5.6}) vanishes for  $\;\theta\,\rightarrow\,\pi/2\,$ (then one can 
no longer
assume the relaxation to be much slower than the precession rate, and hence, 
the averaging over period becomes illegitimate). This is explained in more detail in the next subsection.

One more situation, that does not fall under the auspices of our analysis, is
 when
$\omega\,$ vanishes due to  $(h\,-\,1)\rightarrow 0$. This happens when $c/a$
approaches unity. According to (\ref{5.21}), it will appear that $d\theta /dt
$ remains nonvanishing for $c/a\rightarrow 1$, though on physical 
grounds the alignment rate must decay to zero because, for $c=a$, 
the body simply lacks a major-inertia axis.
 
All in all, (\ref{5.21}) works when $\theta$ is not too close to 
$\pi/2$ and $c/a$ is not too close to unity:
\be
-\;{\dot{\theta}}\;\ll\;(h\;-\;1)\;\frac{J}{I_3}\;\cos \theta \;=
\frac{1\;-\;(c/a)^2}{1\;+\;(c/a)^2}\;\Omega_0\;\cos \theta\;\;\;.
\label{5.24}
\ee
Knowledge of the alignment rate $\;\dot \theta\;$ as a function of the 
precession-cone half-angle $\;\theta\;$ enables one not only to write down a 
typical relaxation time but to calculate the entire dynamics of the process. In
particular, suppose the observer is capable of measuring the precession-cone  
half-angle $\;\theta\;$ with an error $\;\delta\;$. This observer will then 
compute, by means of (\ref{5.21}), the time needed for the body to change its 
residual half-angle from $\;\theta\;$ to $\;\theta\,-\,\Delta \theta\;$, for 
$\;\Delta\,\theta\,>\,\delta\;$. This time will then be compared with the results of 
his further measurements. Below we shall show that such observations will soon 
become possible for spacecraft.

\subsection{{Precession of an Exactly Symmetrical Prolate Body.}}

At the first glance, the dynamics of a freely-spinning elongated body obeys the
same principles as that of an oblate one: the axis of maximal inertia will 
tend to align itself parallel to the angular momentum. If we assume that the
body is (dynamically) prolate (i.e., that $\; I_3 \; = \; I_2 \; > \; I_1 
\;$), it will, once again, be convenient to model it by a 
prism of dimensions $2a \times 2a \times 2c$, though this time half-size $\, c
 \,$ will be larger than $\, a \,$. Then all our calculations will
{\textit{formally}} remain in force, 
up to formula (\ref{5.18}): since the factor $h-1=[1-(c/a)^2]/[1+(c/a)^2]$ is
now negative, the right-hand side in (\ref{5.18}) will change its sign:
\be
\frac{dT_{kin}}{d\theta} \; \; = \; \; 
\frac{J^2}{I_3} \; (h \; - \; 1) \; \; \sin \; \theta \; \; \cos \; \theta 
\; \; =  \; - \; \omega \; J \; \; \sin \; \theta \; \; \; .
\label{77.1}
\ee
Thereby formula (\ref{5.19}) will also acquire a ``minus'' sign in its 
right-hand side:
\be
\frac{d\theta}{dt} \; = \; \left(\frac{dT_{kin}}{d\theta}\right)^{-1} 
\frac{dT_{kin}}{dt} \; = \; - \; \left( \omega \; J \; \; \sin \; \theta 
\right)^{-1} \; \dot{W}
\label{77.2}
\ee
Formula (\ref{5.20}) will remain unaltered. 
Eventually, by using (\ref{77.2}) and (\ref{5.20}), we shall arrive to a 
formula that differs from (\ref{5.21}) only by a sign, provided $\theta$ still
denotes the angle between $\mathbf{\vec J}$ and the body-frame axis 3 
(parallel to dimension $2c$):
\be
d \theta/dt \; = \;\frac{3}{2^4} \; \sin^3 \theta \; \; \frac{63 \; 
(c/a)^4 \; \cot^2 \theta \; + \; 20}{[1+(c/a)^2]^4} \; \; \frac{a^2 \; 
\Omega^3_0 \; \rho}{\mu \; Q}
\label{77.3}
\ee
This looks as if axis 3 tends to stand orthogonal to 
$\mathbf{\vec J}$, which is natural since axis 3 is now not the 
maximal-inertia but the minimal-inertia axis. 

Alas, all this extrapolation is of marginal practical interest, because
even a small difference between $\; I_2 \;$ and $\; I_3 \;$ leads to a 
considerably different type of rotation. This circumstance was pointed out 
by Black et al (1999) and was comprehensively discussed by Efroimsky (2000).
Without getting into excessive mathematical details, here we shall provide a 
simple qualitative explanation of the effect.

\subsection{Triaxial and {\bf{\it{ almost}}} prolate rotators}

Typically, asteroids and comets have elongated shapes, and the above formulae 
derived for oblate bodies make a very crude approximation of the wobble of a 
realistic triaxial body. In the case of a triaxial rotator, with $\;I_3 \geq I_2 
\geq I_1\;$, the solution to the Euler equations is expressable in terms of
elliptic functions. According to Jacobi (1882) and Legendre (1837), it will 
read, for $\;{\bf{ \vec J}}^2\; < \; 2\;I_2 \; T_{\small{kin}}\;$, as 
\begin{eqnarray}
\Omega_1 \;=\;\gamma\;\,{\it{dn}} \left( \omega t , \; k^2 \protect\right) \;\;,
\;\;\;\; \Omega_2 \; = \; \beta \,\; {\it{sn}} \left( \omega t , \; k^2 \protect\right)
\;\;,\;\;\;\; \Omega_3 \; = \; \alpha \;\, {\it{cn}} \left( \omega t,\;k^2
\protect\right) \;\;,\;\;\;
\label{1}
\end{eqnarray}
while for $\;\;{\bf{ \vec J}}^2\;>\;2\;I_2\;T_{{kin}}\;$ it will be: 
\begin{eqnarray}
\Omega_1\;=\;{\gamma}\;\,{\it{cn}}\left({\omega} t,\;{ k}^2
\protect\right)\;\;,\;\;\;\;\Omega_2\;=\;{\beta}\,\;{\it sn}\left({
\omega}t,\;{ k}^2 \protect\right)\;\;,\;\;\;\;\Omega_3\;=\;{\alpha
}\;\,{\it{dn}}\left({ \omega} t,\;{ k}^2 \protect\right)\;\;.\;\;\;
\label{2}
\end{eqnarray}
Here the precession rate $\,\omega\,$ and the parameters $\;\alpha ,\;\beta , \;
\gamma \,$ and $\;{k}\;$ are some algebraic functions of $\;I_{1,2,3}, \;
T_{\small {kin}}\;$ and $\;{\bf  \vec J}^2\;$. For example, $\;{k}\;$ is expressed by
\begin{eqnarray}
k\;=\;\sqrt{\frac{I_3-I_2}{I_2-I_1}\;\frac{{\bf  \vec J}^2-2I_1 T_{kin}}{2I_3 T_{kin}-{\bf  \vec J}^2}}\;\;\;\;\;,\;\;\;\;
\label{211}
\end{eqnarray}
for (\ref{1}), and by
\begin{eqnarray}
k\;=\;\sqrt{\frac{I_2-I_1}{I_3-I_2}\;\frac{2I_3 T_{kin}-{\bf  \vec J}^2}{{\bf  \vec J}^2-2I_1 T_{kin}}}\;\;\;\;\;,\;\;\;\;
\label{212}
\end{eqnarray}
for (\ref{2}). In the limit of oblate symmetry (when $\;I_2/I_1\,\rightarrow\,1\;$), solution
(\ref{2}) approaches (\ref{5.2}), while the applicability region of (\ref{1})
shrinks. Similarly, in the prolate-symmetry limit ($\,(I_3\,-\,I_2)/I_1\,
\rightarrow\,0\,$) the applicability realm of (\ref{2}) will become 
infinitesimally small. The easiest way of understanding this would be to 
consider, in the space $\;\Omega_1\;,\;\Omega_2\;,\;\Omega_3\;,$ the 
angular-momentum ellipsoid $\;\;{\bf{ \vec J}}^2\;=\;I_1^2\;\Omega_1^2\;+\;I_2^2\;
\Omega_2^2 \; + \; I_3^2 \; \Omega_3^2 \;\;$. A trajectory described by the 
angular-velocity vector $\;\bf \Omega\;$ in the space $\;\Omega_1\;,\; \Omega_2
\;,\;\Omega_3\;$ will be given by a line along which this ellipsoid intersects 
the kinetic-energy ellipsoid $\;\;2\;T_{\small{kin}}\;=\;I_1\;\Omega_1^2 
\;+\;I_2\;\Omega_2^2\;+\;I_3\;\Omega_3^2\;\;,$ as on Fig.1. Through the 
relaxation process, 
the angular-momentum ellipsoid remains unchanged, while
the kinetic-energy ellipsoid evolves as the energy dissipates. Thus, the fast 
process, nutation, will be illustrated by the (adiabatically) periodic motion
of $\;\bf  \vec \Omega\;$ along the line of ellipsoids' intersection; the slow 
process, relaxation, will be illustrated by the gradual shift of the moving vector
$\;\bf  \vec \Omega\;$ from one trajectory to another (Lamy \& Burns 1972). On Fig.1,
we present an angular-momentum ellipsoid for an almost prolate body whose 
angular momenta relate to one another as those of asteroid 433 Eros: $\;1\;
\times\;3\;\times\;3.05\;$ (Black et al. 1999). Suppose the initial energy was 
so high that $\;\bf \vec \Omega\;$ was moving along some trajectory close to the point
A on Fig.1. This pole corresponds to rotation of the body about its minor-inertia
axis. The trajectory described by $\;\bf  \vec \Omega\;$ about $\,A\,$ is almost 
circular and remains so until $\;\bf  \vec \Omega\;$ approaches the separatrix\footnote{As we already mentioned above, this 
trajectory on Fig.1 being almost circular does not necessarily mean that the 
precession cone of the major-inertia axis about $\,\bf  \vec J\,$ is circular or almost 
circular.}. This process will be described by solution (\ref{1}). In the vicinity 
of separatrix, trajectories will become noticeably distorted.
After the separatrix is crossed, librations will begin: $\;\bf  \vec 
\Omega\;$ will be describing not an almost circular cone but an elliptic one. This
process will be governed by solution (\ref{2}). Eventually, in the closemost 
vicinity of pole C, the precession will again become almost circular. (This point,
though, will never 
be reached because the alignment of $\;{\bf{\vec \Omega}}\;$ along (or opposite
to) $\;{\bf{\vec J}}\;$ has a vanishing rate for small residual angles: at the 
end of the relaxation process the relaxation rate approaches zero, so that 
small-angle nutations can persist for long times.) 
Parameter $\;k\;$ shows how far the tip of $\;\bf{ \vec \Omega}\;$ is 
from the separatrix on Fig.1: $\;k\;$ is zero in poles
A and C, and is unity on the separatrix. It is defined by (\ref{211}) when $\,\bf 
 \vec \Omega \,$ is between point A and the separatrix, and by (\ref{212}) when $\,\bf
  \vec \Omega \,$ is between the 
separatrix and point C. (For details see (Efroimsky 2000).)

Now suppose that the rotator is almost prolate, i.e., that point C is very closely
embraced by the separatrices. If such a rotator gets even slightly disturbed, 
then, dependent upon the particular value of $\;(I_3\,-\,I_2)/I_3\;$ 
and upon the intensity of the occational disturbance, the vector 
$\;{\bf{\vec \Omega}}\;$ will be either driven away from point $\;C\;$ back to 
the separatrix, without crossing it, or will be forced to ``jump'' over it. 
Crossing of the separatrix may be accompanied
by stochastic flipovers\footnote{The flipovers are unavoidable if dissipation 
of the kinetic energy through one precession cycle is less than a typical 
energy of an occational interaction (a tidal-force-caused perturbation, for 
example).}. If we assume 
that $\;(I_3\,-\,I_2)/I_3\;$ is infinitesimally small, then the separatrix will 
approach point $C$ infinitesimally close, and the faintest tidal
interaction or collision will be able to push the vector $\;{\bf{\vec \Omega}}
\;$ across the separatrix. In other words, an almost prolate body, during the
most part of its history, will be precessing about its minimal-inertia axis.

If in the early stage of relaxation of an almost prolate ($I_3\,\approx\,I_2
$) body the tip of vector $\;\bf{ \vec \Omega}\;$ is near point A, then its slow
departure away from A is governed by formula (9.22) in (Efroimsky 2000):
\ba
\nonumber
\frac{d\,\bra \sin^2 \theta \ket}{dt}\,=\\
\nonumber\\
\nonumber
-\,\frac{4\;\rho^2\;{\bf  \vec J}^2}{\mu\;Q(\omega)}\;\left(I_3\,-\,I_1\right)\;
\left(1\,-\;\bra \sin^2 \theta \ket \,\right)\;\left\{\,\omega\,S_1\,\left[2\;
\bra \sin^2 \theta \ket\;-\;1\;-\right.\right.\\
\nonumber\\
\nonumber
\left.\left.-\;\frac{1}{2}\;\frac{I_3\,-\,I_2}{I_2\,-\,I_1}\;\frac{I_1}{I_3}\;
\left(1\,-\;\bra \sin^2 \theta \ket \,\right)\right]\;-\right.\\
\nonumber\\
\left.-\;\omega\;S_0\;\frac{2\,I_1}{I_3}\;\left(1\,-\;\bra \sin^2\theta \ket \,
\right)\;+\;2\,\omega\,S_2\;\frac{Q(2\omega)}{Q(\omega)}\;\frac{2\,I_1}{I_3}\;
\left(1\,-\;\bra \sin^2 \theta \ket\,\right)\right\}\;.\;\;
\label{6.1}
\ea
where
\be
\omega\;=\;\sqrt{\frac{\left(2\;I_3\;T_{\small{kin}}\;-\;{\bf{ \vec J}}^2
\protect\right)\;\left(I_2\;-\;I_1\protect\right)}{I_1\;I_2\;I_3}}\;\approx\;
\frac{|{\bf{ \vec J}}|}{I_1}\;\sqrt{\frac{\left(I_3\;-\;I_1\protect\right)\;\left(I_2\;-\;I_1\protect\right)}{I_2\;I_3}}\;\sqrt{2\;\bra \sin^2 \theta \ket\;-\;1}\;\;,\;\;
\label{6.2}
\ee
$\theta\;$ is the angle between the angular-momentum vector $\bf  \vec J$ and the 
major-inertia axis $\bf{3}\;\,$; $\;S_{0,1,2}$ are some geometrical factors ($S_0\,=\,0\,$
in the case of $\,I_2\,=\,I_3\,$), and  $\;<...>\;$ symbolises an average over 
the precession cycle. For $\;<cos^2\theta>\;$ not exceeding $\;\approx\,1/7\;$, 
this equation has an exponentially decaying solution. For $\;c/a\,=\,0.6\;$ 
that solution will read:
\be
\Delta t\;\approx\;(-\,\Delta \bra {\theta } \ket )\;\times\;0.08\;\frac{\mu\;Q}{a^2\;
\Omega^3_0 \;\rho}\;\;\;.
\label{6.3}
\ee
Comparing this with (\ref{5.21}), we see that at this stage the relaxation
is about 15 times faster than in the case of an oblate body. 

During the later stage, when $\bf{ \vec \Omega}$ gets close to the 
separatrix, all the higher harmonics will come into play, and our estimate will
become invalid. How do the higher harmonics emerge? Plugging of (\ref{1}) or 
(\ref{2}) into (\ref{5.12}) will give an expression for the acceleration of an 
arbitrary point inside the body. Due to (\ref{4.12}), that expression will
yield formulae for the stresses. These formulae will be similar to 
(\ref{5.13} - \ref{5.16}), but will contain elliptic functions instead of the 
trigonometric functions. In order to plug these formulae for $\;\sigma_{ij}\;$ 
into (\ref{4.7}), they must first be squared and averaged over the precession 
cycle. For a rectangular prizm $\;2a\,\times \,2b\,\times \,2c\;$, a direct 
calculation carried out in (Efroimsky 2000) gives:
\be
\bra \sigma_{xx}^2 \ket = \frac{\rho^2}{4}\,(1-Q)^2\,\beta^4\,(x^2 - a^2)^2
\;\Xi_1 \; \; , \; \; \; \;  
\label{6.4}
\ee
\be
\bra \sigma_{yy}^2 \ket = \frac{\rho^2}{4}\,(S+Q)^2\,\beta^4\,(y^2 - b^2)^2\;
\Xi_1 \;\;\;,\;\;\;\;
\label{6.5}
\ee
\be
\bra \sigma_{zz}^2 \ket = \frac{\rho^2}{4}\,(1-S)^2\,\beta^4\,(z^2 - c^2)^2\;
\Xi_1 \;\;\;,\;\;\;\;
\label{6.6}
\ee
\be
\bra (Tr\,\sigma)^2 \ket =  \frac{\rho^2}{4}\,\beta^4\,\left\{(1-Q)(x^2 - 
a^2)^2\,+\,(S+Q)(y^2 - b^2)\;+\;(1-S)(z^2 - c^2)\right\}^2\;\Xi_1 \;\;\;,\;
\;\;\;
\label{6.7}
\ee
\be
\bra \sigma_{xy}^2 \ket = \frac{\rho^2}{4}\,\left\{(\beta \gamma + \alpha 
\omega k^2)(y^2 - b^2)+(\beta \gamma - \alpha \omega k^2)(x^2 - a^2)\right\}^2
\;\Xi_2 \;\;\;,\;\;\;\;
\label{6.8}
\ee
\be
\bra \sigma_{xz}^2 \ket = \frac{\rho^2}{4}\,\left\{(\beta \omega + \alpha 
\gamma)(z^2 - c^2)  + (\beta \omega - \alpha \gamma)(x^2 - a^2) \right\}^2\;
\Xi_3 \;\;\;,\;\;\;\;
\label{6.9}
\ee
\be
\bra \sigma_{yz}^2 \ket = \frac{\rho^2}{4}\,
\left\{(\alpha \beta + \omega \gamma)(z^2 - c^2)  + (\alpha \beta - \omega 
\gamma)(y^2 - b^2) \right\}^2\;\Xi_4 \;\;\;,\;\;\;\;
\label{6.10}
\ee
where $Q$ and $S$ are some combinations of $\,I_1, \,I_2,\,I_3\,$, defined by
formula (2.8) in (Efroimsky 2000). Factors $\,\Xi_{1,2,3,4}\,$ stand for 
averaged powers of the elliptic functions:
\ba
\nonumber
\Xi_1\;\equiv\;\bra\,\;\left(\,{\it sn}^2(u,\,k^2)\;-\;\,<\,{\it sn}^2(u,\,k^2)
\,>\,\;\right)^2\,\;\ket\;=\\
\nonumber\\
=\;< \,{\it sn}^4(u,\,k^2)\,> \;-\;< \,{\it sn}^2(u,\,k^2)\,>^2\;\;\;,\;\;\;
\label{6.11}
\ea
\ba
\nonumber
\Xi_2\;\equiv\; \bra \,\;\left( \, \sn (u,\,k^2)\; \cn (u,\,k^2)\;-\; \bra \,\;
\sn(u,\,k^2)\; \cn (u,\,k^2)\,\ket \,\right)^2\,\; \ket \;=\\
\nonumber\\
=\;<\,{\sn}^2(u,\,k^2)\;{\cn}^2(u,\,k^2)\,>\;-\;<\,\sn (u,\,k^2)\;\cn (u,\,k^2)
\,>^2\;\;\;,\;\;\;
\label{6.12}
\ea
\ba
\nonumber
\Xi_3\;\equiv\;\bra\,\;\left(\;{\it cn}(u,\,k^2)\;{\it dn}(u,\,k^2)\;-\;\bra\,\;
{\it cn}(u,\,k^2)\;{\it dn}(u,\,k^2)\,\;\ket\,\;\right)^2\,\;\ket\;=\\
\nonumber\\
=\;<\,{\it cn}^2(u,\,k^2)\;{\it dn}^2(u,\,k^2)\,>\;-\;<\,{\it cn}(u,\,k^2)\;
{\it dn}(u,\,k^2)\,>^2\;\;\;,\;\;\;
\label{6.13}
\ea
\ba
\nonumber\\
\nonumber
\Xi_4\;\equiv\;\bra\,\;\left(\,{\sn}(u,\,k^2)\;{\dn}(u,\,k^2)\;-\;\bra\,\;
{\sn}(u,\,k^2)\;{\dn}(u,\,k^2)\,\ket\,\;\right)^2\,\;\ket\;=\\
\nonumber\\
=\;<\,{\sn}^2(u,\,k^2)\;{\dn}^2(u,\,k^2)\,>\;-\;<\,{\sn}(u,\,k^2)\;{\dn}(u,\,k^2)\,>^2\;\;\;,\;\;\;
\nonumber\\
\label{6.14}
\ea
where averaging implies:
\ba
< ... > \;\equiv\;\frac{1}{\tau}\;\int_{0}^{\tau}\;\,.\,.\,.\,\;du\;\;\;,\;\;\;
\label{averaging}
\ea
$\tau\;$ being the mutual period of $\,\sn\,$ and $\,\cn\,$ and twice the period of $\;\dn\,$:
\be
\tau\;=\;4\;K(k^2)\;\equiv\;4\;\int_{0}^{\pi/2}\;
(1\,-\,k^2\,\sin^2 \psi )^{-1/2}\;d\,\psi\;\;\;.
\label{6.15}
\ee
The origin of expressions (\ref{6.11} - \ref{6.14}) can be traced from formulae
(8.4, ~8.6 - 8.13) in (Efroimsky 2000). For example, expression (\ref{5.11}), 
that gives acceleration of an arbitrary point inside the 
body, contains term $\;\sn^2(\omega t, \,k^2)\;$. (Indeed, one of the 
components of the angular velocity is proportional to $\,\sn (...)\,$, while 
the centripetal part of the acceleration is a quadratic form of the 
angular-velocity components.) The term $\;\sn^2(\omega t, \,k^2)\;$ in the 
formula for acceleration yields a similar term in the expression for $\;
\sigma_{xx}\;$. For this reason expression (8.6) in (Efroimsky 2000), that 
gives the {\bf{~time-dependent part}} of $\;\sigma_{xx}\;$, contains $\;\sn^2(.
..)\,-\,<sn^2(...)>\;$, wherefrom (\ref{6.11}) ensues.

Now imagine that in the formulae (\ref{6.4} - \ref{6.10}) the elliptic 
functions are presented by their series expansions over sines and cosines
(Abramovitz \& Stegun 1965):
\ba
{\sn}(\omega t,\,k^2)\;=\;\frac{2\pi}{k\,K}\sum_{n=1}^{\;\;\;\;\;\infty\;\;\;
*}\frac{q^{n/2}}{1\,-\,q^{n}}\;\sin\left(\omega_n\,t\right) \;\;\;\;
\;,\;\;\;\;\;
\label{6.16}
\ea
\ba
{\cn}(\omega t,\,k^2)\;=\;\frac{2\pi}{k\,K}\sum_{n=1}^{\;\;\;\;\;\infty\;\;\;
*}\frac{q^{n/2}}{1\,+\,q^{n}}\;\cos\left(\omega_n\,t\right)\;\;\;
\;\;,\;\;\;\;\;
\label{6.17}
\ea
\ba
{\dn}(\omega t,\,k^2)\;=\;\frac{\pi}{2\,K}\;+\;\frac{2\pi}{K}{\sum_{n=0}^{\;
\;\;\;\;\infty\;\;\;**}}\;\frac{q^{{n}/{2}}}{1\,+\,q^{n}}\;\cos\left(\omega_n
\, t\right) \;\;\;\;\;, \;\;\;\;\;
\label{6.18}
\ea
where 
\be
\omega_n\,=\,n\,\omega\,\pi/(2K(k^2))\;\;\;,\;\;\;q\,=\,
\exp(-\pi K({k'}^2)/K(k^2))\;\;\;,\;\;\;{k'\,}^2\,\equiv\,1\,-\,k^2\;\; 
\label{3}
\ee
and the function $\,K(k^2)\,$ is the complete elliptic integral of the first 
kind (see (\ref{6.15}) or (\ref{6.24})). A star in the superscript denotes a
sum over odd $n$'s only; a double star stands for a sum over even $n$'s. 
Insertion of 
(\ref{6.16}-\ref{6.18}) into (\ref{6.4}-\ref{6.10}) will produce, after 
squaring of $\;\sn , \, \cn , \, \dn\;$, an infinite amount of terms like $\;
\sin^2(\omega_n t)\;$ and $\;\cos^2(\omega_n t)\;$, along with an infinite 
amount of 
cross terms. The latter will be removed after averaging over the 
precession period, while the former will survive for all $n$'s and will average
to 1/2. Integration over the volume will then lead to an expression 
like (\ref{4.9}), with an infinite amount of contributions $\;\bra W_n \ket\;$ 
originating from all $\,\omega_n\,$'s, $\;n\,=\,1,\,.\,.\,.\,,\,\infty$. 
This is how an infinite amount of overtones comes into play. These overtones 
are multiples not of precession rate $\,\omega\,$ but of the "base frequency"
$\,\omega_1\,\equiv\,\omega \pi/(2K(k^2))\,$ which is lower than $\,\omega\,$.
Hence the stresses and strains contain not only Fourier components oscillating at  
frequencies higher than the precession rate, but also components oscillating at frequencies
lower than $\,\omega\,$. This is a very unusual and counterintuitive phenomenon.

The above series (called ''nome expansions'') typically converge very 
quickly, for $\,q\,\ll\,1\,$. Note, however, that $\,q\,\rightarrow\,1\,$ 
at the separatrix. Indeed, on approach to the separatrix we have: 
$\,k\,\rightarrow\,1\,$, wherefrom $\,K(k^2)\,\rightarrow\,\infty\,$; therefore 
$\,q\,\rightarrow\,1\,$ and  $\,\omega_n \,\rightarrow\,0\,$ (see eqn. 
(\ref{3})). The period of rotation (see (\ref{6.15})) becomes infinite. 
(This is the reason why near-separatrix states can mimic the principal one.)

Paper (Efroimsky 2000), addressed relaxation in the vicinity of 
poles. This case corresponds to $\;k\,\ll\,1\,$. For this reason we used, 
instead of (\ref{6.16} - \ref{6.18}), trivial approximations $\; \omega_1\,\approx \, \omega\; ,\;\sn (\omega 
t,\,k^2)\,\approx\,\sin (\omega t)\;,\;\cn (\omega t,\,k^2)\,\approx\,\cos (
\omega t)\;,\;\dn (\omega t,\,k^2)\,\approx\,$1. These approximations, along with
(\ref{6.4} - \ref{6.14}) enabled us to assume that the terms $\,\sigma_{xz}^2\,
$ and $\,\sigma_{yz}^2\,$ in (\ref{5.6}) are associated with the principal 
frequency $\,\omega\,$, while $\,<\sigma_{xx}^2>\,$, $\,<\sigma_{yy}^2>\,$, $\,
<\sigma_{zz}^2>\,$, $\,<(Tr\,\sigma )^2>\,$ and $\,\sigma_{xy}^2\,$ are 
associated with the second harmonic $\;2\,\omega\,$. No harmonics higher than
second appeared in that case. However, if we move away from the poles, 
parameter $\,k\,$ will no longer be small (and will be approaching unity as we 
approach the separatrix). Hence we shall have to take into account all terms in
(\ref{6.16} - \ref{6.18}) and, as a result, shall get an infinite amount of 
contributions from all $\;\omega_n\,$'s in (\ref{4.7} - \ref{4.9}). Thus we see
that the problem is very highly nonlinear. It is nonlinear even though the
properties of the material are assumed linear (strains $\;\epsilon\;$ are 
linear functions of stresses $\;\sigma\;$). Retrospectively, the nonlinearity 
originates because the dissipation rate (and, therefore, the relaxation rate) 
is proportional to the averaged (over the cycle) elastic energy stored in the 
body experiencing precession-caused alternating deformations. The average 
elastic energy is proportional to $\;<\sigma\,\epsilon>\;$, i.e., to  
$\;<\sigma^2>\;$. The stresses are proportional to the components of the 
acceleration, that are quadratic in the components of the angular velocity
(\ref{1} - \ref{2}). All in all, the relaxation rate is a quartic form of the 
angular-velocity components that are expressed by the elliptic functions 
(\ref{6.16} - \ref{6.18}). 

A remarkable fact about this nonlinearity is that it produces oscillations of
stresses and strains not only at frequencies higher than the precession 
frequency $\;\omega\;$ but also at frequencies lower than $\,\omega$. This is 
evident from formula (\ref{3}): 
the closer we get to the separatrix (i.e., the closer $\,k^2\,$ gets to unity),
the smaller the factor $\;\pi/(2K)\;$, and the more lower-than-$\omega$ 
frequencies emerge.

A quantitative study of near-separatrix wobble will imply attributing extra 
factors of $\;\omega_n/Q(\omega_n)\;$ to each term of the series (\ref{4.9})
and investigating the behaviour of the resulting series (\ref{4.10}). This
study will become the topic of our next paper. Nevertheless, some qualitative 
judgement about the near-separatrix behaviour can be made even at this point.

For the calculation of the dissipation rate (\ref{4.10}), the value of the 
average elastic energy $\,<W>\,$ given by the sum (\ref{4.9}) is of no use
(unless each of its terms is multiplied by $\;\omega_n/Q (\omega_n)\;$ and 
plugged into (\ref{4.10})). For this reason, the values of the terms $\,<
\sigma_{ij}^2>\,$ entering (\ref{4.7}) are of no practical value either; only 
their expansions obtained by plugging (\ref{6.16} - \ref{6.18}) into (\ref{6.4}
 - \ref{6.14}) do matter. Nonetheless, let us evaluate $\,<W>\,$ near the 
separatrix. To that end, one has to calculate all $\,<\sigma_{ij}^2>\,$'s by 
evaluating (\ref{6.11} - \ref{6.14}). Direct integration in (\ref{6.11} - 
\ref{averaging}) leads to:
\ba
\Xi_1\;=\;\frac{1}{3\,k^4}\;\left\{k^2\;-\;1\;+\;\frac{2\,E}{K}\;\left(2\,-\,
k^2\right)\;-\;3\;\left(\frac{E}{K}\right)^2\right\}\;\;\;\;,\;\;\;
\label{6.19}
\ea
\ba
\Xi_2\;=\;\frac{1}{3\,k^4}\;\left\{2\;\left(k^2\;-\;1\right)\;+\;\frac{E}{K}\;
\left(\,-\,2\,-\,5\,k^2\right)\right\}\;\;\;\;,\;\;\;
\label{6.20}
\ea
\ba
\Xi_3\;=\;\frac{1}{3\,k^2}\;\left\{\frac{E}{K}\;\left(1\;+\;k^2\right)\;+\;
\left(k^2\;-\;1\right)\right\}\;\;\;\;,\;\;\;
\label{6.21}
\ea
\ba
\nonumber\\
\Xi_4\;=\;\frac{1}{3\,k^2}\;\left\{\frac{E}{K}\;\left(2\;k^2\;-\;1\right)\;+\;
\left(1\;-\;k^2\right)\right\}\;\;\;\;,\;\;\;
\label{6.22}
\ea
$K\,$ and $\;E\;$ being abbreviations for the complete elliptic integrals of the 1st and 2nd kind:
\ba
\nonumber
K\;\equiv\;K(k^2)\;\equiv\;\int_{0}^{\pi/2}\;
(1\,-\,k^2\,\sin^2 \psi )^{-1/2}\;d\,\psi\;\;\;,\\
\label{6.24}
\\
\nonumber
E\;\equiv\;E(k^2)\;\equiv\;\int_{0}^{\pi/2}\;
(1\,-\,k^2\,\sin^2 \psi )^{1/2}\;d\,\psi\;\;\;.\;\;\;
\ea
In the limit of $\,k\,\rightarrow\,1\,$, the expression for $\;K\;$ will 
diverge and all $\;\Xi_i\;$ will vanish. Then all $\,<\sigma_{ij}^2>\,$ will 
also become nil, and so will $\,<W>\,$. As all the inputs $\,<W(\omega_n)>\,$ 
in (\ref{4.10}) are nonnegative, each of them will vanish too. Hence the 
relaxation slows down near the separatrix. Moreover, it appears to completely 
halt on it. How trustworthy is this conclusion? On the one hand, it might have 
been guessed simply from looking at (\ref{6.15}): since for $\,k\,\rightarrow\,
1\,$ the period $\,4\,K(k^2)\,$ diverges (or, stated differently, since the 
frequencies $\,\omega_n\,$ in (\ref{3}) approach zero for each fixed n, then 
all the averages may vanish). On the other hand, though, the divergence of the 
period undermines the entire averaging procedure: for $\;\tau\;\rightarrow\;
\infty\;$, expression (\ref{2.3}) becomes pointless. Let us have a look at the 
expressions for the angular-velocity components near the separatrix. According 
to (Abramovits \& Stegun 1965), these expressions may be expanded into series 
over small parameter $\;(1-k^2)\;$:
\begin{eqnarray}
\nonumber
\Omega_1\;=\;\gamma\;\,{\it{dn}}\left(\omega t ,\; k^2 \protect\right)\;=\;
\gamma\;\,\left\{{\it{sech}}\left(\omega t \protect\right)^{\left.~ \right.}_{\left.~ \right.}\;+\right.\;\;\;\;\;\;\;\;\;\;\;\;\;\;\;\\
\nonumber\\
\left.+\;\frac{1}{4}\;(1\,-\,k^2)\;\left[{\mathit{sinh}}(\omega t)\;{\mathit
{cosh}}(\omega t)\;+\;\omega\,t\right]\;{\mathit{sech}}(\omega t)\;{\mathit{
tanh}}(\omega t) \right\}\;+\;O\left((1-k^2)^2\right)\;\;\;,\;\;
\label{6.224}
\end{eqnarray}
\ba
\nonumber
\Omega_2 \; = \; \beta \, \; {\sn} \left(\omega t , \; k^2 \protect\right)\;=
\;\beta\;\,\left\{{\mathit{tanh}}^{\left. \right.}_{\left.\right.}(\omega t)\;+\;\right.\;\;\;\;\;\;\;\;\;\;\\
\nonumber\\
\left.+\;\frac{1}{4}\;(1\,-\,k^2)\;\left[{\mathit{sinh}}(\omega t)\;{\mathit
{cosh}}(\omega t)\;-\;\omega\,t\right]\;{\mathit{sech}}^2(\omega t) \right\}\;+
\;O\left((1-k^2)^2\right)\;\;\;,\;\;
\label{6.25}
\end{eqnarray}
\ba
\Omega_3 \; = \; \alpha\;\,{\it{cn}}\left(\omega t,\;k^2\protect\right)\;=
\alpha\;\,\left\{{\mathit{sech}}(\omega t)\;-\;\right.\;\;\;\;\;\;\;\;\;\;\\
\nonumber\\
\left.-\;\frac{1}{4}\;(1\,-\,k^2)\;\left[{\mathit{sinh}}(\omega t)\;{\mathit
{cosh}}(\omega t)\;-\;\omega\,t\right]\;{\mathit{sech}}(\omega t)\;{\mathit{
tanh}}(\omega t) \right\}\;+\;O\left((1-k^2)^2\right)\;\;\;.\;\;
\label{6.26}
\end{eqnarray}
These expansions will remain valid for small $\;k^2\;$ up to the point $\;k^2\,
=\,1\;$, inclusively. It doesn't mean, however, that in these expansions we 
may take the limit of $\;t\,\rightarrow\,\infty\;$. (This difficulty arises
because this limit is not necessarily interchangeable with the infinite sum of 
terms in the above expansions.) Fortunately, though, for $\;k^2\,=\,1\;$, the 
limit expressions
\begin{eqnarray}
\Omega_1\;=\;\gamma\;\,{\it{dn}}\left(\omega t ,\; 1 \protect\right)\;=
\;\gamma\;\,{\it{sech}}\left(\omega t \protect\right)\;\;\;\;,\;\;
\label{6.27}
\end{eqnarray}
\ba
\Omega_2 \; = \; \beta \, \; {\sn} \left(\omega t , \; 1 \protect\right)\;=
\;\beta\;\,{\mathit{tanh}}(\omega t)\;\;\;\;,\;\;
\label{6.28}
\end{eqnarray}
\ba
\Omega_3 \; = \; \alpha\;\,{\it{cn}}\left(\omega t,\; 1 \protect\right)\;=\;
\alpha\;\,{\mathit{sech}}(\omega t)\;\;\;\;\;\;
\label{6.29}
\end{eqnarray}
make an {\underline{exact}} solution to (\ref{22.4}). Thence we can see what 
happens to vector $\;\bf{ \vec \Omega}\;$ when its tip is right on the separatrix.
If there were no inelastic dissipation, the tip of vector $\;\bf{ \vec \Omega}\;$ 
would be slowing down while moving along the separatrix, and will come to halt at 
one of the middle-inertia homoclinic unstable poles (though it would formally take
$\;\bf{ \vec \Omega}\;$ an infinite time to get there, because $\,\Omega_1\,$ and 
$\,\Omega_3\,$ will be approaching zero as $\;\sim\,\exp (-\omega t)\;\,$). When $
\;\bf{ \vec \Omega}\;$ 
gets sufficiently close to the homoclinic point, the precession will slow down so 
that an observer would get an impression that the body is in a simple-rotation 
state. In reality, some tiny dissipation will still be present even for very 
slowly evolving $\;\bf{ \vec \Omega}\;$. It will be present because this slow 
evolution will cause slow changes in the stresses and strains. The dissipation 
will result in a further decrease of the kinetic energy, that will lead to a 
change in the value of $\;k^2\;$ (which is a function of energy; see (\ref{211}) 
and (\ref{212})). A deviation of $\;k^2\;$ away from unity will imply a shift of $
\;\bf{ \vec \Omega}\;$ away from the separatrix towards pole C. So, the separatrix
eventually {\underline{will}} be crossed, and the near-separatrix slowing-down 
does NOT mean a complete halt.

This phenomenon of near-separatrix slowing-down (that we shall call 
{\bf{lingering effect}}) is not new. In a slightly different context, it was 
mentioned by Chernous'ko (1968) who investigated free precession of a tank filled 
with viscous liquid and proved that, despite the apparent trap, the separatrix is
crossed within a finite time. Recently, the capability of near-intermediate-axis 
rotational states to mimic simple rotation was pointed out by Samarasinha, Mueller
\& Belton (1999) with regard to comet Hale-Bopp. 

We, thus, see that the near-separatrix dissipational dynamics is very subtle, 
from the mathematical viewpoint. On the one hand, more of the higher overtones 
of the base frequency will become relevant (though the base frequency itself will 
become lower, approaching zero as the angular-velocity vector approaches the 
separatrix). On the other hand, the separartrix will act as a (temporary) trap,
and the duration of this lingering is yet to be estimated.

One should, though, always keep in mind that a relatively weak push can help the 
spinning body to cross the separatrix trap. So, for many rotators (at least, for 
the smallest ones, like cosmic-dust grains) the observational reality near 
separatrix will be defined not so much by the mathematical sophistries but rather 
by high-order physical effects: the solar wind, magnetic field effects, etc... In 
the case of a macroscopic rotator, a faint tidal interaction or a collision with a
smaller body may help to cross the separatrix.

\section{\underline{APPLICATION TO WOBBLING ASTEROIDS}}

\subsection{Relaxation Rates of Comets and Asteroids}

Knowledge of the alignment rate $\;\dot \theta\;$ as a function of the 
precession-cone half-angle $\;\theta\;$ enables one not only to write down a 
typical relaxation time but to calculate the entire dynamics of the process. In
particular, suppose the observer is capable of measuring the precession-cone  
half-angle $\;\theta\;$ with an error $\;\delta\;$. This observer will then 
compute, by means of (\ref{5.21}), the time needed for the body to change its 
residual half-angle from $\;\theta\;$ to $\;\theta\,-\,\Delta \theta\;$, for 
$\;\Delta\,\theta\,>\,\delta\;$. This time will then be compared with the 
results of 
his further measurements. Below we shall show that such measurements will soon 
become possible for spacecraft-based observation means.

First, let us find a typical relaxation time, i.e., a time span 
necessary for the major-inertia axis to shift considerably toward alignment 
with $\bf J\;$. This time may be defined as:
\begin{equation}
t_r \; \;  \; \; \equiv \;\; \int_{\theta_0}^{\delta} \; 
\frac{d \theta}{d \theta/dt} \; \; \; , \;\;\;
\label{5.25}
\end{equation}
$\theta_0$ being the initial half-angle of the precession cone ($\theta_0<\pi/2
$), and $\;\delta\;$ being the minimal experimentally-recognisable value of $\;
\theta\;$. A finite $\;\delta\;$ will prevent the ``slow-finish'' divergency. A
particular choice of $\;\theta_0\;$ and $\;\delta\;$ will lead to an 
appropriate numerical factor in the final expression for $\;t_r\;$. As 
explained in (Efroimsky 2001), $\;t_r\;$ is not very sensitive to the choice of
angle $\;\theta_0\;$, as long as this angle is not too small. This weak 
dependence upon the initial 
angle is natural since our approach accounts for the divergence at small angles
(``exponentially slow finish'') and ignores the ``slow start''. Therefore one can take, for a crude estimate, 
\be
\theta_0\,=\,\pi/2\;\;\;.\;\;
\label{5.26}
\ee
For $\;t_r\;$ it would give almost the same result as, say, $\;\pi/3\;$ or $\;
\pi/4\;$. A choice of $\;\delta\;$ must be determined exclusively by the 
accuracy of the observation technique: $\;\delta\;$ is such a minimally 
recognizable angle that precession within a cone of half-angle $\;\delta\;$ or 
less cannot be detected. Ground-based photometers measure the 
lightcurve-variation amplitude that is approximately proportional to the 
variation in the cross-sectional area of the wobbling body. In such sort of 
experiments the relative error is around {\it ~0.01}. In other words, only 
deviations from one revolution to the next exceeding {\it ~0.01 mag} may be 
considered real. This corresponds to precession-cone half-angles $\;\delta\,
\approx\,10^o\;$ or larger (Steven Ostro, private communication). Ground-based 
radars have a much sharper resolution and can grasp asteroid-shape details as 
fine as $\;10\;m\;$. This technique may reveal precession at half-angles of 
about 5 degrees. NEAR-type missions potentially may provide an accuracy of $
\,0.01^o\;$ (Miller et al. 1999). For a time being, we would lean to a 
conservative estimate
\be
\delta\;=\;6^o\;\;\;\;,\;\;\;
\label{5.27}
\ee
though we hope that within the coming years this limit may be reduced  
by three orders due to advances in the spacecraft-borne instruments.

Remarkably, $\;t_r\;$ is not particularly sensitive to 
the half-sizes' ratio $\;c/a\;$ either, when this ratio is between 0.5 - 0.9 (which is 
the case for realistic asteroids, comets and many spacecraft). Our formulae 
give:
\begin{eqnarray}
\nonumber
t_{(our \; result)} \; \approx \; (1\;-\;2) \; \frac{\mu \; Q}{
\rho \; 
a^2\; \Omega^3_0} \;\;\;\;\;\; for \;\;\;\;\;\; \theta_0\,\approx\,(2\;-\;3)\, \delta
\;=\;12\;-\;18^o\;\;\;\;\;\;;
\ea
\ba
t_{(our \; result)}\;\approx \;(3\;-\;4) \;\frac{\mu \; Q}{\rho 
\;a^2 \; \Omega^3_0} \; \; \; \; \;\;for \;\;\;\;\;\; \theta_0\,\approx\,\pi/4\;\;\;\;\;\;\;\;;\;\;\;\;
\label{5.28}
\ea
\ba
\nonumber
t_{(our \; result)} \; \approx \; (4\;-\;5) \; \frac{\mu \; Q}{
\rho \; a^2 \; \Omega^3_0} \;\;\;\;\;\; for \;\;\;\;\;\; \theta_0\,\stackrel{<}{\sim}\,\pi/2\;\;\;\;\;\;.\;\;\;\;
\end{eqnarray}
(Mind though that, according to (\ref{5.24}), $\theta$ should not approach 
$\pi/2$ too close.) To compare our results with a preceding 
study, recall that according to Burns \& Safronov (1973)
\begin{eqnarray}
t_{(B \; \& \;S)} \; \approx \; 100 \; \frac{\mu \; Q}{\rho \; a^2 \; 
\Omega^3_0} \; \; \; .\;\;\;
\label{5.29}
\end{eqnarray}
The numerical factor in Burns \& Safronov's formula is about $100$ for objects 
of small oblateness, i.e., for comets and for many asteroids. (For objects of 
irregular shapes Burns and Safronov suggested a factor of about $\,20$
in place of $\,100$.)

This numerical factor is the only difference between our formula and that of 
Burns \& Safronov. This difference, however, is quite considerable: for small
residual half-angles $\;\theta\;$, our value of the relaxation time is  
two orders smaller than that predicted by Burns \& Safronov. For larger 
residual half-angles, the times differ by a factor of several dozens. We see 
that the effectiveness of the inelastic relaxation was much underestimated by 
our predecessors. There are three reasons for this 
underestimation. The first reason is that our calculation was based on an
improved solution to the boundary-value problem for stresses. Expressions 
(\ref{5.13}) - (\ref{5.16}) show that an overwhelming share of the 
deformation (and, therefore, of the inelastic dissipation) is taking place in 
the depth of the body. This is very counterintuitive, because on a heuristic 
level the picture of precession would look like this: a centrifugal bulge, with
its associated strains, wobbles back and forth relative to the body as $\;
\Omega\;$ moves through the body during the precession period. This naive 
illustration would make one think that most of the dissipation is taking place 
in the shallow regions under and around the bulge. It turns out that in reality
most part of the deformation and dissipation takes place deep beneath the 
bulge (much like in the simple example with the liquid planet, that we provided
in subsection 1.5. The second, most important, reason for our formulae 
giving smaller 
values for the relaxation time is that we have taken into account the second 
harmonic. In many rotational states this harmonic turns to be a provider of the
main share of the entire effect. In the expression $\;(63 (c/a)^4\,\cot^2 
\theta + 20) \;$ that is a part of formula (\ref{5.21}), the term $\; 63 
(c/a)^4 \,\cot^2 \theta\;$ is due to the principal frequency, while the term $
\;20\;$ is due to the second harmonic\footnote{For calculational details, see 
Lazarian \& Efroimsky (1999).}. For $\;c/a\;$ belonging to the realistic 
interval $\;0.5\,- \,0.9\;$, the second harmonic contributes (after integration
from $\;\theta_0\;$ through $\;\delta\;$) a considerable input in the 
entire effect. This input will be of the leading order, provided the initial 
half-angle $\;\theta_0\;$ is not too small (not smaller than about $\,30^o\,$).
In the case of a small initial half-angle, the contribution of the second mode 
is irrelevant. Nevertheless it is the small-angle case where the discrepancy 
between our formula and (\ref{5.29}) becomes maximal. The estimate (\ref{5.29})
for the characteristic time of relaxation was obtained in Burns \& Safronov 
(1973) simply as a reciprocal to their estimate for $\;\dot \theta\;$; it 
ignores any dependence upon the initial angle, and thus gives too long 
times for small angles. The dependence of the dissipation rate of the values of
$\;\theta\;$ is the third of the reasons for our results being so different 
from the early estimate (\ref{5.29}).

Exploration of this, third, reason may give us an important handle on  
observation of asteroid relaxation. It follows from (\ref{5.21}) that a small 
decrease in the precession-cone half-angle, $\;-\,\Delta \theta\;$, will be 
performed during the following period of time:
\ba
\Delta t\;=\;(-\,\Delta \theta)\;\frac{2^4}{3}\;\frac{[1+(c/a)^2]^4}{63\;
(c/a)^4\;\cot^2 \theta \;+\;20}\;\frac{1}{\sin^3 \theta}\;\frac{\mu\;Q}{a^2\;
\Omega^3_0 \;\rho}\;\;\;.
\label{5.30}
\ea 
For asteroids composed of solid silicate rock, the density 
may be assumed $\;\rho\,\approx\,2.5\,\times\,10^3\;kg/m^3\;$, while the 
product in the numerator should be $\;\mu\,Q\,\approx\,1.5\,\times\,10^{13}\;
dyne/cm^2\,=\,1.5\,\times\,10^{12}\,Pa\;$ as explained in Efroimsky \& Lazarian
(2000). Burns \& Safronov suggested a much higher value of $\;3\,\times \,
10^{14}\;dyne/cm^2\,=\,3\,\times\,10^{13}\,Pa\;$, value acceptable within the  
terrestial seismology but, probably, inapplicable to asteroids.

For asteroids composed of friable materials, Harris (1994) suggests the 
following values: $\;\rho\,\approx\,2\,\times\,10^3\;kg/m^3\;$ and $\;\mu\,Q\,
\approx\,5\,\times\,10^{12}\;dyne/cm^2\,=\,5\,\times\,10^{11}\,Pa\;$. 
Naturally, this value is lower than those appropriate for solid rock (Efroimsky
\& Lazarian 2000), but in our opinion it is still too high for a friable 
medium. Harris borrowed the aforequoted value from preceding studies of Phobos
(Yoder 1982). Mind, though, that Phobos may consist not {\textit only} of rubble:
it may have a solid component in the centre. In this case, a purely rubble-pile
asteroid may have a lower $\;\mu\,Q\;$ than suggested by Harris. Anyway, as 
a very conservative estimate for a rubble-pile asteroid, we shall take the 
value suggested by Harris.

As for the geometry, let, for example, $\;\theta\,=\,\pi/3\;$ and $\;c/a\,=\,
0.6\;$. Then
\be
\Delta t\;=\;(-\,\Delta \theta)\;1.2\;\frac{\mu\;Q}{a^2\;\Omega^3_0 \;\rho}\;\;
\;.
\label{5.31}
\ee 
If we measure time $\;\Delta t\;$ in years, the revolution period $\;T\,=\,2\,
\pi/\Omega_0\;$ in hours, the maximal half-size $\;a\;$ in kilometers, and $
\;\theta\;$ in angular degrees ($\Delta \theta\,=\,\Delta \theta^o\,\times\,
1.75\,\times\,10^{-2}$), our formula (\ref{5.30}) will yield:
\be
\Delta t_{(years)}\;=\;(-\,\Delta \theta^o)\;\times\;1.31\;\times\;10^{-7}\;
\frac{\mu\;Q}{\rho}\;\frac{T^3_{(hours)}}{a^2_{(km)}}\;=\;0.33\;\frac{T^3_{(
hours)}}{a^2_{(km)}}\;\;\;,
\label{5.32}
\ee
where we accepted Harris' values of $\;\mu\,Q\,=\,5\,\times\,10^{11}\,Pa\;$ and
$\rho\,=\,2\,\times\,10^3\;kg/m^3\;$, and the angular resolution of 
spacecraft-based devices was assumed to be as sharp as $\;|\Delta \theta|\,=\,
0.01^o\;$, according to Miller et al. (1999).

\subsection{Parameters of Well-Consolidated Asteroids}

Values of the parameters that appear in the above formulae depend both upon the
body temperature and upon the wobble frequency. The temperature-, pressure- and
frequency-caused variations of the density $\; \rho \;$ are tiny and may be
neglected. This way, we can use the (static) densities appropriate to the 
room temperature and pressure: $\rho^{(silicate)} \; \approx \; 2500 \; kg \; 
m^{-3} \,$ and $\rho^{(carbon)} \; \approx \; 2000 \; kg \; m^{-3}$. 

As for the adiabatic shear modulus $\; \mu \;$, tables of 
physical quantities would provide its values at room temperature and 
atmospheric pressure, and for quasistatic regimes solely. As for the possible 
frequency-related effects in materials (the so-called ultrasonic attenuation),
these become noticable only at frequencies higher than $\; 10^8 \; Hz \;$ 
(see section 17.7 in Nowick and Berry 1972). Another fortunate circumstance is
that the pressure-dependence of the elastic moduli is known to be weak 
(Ahrens 1995). Besides, the elastic moduli of solids are known to be 
insensitive to temperature variations, as long as these variations are far 
enough from the melting point. The value of $\; \mu \;$ may increase by 
several percent when the temperature drops from room temperature to $\;
10 \, K \;$. Dislocations don't affect the elastic moduli either. Solute 
elements have very little effect on moduli in quantities up to a few 
percent. Besides, the moduli vary linearly 
with substitutional impurities (in which the atoms of the impurity replace 
those of the hosts). However hydrogen is not like that: it enters the 
interstices between the atoms of the host, and has marginal effect on the 
modulus. As for the role of the possible porosity, the elastic moduli scale 
as the square of the relative density. For porosities up to about $20\;\%~\,$,
 this is not of much relevance for our estimates.

According to Ryan and Blevins (1987), for both carbonaceous and silicate rocks
one may take the shear-modulus value  $\mu \; \approx \; 10^{10} \; Pa \; \;$.

Theoretical estimation of the $\;Q-$factor for asteroids is difficult. As well known from seismology, the $\; Q-$factor bears 
a pronounced dependence upon: the chemical composition, graining, frequency, 
temperature, and confining pressure. It 
is, above all, a steep function of the humidity which presumably affects the 
interaction between grains. The  $\; Q-$factor is less sensitive to the
porosity (unless the latter is very high); but it greatly depends upon the
amount and structure of cracks, and generally upon the mechanical nature of 
the aggregate. Whether comets and asteroids are loose aggregates or solid 
chunks remains unknown. We shall address this question in the next subsection.
For now,  we shall assume that the body is not a loosely 
connected aggregate but a solid rock (possibly porous but nevertheless 
well-consolidated), and shall try to employ some knowledge available on  
attenuation in the terrestial and lunar crust. 

We are in need of the values of the quality factors for silicate and 
carbonateous rocks. We need these at the temperatures about 150 K,  , 
zero confining pressure, frequencies appropriate to asteroid precession 
($10^{-6} \; - \; 10^{-4} \; Hz$), and (presumably!) complete lack of moisture.

Much data on the behaviour of 
$\;Q-$factors is presented in the seismological literature. Almost all
of these measurements have been made under high temperatures (from
several hundred up to 1500 Celsius), high confining 
pressures (up to dozens of MPa), and unavoidably in the presence of 
humidity. Moreover, the frequencies were typically within the range from  
dozens of $kHz$ up to several $MHz$. Only a very limited number of measurements
have been performed at room pressure and temperature, while 
no experiments at all have been made thus far with rocks at low temperatures
(dozens of $\;K\;$). The information about the role of humidity is 
extremely limited. Worst of all, only a few experiments were made with rocks at
the lowest seismological frequences ($10^{-3} \; - \; 10^{-1} \; Hz$), and 
none at frequencies between $\; 10^{-6} \;$ and $\; 10^{-2} \; Hz$,  
though some indirectly achieved data are available (see Burns 1977, Burns 
1986, Lambeck 1980, and references therein).  

It was shown by Tittman et al. (1976) that the $\; Q-$factor of about $\; 60 
\;$ measured under ambient conditions on an as-received lunar basalt was 
progressively increased ultimately to about $\; 3300 \;$ as a result of 
outgassing under hard vacuum. The latter number will be our starting point. 
The measurements were performed by Tittman et al. at $\; 20 \; kHz \;$ 
frequency, room temperature and no confining pressure. How might we estimate
the values of $\; Q \;$ for the lunar basalt, appropriate to the
lowest frequencies and temperatures? 

As for the frequency-dependence, it is a long-established fact (e.g., 
Jackson 1986, Karato 1998) that $\; \; Q \; \sim \omega^{\alpha} \;$ with $\;
 \alpha \;$ around 0.25. This dependence reliably holds for all rocks 
within a remarkably broad band of frequencies: from hundreds of $\; kHz \;$ 
down to $\;10^{-1} Hz$. Very limited experimental data are avaliable for 
frequencies down to $\;10^{-3} Hz$, and none below this
threshold. Keep in mind that $\; \alpha \;$ being close to 0.25
holds well only at temperatures of several hundred Celsius and higher, 
while at lower temperatures $\; \alpha \;$ typically decreases to 0.1
and less.

As regards the temperature-dependence, there is no consensus on this point in
the geological literature. Some authors (Jackson 1986) use a simple rule:
\be
\; Q \; \sim \; \omega^{\alpha} \; \exp(A^*/RT) \; \;\;,     
\label{wrong}
\ee
$\;A^*\;$ being the apparent activation energy. A more refined 
treatment takes into account the interconnection between the 
frequency- and temperature-dependences. Briefly 
speaking, since the quality factor is dimensionless, it must retain this 
property despite the exponential frequency-dependence. This may be achieved 
only in the 
case that $\; Q \;$ is a function not of the frequency {\it{per se}} but of 
a dimensionless product of the frequency by the typical time of 
defect displacement. The latter exponentially depends upon the activation 
energy, so that the resulting dependence will read (Karato 1998): 
\be
\; Q \; \sim \; \left[ \omega \; \exp(A^*/RT) \protect\right]^{\alpha} \;\;\;, 
\label{right}
\ee
where $\; A^* \; $ may vary from 150 - 200 $\; kJ/mol \;$ (for dunite and 
polycristalline forsterite) up to 450 $\; kJ/mol \;$ (for olivine). 
This interconnection between the frequency- and temperature-dependences 
tells us that whenever we lack a pronounced frequency-dependence, the 
temperature-dependence is absent too. It is known, for example 
(Brennan 1981) that at room temperature and pressure, at low 
frequencies ($10^{-3} \; - \; 1 \; Hz$) the shear $\; Q-$factor is 
almost frequency-independent for granites and (except 
some specific peak of attenuation, that makes $Q$ increase twice) for basalts.
It means that within this range of frequencies $\; \alpha \;$ is 
small (like 0.1, or so), and $\;Q\;$ may be assumed almost temperature-independent too. 

Presumably, the shear Q-factor, reaching several thousand at $\; 20 \;kHz 
\;$, descends, in accordance with (\ref{right}), to several hundred when  
the frequency decreases to several $\; Hz$. Within this band of frequencies, 
we should use the power $\;\alpha\,\approx\,0.25\;$, as well known from 
seismology. When we go to lower 
frequencies (from several $\; Hz$ to the desirable $10^{-6} \; - \; 10^{-4} \;
Hz$), the Q-factor will descend at a slower pace: it will obey (\ref{right}) 
with $\;\alpha\,<\,0.1\;$. Low values of $\; \alpha \;$ at low  
frequencies are mentioned in Lambeck (1980) and in Lambeck 
(1988)\footnote{See also Knopoff (1963) where a very slow and smooth frequency-dependence of $Q$ at 
low frequencies is pointed out.}. The book by Lambeck containes much material 
on the $Q$-factor of the Earth. Unfortunately, we cannot employ the numbers 
that he suggests, because in his book the quality factor is defined for the 
Earth as a whole. Physically, there is a considerable difference between the 
$Q$-factors emerging in different circumstances, like for example, between 
the effective tidal 
$Q$-factor\footnote{A comprehensive study of the effective tidal $Q$-factors 
of the planets was performed by Goldreich and Soter (1966)}
and the $Q$-factor of the Chandler wobble). In regard to the latter, Lambeck 
(1988) refers, on page 552, to Okubo (1982) who suggested that for the 
Chandler 
wobble $\;50\,<\,Q\,<\,100\;$. Once again, this is a value for the Earth as a 
whole, with its viscous layers, etc. We cannot afford using these numbers 
for a fully solid asteroid.
               
Brennan (1981) suggests for the shear $\; Q-$factor the following 
values\footnote{Brennan mentions the decrease of $Q$ with humidity, but 
unfortunately does not explain how his specimens were dried.}: 
$Q^{(granite)}_{(shear)} \; \approx \; 250  \; $, $\; Q^{(basalt)}_{(shear)} 
\; \approx \; 500 \;$.
It would be tempting to borrow these values\footnote{These
data were obtained by Brennan for strain amplitudes within the linearity 
range. Our case is exactly of this sort since the typical strain in a
tumbling body will be about $\; \sigma/ \mu \; \approx \; \rho \Omega^2
\; a^2/ \mu \,.$ For the size $\; a \; \approx \; 1.5 \, \times \,
10^4 \, m$ and frequency not exceeding $\; 10^{-4} Hz$, this strain is
less than $\; 10^{-6} \,$ which is a critical threshold for linearity.}, if 
not for one circumstance: 
as is well known, absorption of only several monolayers of a saturant may 
dramatically decrease the quality factor. We have already mentioned
this in respect to moisture, but the fact is that this holds also for 
some other saturants\footnote{like, for example, ethanol (Clark et al. 1980)}. 
Since the asteroid material may be well saturated with hydrogen (and possibly 
with some other gases), its $\,Q$-factor may be much affected. 

It may be good to perform experiments, both on carbonaceous and
silicaceous rocks, at low frequencies and temperatures, 
and with a variety of combinations of the possible saturants. These 
experiments should give us the values for
both shear and bulk quality factors. The current lack of experimental data 
gives us no choice but to start with the value 3300 obtained by Tittman for 
thoroughly degassed basalts, and then to use formula (\ref{right}). This will 
give us, at $\;T\,=\,150\;K\;$ and $\;\omega\,=\,10^{-5}\;Hz\;$:
\be
Q^{(basalt)}_{(shear)} \; \approx \; 100 
\label{formula}
\ee 
This value of the shear $Q$-factor for granites and basalts differs from the one chosen in 
Burns and Safronov (1979) only by a factor of 3. For carbonateous materials 
$\; Q \;$ must be surely much less than that of silicates, due to weaker 
chemical bonds. So for carbonaceous rocks, it is for sure that
\be
Q^{(carb)} \; < \; 100 \; \; \; ,
\label{9.2}
\ee
though this upper boundary is still too high.

Consider asteroid 4179 Toutatis. This is an S-type asteroid analogous to stony 
irons or ordinary chondrites, so the solid-rock value of $\;\mu\,Q\;$ may be 
applicable to it:  $\;\mu\,Q\,\approx\,
1.5\,\times\,10^{13}\;dyne/cm^2\,=\,1.5\,\times\,10^{12}\,Pa\;$. Its density 
may be roughly estimated as $\;\rho\,=\,2.5\,\times\,10^3\;kg/m^3\;$ (Scheeres 
et al. 1998). Just as 
in the case of (\ref{5.32}), let us measure the time $\;\Delta t\;$ in years, 
the revolution period $\;T\;$ in hours ($T_{(hours)}\,=\,175$), the maximal 
half-radius $\;a\;$ in kilometers ($\;a_{(km)}\,=\,2.2\;$), and $\;\theta\;$ in
angular degrees ($|\Delta \theta^o|\,=\,0.01$). Then (\ref{6.3}) will yield:
\be
\Delta t_{(years)}\;\approx\;5.1\;\times\;10^{-2}\;\frac{T^3_{(hours)}}{a^2_{
(km)}}\;=\;5.6\;\times\;10^4\;years
\label{7.1}
\ee
Presently, the angular-velocity vector $\,\bf \Omega\,$ of Toutatis is at the 
stage of precession about $\,A\;$ (see Fig.1). However its motion does not obey the 
restriction $\;{\bra \cos^2 \theta  \ket}\,<\,1/7\,$ under which (\ref{6.3}) works 
well. A laborious calculation based on equations (2.16) and (A4) from Efroimsky (2000)
and on formulae (1), (2) and (11) from Scheeres et al (1998) shows that in the case of 
Toutatis $\;{\bra \cos^2 \theta  \ket}\,\approx\,2/7\,$. Since the violation is not 
that bad, one may still use (\ref{7.1}) as the zeroth approximation. Even if it is a 
two or three order of magnitude overestimate, we still see that the chances for 
experimental observation of Toutatis' relaxation are slim.
 
This does not mean, though, that one would not be able to observe asteroid 
relaxation at all. The relaxation rate is sensitive to the parameters of the body (size and 
density) and to its mechanical properties ($\,\mu Q\,$), but the precession period is certainly
the decisive factor.  Suppose that some asteroid is loosely-connected 
($\;\mu\,Q\,=\,5\,\times\,10^{12}\,dyne/cm^2\,=\,5\,\times\,10^{11}\,Pa\;$ and 
$\rho\,=\,2\,\times\,10^3\;kg/m^3\;$), has a maximal half-size 17 km, and is 
precessing with a period of 30 hours, and {\it{is not too close to the separatrix}}. 
Then an optical resolution of $|\Delta \theta^o|\,=\,0.01$ degrees will lead to 
the following time interval during which a $\;0.01^o\;$ change of the precession-cone 
half-angle will be measurable:
\be
\Delta t_{(years)}\;\approx\;2.12 \;\times\;10^{-2}\;\frac{T^3_{(hours)}}{a^2_{
(km)}}\;=\;2\;\;years\;\;\;\;
\label{7.2}
\ee
which looks most encouraging. In real life, though, it may be hard to observe precession
relaxation of an asteroid, for one simple reason: too few of them are in the states when
the relaxation rate is fast enough. Since the relaxation rate is much faster than believed 
previously, most excited rotators have already relaxed towards their principle states and 
are describing very narrow residual cones, too narrow to observe. The rare exceptions are 
asteroids caught in the near-separatrix ''trap''. These are mimicing the principal state. 

\subsection{Loosely-Connected Asteroids}

Above we mentioned that the mechanical structure of the small bodies is still
in question.  At present, most astronomers lean toward the so-called 
rubble-pile hypothesis, in regard to both asteroids and comets. The hypothesis 
originated in mid-sixties (${\ddot{O}}$pik 1966) and 
became popular in the end of the past century (Burns 1975; 
Weidenschilling 1981; Asphaug \& Benz 1994; Harris 1996; Asphaug \& Benz 1996; 
Bottke \& Melosh 1996a,b; Richardson, Bottke \& Love 1998; Bottke 1998, Bottke,
Richardson \& Love 1998; Bottke, Richardson, Michel \& Love 1999, Pravec \& 
Harris 2000). 

This hypothesis rests on several arguments the main of which is the 
following: the large fast-rotating asteroids are near the rotational breakup 
limit for aggregates with no tensile strength. This is a strong argument, and 
one would find difficulty to object to it. Still, we would object to two 
other
arguments often used in support this hypothesis. One such dubious argument is 
the low density of asteroid 253 Mathilde (about 1.2 $\; g/cm^3$) . This low 
density (Veverka et al 1998, Yeomans et al 1998) may be either interpreted in 
terms of the rubble-pile hypothesis (Harris 1998), 
or be put down to Mathilde being perhaps mineralogically akin to low-density 
carbonaceous chondrites, or be explained by a very high porosity. However, in 
our opinion, the word ''porous'' is not necessarily a synonim to 
''rubble-pile'', even though in the astronomical community they are often used 
as synonims. In fact, a material may have high porosity and, at the same time, 
be rigid. 

Another popular argument, that we would contest, is the one about crator 
shapes. Many colleagues believe that a rigid body would be shattered into 
smitherines by collisions; therefrom they infer that the asteroids must be 
soft, i.e., rubble. In our opinion, though, a highly porous but still 
consolidated material may stand very energetic collisions without being 
destroyed, if its porous structure damps the impact. It is know from the 
construction engineering that some materials, initially friable, become 
relatively rigid after being heated up (like, for example, asphalt).  They 
remain porous and may be prone to creep, but they are, nevertheless, 
sufficently rigid and well connected.

We would also mention that, in our opinion, the sharply-defined craters on the 
surfaces of some asteroids witness {\bf{against}} the application of the 
rubble-pile hypothesis to asteroids. 

For these reasons, we expressed in Efroimsky \& Lazarian (2000) our 
conservative opinion on the subject:
{\it{at least some asteroids are well-connected, though we are 
uncertain whether this is true for all asteroids.}} This opinion met a cold 
reaction from the community. However, it is supported by the recentmost 
findings. The monolithic nature of 433 Eros is the most important of these 
(Yeomans 2000).
Other include 1998KY26 studied in 1999 by Steven Ostro and his team: from the 
radar and optical observations (Ostro et al 1999), the team inferred that this 
~ 30-meter-sized body, as well as several other objects, is monolithic\footnote{We should 
mention here Vesta as a reliable example of an asteroid being a 
solid body of a structure common for terrestial planets: Hubble images of 
Vesta have revealed basaltic regions of solidified lava flows, as well as a 
deep impact basin exposing solidified mantle.}. 

Still, despite our conservative attitude toward the rubble-pile hypothesis,
we have to admit that the main argument used by its proponents (the absence of 
large fast rotators) remains valid, and the question why the large 
fast-spinning asteroids are near the rotational breakup limit for loose 
aggregates is still awaiting its answer.

As explained in the preceding subsection, to register relaxation of a 
solid-rock monolith may take thousands of years. 
However, if the body is loosely connected, the inelastic 
dissipation in it will be several orders faster and, appropriately, its 
relaxation rate will be several orders higher. 

\subsection{Application to Asteroid 433 Eros in Light of Recent Observations}

As already mentioned above, asteroid 433 Eros is in a spin 
state that is either principal one or very close to it. This differs from
the scenario studied in (Black et al 1999). According to that scenario, an 
almost prolate body would be spending most part of its history wobbling about 
the minimal-inertia axis. Such a scenario was suggested because the gap between
the separatrices embracing pole C on Fig.1 is very narrow, for an almost 
prolate body, and therefore, a very weak tidal interaction or impact would 
push the asteroid's angular velocity vector $\bf{\Omega}$ across the 
separatrix, away from pole C. This scenario becomes even more viable due to the
''lingering effect'' described in subsection 2.2, i.e., due to the relative 
slowing down of the relaxation in the closemost vicinity of the separatrix. 
 
Nevertheless, this scenario has not been followed by Eros. This could have 
happened for one of the following reasons: either the dissipation rate in the 
asteroid is high enough to make Eros well relaxed after the recentmost 
disruption, or the asteroid simply has not experienced impacts or tidal 
interactions for hundreds of millions of years. 

The latter option is very 
unlikely: currently Eros is at the stage of leaving the main belt; it comes 
inside the orbit of Mars and approaches that of the Earth. It is then probable 
that Eros during its recent history was disturbed by the tidal forces that 
drove it out of the principal spin state.
Hence we are left with the former option, one that complies with our theory
of precession relaxation. The fact that presently Eros is within less than 0.1 
degree from its principal spin state means that the precession relaxation 
is a very fast process, much faster than believed previously\footnote{Note that
the complete (or almost complete) relaxation of Eros cannot be put down to the 
low values of the quality factor of a rubble pile, because this time we are 
dealing with a rigid monolith (Yeomans et al. 2000).}.

\subsection{Unresolved issues}

Our approach to calculation of the relaxation rate is not without its 
disadvantages. Some of these are of mostly aesthetic nature, but at least one 
is quite alarming.

As was emphasised in Section 3, our theory is adiabatic, in that it assumes 
the presence of two different time scales or, stated differently, the 
superposition of two 
motions: slow and fast. Namely, we assumed that the relaxation rate is much 
slower than the 
body-frame-related precession rate $\,\omega\,$ (see formulae (\ref{22.5}) and 
(\ref{22.6})). 
This enabled us to conveniently substitute the dissipation rate by its average 
over a 
precession cycle. The adiabatic assertion is not necessarily fulfilled when 
$\,\omega\,$ 
itself becomes small. This happens, for example, when the dynamical oblateness 
of an oblate ($\,I_3\,>\,I_2\,=\,I_1\,\equiv\,I\,$) body is approaching zero:
\be
(h\,-\,1)\,\rightarrow\,0\;\;\;,\;\;\;\;\;\;\;\;\;\;\;h\;\equiv\;I_3/I\;\;\;\;.
\;\;\;
\label{14.1}
\ee
Since in the oblate case $\,\omega\,$ is proportional to the oblateness 
(see (5.4)), it too 
will approach zero, making our adiabatic calculation inapplicable. This is the 
reason why one cannot and shouldn't compare our results, in the limit of $\,(h
\,-\,1)\,\rightarrow \,0\,$, with the results obtained by Peale (1973) for an 
almost-spherical oblate body. 

Another minor issue, that has a lot of mathematics in it but hardly bears any 
physical significance, is our polynomial approximation (\ref{5.13} - \ref{5.16}
$\,$,$\,$ \ref{6.4} - \ref{6.10}) to the stress tensor. As explained in Section
4, this approximation keeps the symmetry $\,\sigma_{ij}\,=\,\sigma_{ji}\,$ and 
exactly satisfies (\ref{4.12}) with (\ref{5.12}) plugged in. The boundary 
conditions are fulfilled exactly for the diagonal components of the tensor and 
approximately for the off-diagonal elements. In the calculation of the 
relaxation rate, this approximation will result in some numerical factor, and 
it is highly improbable that this factor differs much from unity.

A more serious difficulty of our theory is that it cannot, without further 
refinement, give a reasonable estimate for the duration of the near-separatrix 
slowing-down mentioned in subsection 2.2. On the one hand, many (formally, 
infinitely many) overtones of the base frequency $\,\omega_1\,$ come into play 
near the separatrix; on the other hand, the base frequency approaches zero. 
Thence, it will take some extra work to account for the dissipation associated
with the stresses oscillating at $\,\omega_1\,$ and with its lowest overtones. 
(The dissipation due to the stresses at these low frequency cannot be averaged 
over their periods.)  

There exists, however, one more, primary difficulty of our theory. Even though 
our calculation predicts a much faster relaxation rate than believed 
previously, it still may fail to account for the observed relaxation which 
seems to be even faster than we expect. According to the results obtained by 
NEAR, the upper limit on non-principal axis rotation is better than 0.1 angular
degree. How to interpret such a tough observational limit on Eros' residual 
precession-cone width? Our theory does predict very swift relaxation, but it 
also shows that the relaxation slows down near the separatrix and, especially, 
in the closemost vicinity of points $\,A\,$ and $\,C\,$ on Fig.1. Having 
arrived to the close vicinity of pole $\,C\,$, the angular-velocity vector $\,
\Omega\,$ must exponentially slow down its further approach to $\,C\,$. For 
this reason, a body that is monolithic (so that its $\,\mu Q\,$ is not too low)
and whose motion is sometimes influenced by tidal or other interactions,  must 
demonstrate to us at least some narrow residual precession cone. As already 
mentioned, for the past million or several millions of years Eros has been at 
the stage of leaving the main belt. It comes inside the Mars orbit and 
approaches the Earth. Most probably, Eros experienced a tidal interaction 
within the said period of its history. Nevertheless it is presently in or 
extremely close to its principal spin state. The absence of a visible residual
precession indicates that our theory may still be incomplete. In particular, 
our $\,Q$-factor-based empirical description of attenuation should become the
fair target for criticisms, because it ignores several important physical 
effects.

One such effect is material fatigue. It shows itself whenever a rigid material 
is subject to repetitive load. In the case of a wobbling asteroid or comet, the stresses are tiny, 
but the amount of repetitive cycles, accumulated over years, is huge. At each cycle, the picture of emerging stresses is virtually the same. Moreover, beside the periodic stresses, there exists a constant component 
of stress. This may lead to creation of ''weak 
points'' in the material, points that eventually give birth to cracks or other defects. This may 
also lead to creep, even in very rigid materials. The creep will absorb some of the excessive energy
associated with precession and will slightly alter the shape of the body. The alteration will be such
that the spin state becomes closer to the one of minimal energy. It will be achieved through the slight 
change in the direction of the principal axes in the body. If this shape alteration is due to the emergence 
of a considerable crack or displacement, then the subsequent damping of precession will be performed by a finite step, not gradually. 

Another potentially relevant phenomenon is the effect that a periodic forcing (such as the solar gravity gradient) would have on the evolution and relaxation of the precession dynamics. It is possible that this 
sort of forcing could influence the precessional dynamics of the body\footnote{We are thankful to Daniel Scheeres who drew our attention to this effect.}.

\section{\underline{APPLICATION TO WOBBLING COMETS}}
\medskip

\subsection{Excitation of the Nucleus' Rotation by Reactive Torques}

According to the widely accepted Whipple's model, comet nuclei are
conglomerates of ice and dust, also called ``dirty snowballs'' (Whipple 1951). 
TV-images of P/Halley and P/Borrelly nuclei, obtained by spacecraft
{\it Vega-1,2, GIOTTO} and {\it DEEP SPACE-1}, respectively, allow us
to suppose that, generally,  cometary nuclei have irregular nonconvex shapes
with typical sizes $R_*$ within the range of 1 to 10 kilometers. Numerically 
generated images of comets P/Halley and P/Borelly are presented on Figure 2 and
Figure 3. A typical
comet rotates slower than most asteroids of a like size. 
Rotational periods of most comets lie within the interval from several hours 
to several days.
  
Solar radiation instigates matter sublimation from the comet-nuclei surfaces;
this process is especially intensive at heliocentric distances $r < 3$ AU.
Sublimated volatiles ($H_2 O\,$, $\,CO\,$, $\,CO_2\,$, $\,\dots$) and emansipated dust 
form an expanding atmosphere called coma. As a result, an Earth-based observer
will not only register periodic brightness variations of the comet (feature
common to spinning comets and asteroids), but will also observe effects 
specific to comets solely. These will be the morphological features of the 
coma.  As an obvious
\begin{figure}
\plotone{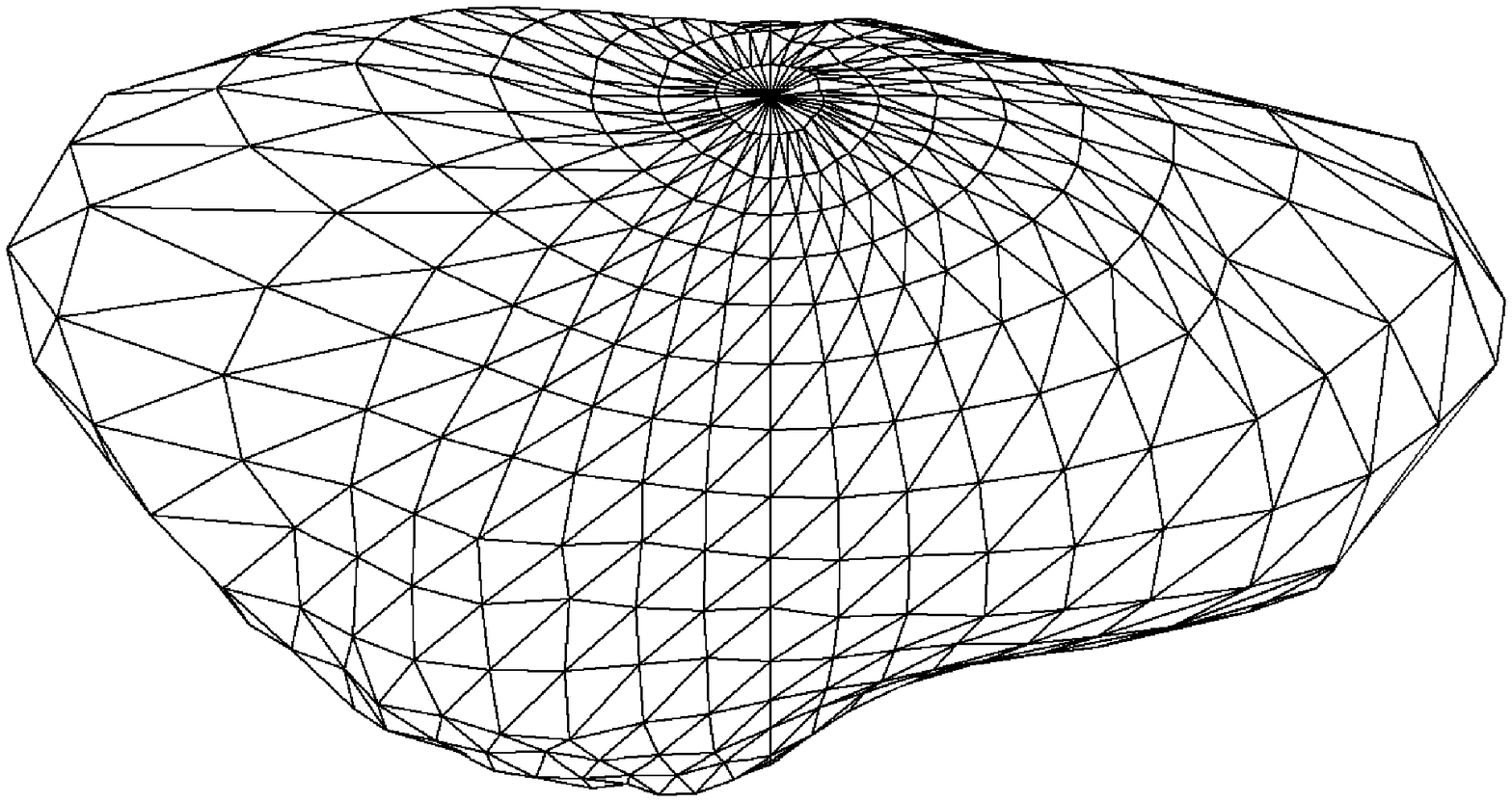}
\caption{The general view of P/Halley nucleus, based on the numeric-shape model
         developed by Stooke and Abergel (1991) who used the Vega - 1, 2 and 
         Giotto images.}
\end{figure}
\clearpage
\noindent
example, we would mention the jets: lengthy narrow 
flows of matter, initiating on the nucleus' surface or close to it. A more 
subtle phenomenon will be twisting of such jets. As was pointed out by Whipple (Whipple 1951), evaporating matter should
produce a reactive torque, $\taubold$, that acts on the nucleus and alters its spin
state\footnote{There are analogies between the spin-up of comets due to
matter ejection and spin-up of grains due to ejection of H$_2$ atoms
at the catalytic sites on grain surface. The theory of grain rotation due
to such torques is discussed in Purcell (1979), Spitzer \& McGlynn (1979),
Lazarian \& Draine (1997).}. To calculate it, one can use formula 
\ba
\taubold = - \sum_{j=1}^{N} Q_j(\Rv_j \times \vv_j),
\label{66.1}
\ea
where $N$ is the number of faces of the polyhedron approximating the 
nucleus' shape, $Q_j$ is the rate of mass ejection by the $j$-th face,
$\Rv_j$ is the radius vector of the face's centre in the body's principal 
frame of reference, and $\vv_j$ is the effective velocity of the ejected 
matter.
\begin{figure}
\plotone{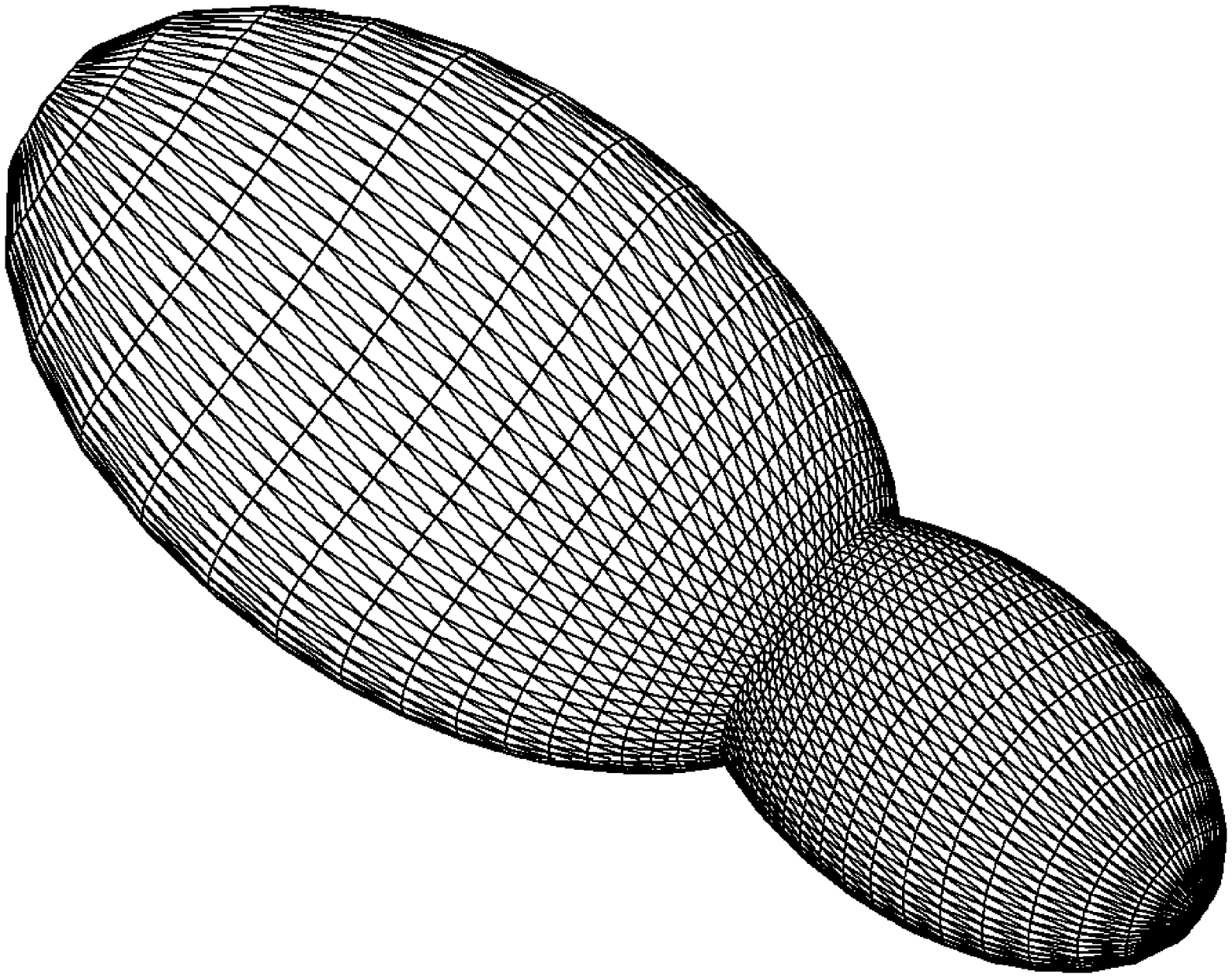}
\caption{P/Borelly nucleus: rough reconstruction based on Deep Space - 1 
images. According to (Neishtadt, Scheeres, Sidorenko, Stooke \& Vasiliev 2002),
this nucleus can be approximated by a combination of two ellipsoids with major 
semiaxes $1.6 \;km$, $1.8\; km$, 
$3.0 \;km$, and $0.96 \;km$, $1.08\; km$, $1.8 \;km$, the distance between 
their centres being $3.7 \;km$.}
\end{figure}
The process of mass ejection depends on local illumination conditions
and on the current heliocentric distance. Its accurate description is
a subject of a rather involved research (Crifo \& Rodionov 1999). Various  
empirical 
models have been employed to study rotation evolution of cometary nuclei.
For example, to model the relative dependence of the mass ejection
rate on the helicentric distance, the following expression was suggested
in (Marsden, Sekanina \& Yeomans 1973) and has been widely used since then:
\ba
g(r)=g_0 
\left(
\frac{r}{r_0}
\right)^{-2.15}
\left[
1 + \left( \frac{r}{r_0} \right)^{5.093}
\right]^{-4.6142}.
\label{66.2}
\ea
Here $r_0=2.808$ AU, $g_0$ is a normalising multiplier (whose
its value can be chosen in such a way that $g(q)=1$, where $q$ stands for the 
perihelion distance). 

In most publications on the topic, spin evolution of a comet nuclei due to 
reactive torques was studied through numeric integration of the equations of  
rotation (Wilhelm 1987, Julian 1990, Peale \& Lissauer 1989, Samarasinha \& 
Belton 1995, Jorda \& Licandro 2002). A more systematic approach to the problem
was devoloped in (Neishtadt, Scheeres, Sidorenko \& Vasiliev 2002),  where
possible secular phenomena were studied by means of an accurate averaging 
procedure (Arnold 1978), and relevant physical parameters, that control the
evolution of cometary rotation, were extracted.

Numeric and analytic investigations have revealed certain patterns in the spin
evolution. For nuclei with fixed active regions, the direction of the angular
momentum vector $\Lv$ typically spirals either toward the orbital direction of
peak outgassing or opposite to it (Neishtadt, Scheeres, Sidorenko \& Vasiliev 2002; Samarasinha \& Belton 1995). The timescale $t_R$
for the reactive torque to change the nucleus spin state can be estimated
as
\ba
t_R \sim
\frac{L_*}{(D_* R_*) Q_* V_* \Phi_*} ~~,
\label{66.3}
\ea
$L_*$ being the nucleus angular momentum, $Q_*$ being the total mass
ejection rate at perihelion, and $V_*$ being the effective exhaust velocity.
The product $(D_* R_*)$ can
be interpreted as the torque's arm. Here the dimensionless parameter $D_*$ 
characterises the dependence of reactive torque on active zones'
distribution over the nucleus surface, while the dimensionless parameter $\Phi_*$
is a certain integral characteristic of the nucleus' heating on its
orbit:
\ba
\Phi_* = \frac{1}{T} \int_0^T g(r(t))dt =
\frac{(1-e^2)^{3/2}}{\pi}
\int_0^\pi \frac{g(r(\nu))d\nu}{(1 + e \cos \nu)^2}.
\label{66.4}
\ea
Here $T$ and $e$ are comet's orbital period and eccentricity, while $\nu$ is the 
nucleus true anomaly.

Although the long-term history of the nucleus spin state is dominated by
reactive torques, we should not forget that nuclei are not perfectly rigid
bodies. Splitting of comets' nuclei by tidal stress, phenomenon that has 
been observed many a time by astronomers, has inspired some theoreticians to 
speculate that nuclei consist of weakly bonded components. This so-called 
``rubble-pile hypothesis'' has been extended by some authors even to asteroids.
As we already mentioned in one of the preceding sections of this review,
this hypothesis has its {\it{pro}}'s and {\it{con}}'s and, therefore,
both proponents and critics. What is certain thus far, is that at least some
comets are indeed weak. For example, two unequal components of comet P/Borrelly
are connected with an intermediate zone covered by a complicated
network of cracks. This structure, well visible on the available images, has
lead to a hypothesis that the P/Borrelly nucleus is assembled of two pieces in 
loose contact.
 
When the comet is not too far from the Sun, time  $t_R$ is, typically, much 
shorter than the internal-relaxation time. Nevertheless, it is possible that 
some qualitative properties of the nucleus' rotation are associated with  
non-rigidity. In particular, since the timescale for wobble excitation by 
$\taubold$ is often shorter than the damping time, it is reasonable to suppose 
that many cometary nuclei should be in excited rotational states after passing 
their perihelia (and sometimes even before approaching the perihelia). So far 
such an excited state has been reliably established for P/Halley 
only\footnote{According 
to (Belton, Julian, Anderson \& Mueller 1991), the long axis
of P/Halley nucleus is inclined to the angular-momentum vector $\Lv$
by $\approx 66^\circ$ and is rotating about $\Lv$ with a period of $3.69$ days.
The spin component about the long axis has a period of $7.1$ days.}. However, 
as we mentioned above there are several more comets whose spin state may be 
complex.

The small amount of tumbling comets observed thus far 
may be put down to one or both of the following reasons. The first is the 
still insufficient precision of our observation techniques. The second reason
is the possibility of relaxation rate being much higher than believed 
previously.

\subsection{Parameters of the Comets}
 
For the dirty ice Peale and Lissauer (1989) suggest,  for Halley's 
comet, $\,\mu \, \approx \, 10^{10} \ dyne/cm^2$ $= \; 10^{9}\; Pa \;$ while 
$\; Q \; < \; 100 \;$. The latter number, $100$, characterises attenuation in 
the solid ice. The question arises whether it is applicable to comets. In the
preceding section we mentioned, in regard to asteroids, the ``rubble-pile'' 
hypothesis. Despite that hypothesis being fortified by various theoretical 
explanations, our attitude to it still remains conservative. 

In the case of comets, though, this hypothesis may work. It is a 
well-established fact that comets sometimes get shattered by tidal forces. For 
example, Shoemaker-Levy 9 broke into 21 pieces on the perijove preceding the 
impact (Marsden 1993).  In 1886 Comet Brooks 2 was rent into pieces by the 
jovian gravity (Sekanina 1982). Comet West disintegrated in 1976 (Melosh and 
Schenk 1993). It seems that {\it {at least some comets are loosely consolidated
aggregates, though we are unsure if this is the case for all comets}}. 
Hopefully, our understanding of the subject will improve after the Deep Impact 
mission reaches its goal.

Loosely-connected comets, presumably, consist not so much of solid ice but 
more of firn (coarse-grained snow), material whose quality factor is of order unity. Hence the estimate
\be
Q^{(comet)}\;\approx\;1
\label{comet}
\ee
is, in our opinion, more realistic. So, let us assume $\mu Q \approx 10^{10}\,dyne/cm^2 \, = \, 10^{9}\,Pa$. As for the average density of a comet, most probably it does not deviate much from $1.5\,\times
\,10^3\,kg/m^3$. Indeed, on 
the one hand, the major part of the material may have density close to that of 
firn, but on the other hand a typical comet will carry a lot of crust and dust
on and inside itself. Now, consider a comet of a maximal half-size 7.5 km (like
that of Halley comet (Houpis and Gombosi 1986)) precessing with a period of 3.7
days $\approx$ 89 hours (just as Halley does\footnote{Belton et al 1991, 
Samarasinha and A'Hearn 1991, Peale 1992}). If we once again 
assume the angular resolution of the spacecraft-based equipment to be  $|\Delta
\theta^o|\,=\,0.01$, it will lead us to the following damping time:
\be
\Delta t_{(years)}\;\approx\;5.65 \;\times\;10^{-5}\;\frac{T^3_{(hours)}}{a^2_{
(km)}}\;=\;0.7\;\;year\;\;\;.\;
\label{7.3}
\ee
This means that the cometary-relaxation damping may be measurable.
It also follows from (\ref{7.3}) that, to maintain the observed tumbling state 
of the Comet P/Halley, its jet activity should be sufficiently high. 

\subsection{Missions to Comets, and the Possible Nuclei-Dynamics Studies}

One of the main results of our research is that, according to (\ref{7.3}), 
wobble damping of comets may be registered within a one-year time 
span or so, provided the best available spacecraft-based devices are 
used and the comet's spin state is not too close to the separatrix. 
In the coming years a new generation of comet-aimed missions
will increase the amount and quality of observations. 
Forthcoming enlargement of the comet database will 
result in deepening of our understanding of comet nuclei rotational dynamics.

The Stardust mission, 
that is to 
visit comet P/Wild 2, will be a fly-by one, and thus will not be able to trace
the spin-state evolution. Spececraft Deep Impact will approach comet 
9P Tempel 1 on the 3 of July 2005, and will shoot a 500-kg impactor at 10 km/s
speed, to blast a crater into the nucleus, to reveal its interior. Sadly,
though, this encounter too will be short.

Rosetta orbiter is designed to approach 46/P Wirtanen and to 
escort it for about 2.5 years (Hubert \& Schwehm 1991). Wirtanen seems to be a 
wobbling comet (Samarasinha, Mueller \& Belton 1996), (Rickman \& Jorda 1998), 
and one 
of the planned experiments is observation of its rotation state, to be carried 
out by the OSIRIS camera (Thomas et al 1998). The mission will start about 1.5 
year before the perihelion, but the spin state will be observed only once,
about a 
year before the perihelion, at a heliocentric distance of 3.5 AU. Three 
months later the comet should come within 3 AU which is to be a crucial 
threshold for its jetting activity: at this distance outgassing of water will 
begin. The strongest non-gravitational torques emerge while 
the comet is within this distance from the Sun. It is predominantly during 
this period that wobble is instigated. Hence, it may be good to expand the 
schedule of the spin-state observations: along with the measurement currently 
planned for 3.5 AU before perihelion, another measurement, at a similar 
heliocentrical distance after the perihelion, would be useful. The comparison 
of these observations will provide information of the precession increase 
during the time spent by the comet within the close proximity to the Sun. 
Unfortunately, the Rosetta programme will be over soon afterwards, and a third 
observation at a larger distance will be impossible. The third observation 
performed well outside the 3 AU region might reveal the wobble-damping rate, 
and thereby provide valuable information about the composition and inner 
structure of the comet nucleus. It would be most desirable to perform such a 
three-step observation by the future escort missions to comets. What are the 
chances of success of such an experiment? On the one hand, the 
torques will not be fully eliminated after the comet leaves the 3 AU
proximity of the Sun; though the outgassing of water will cease, some 
faint sublimation of more volatile species (like CO, CO$_2$, CH$_3$OH) will 
persist for long. On the other hand,  the damping rate is high enough and a 
comparison of two observations separated by 
about a year will have a very good chance of registering wobble relaxation.

\section{\underline{APPLICATION TO COSMIC-DUST ALIGNMENT}}

Inelastic relaxation is also important for cosmic-dust grains of sizes within
the interval from $3\times 10^{-7} \;cm$ through $10^{-4} \;cm$. Such 
particles get their long axes aligned with respect to the interstellar magnetic
field (see review by Lazarian 2000 for description of alignment mechanisms).
Absorption and emission of light by such particles is charachterised by
appropriate effective cross sections which depend upon the direction from which
the observer looks at the particle. The difference between cross sections
appropriate to the directions parallel to the long and short axes of the 
granule yields polarisation of the observed starlight, in case this light 
passes through clouds of aligned dust. Dust particles tend to align their 
orientation relative to the interstellar magnetic field, and it turns out that
the polarisation of light depends upon the orientation and magnitude of
the field. This circumstance gives birth to a very 
powerful technique for magnetic field studies, technique that rests on our 
understanding of the grain alignment, and on our ability to observe this 
alignment through measuring the polarisation of starlight passing through dust 
clouds. 

Dust interact with the ambient media and rotate fast. There are a number
of processes that allow grains to spin at a rate much larger than the
typical rate of thermal rotation $\omega_{th}\sim 
(kT_{gas}/J_{grain})^{1/2}$, where $J_{grain}$ is a typical angular momentum
of the grain.
For instance, formation of H$_2$ molecules over catalytic sites on grain
surface gives rise to the so-called Purcell rockets (Purcell 1979),
which can spin up grains by ejecting nascent
H$_2$ molecules.  Under such uncompensated torques the grain will spin-up 
to velocities much higher than $\omega_{th}$, and Purcell termed those
velocities ``suprathermal.''

It is obvious that grain wobble   limits the degree to which the
grain can be aligned. The ideas on the role of this wobble have been evolving
through years.
Initially it was assumed that the distribution of grain rotation
among the axes is thermal $I_i \omega^2_i\sim kT_{grain}$ (see
Jones \& Spitzer 1967). Later Purcell (1979) realised that internal 
relaxation tends to bring  grain angular momentum
${\bf \vec J}$ parallel to the axis of maximal inertia. He discussed
both inelastic relaxation and a more subtle mechanism that he termed
Barnett relaxation. 
 
The Barnett effect is converse of the Einstein-Haas effect.
The Einstein-Haas effect is rotation acquired by a paramagnetic body 
subject to remagnetisation. This happens because the flipping 
electrons transfer the angular momentum (associated with their spins)
 to the lattice. In the case of Barnett effect, 
the rotating body shares a part of its angular momentum with the electron
subsystem; this sharing entails magnetisation. 
A typical value
of the Barnett-induced magnetic moment is ${M}\approx 
10^{-19}\omega_{(5)}$~erg
gauss$^{-1}$ (where $\omega_{(5)}\equiv \omega/10^5{\rm s}^{-1}$).
Purcell offered the following illustrative
explanation of the phenomenon. If a rotating body contains equal amount 
of spin-up and spin-down unpaired electrons, its magnetisation is
nil. Its kinetic energy would decrease, with the total angular
momentum remaining unaltered, if some share of the entire 
angular momentum could be transferred to the spins by turning some of
the unpaired spins over (and, thus, by dissipating some energy). This 
potential possibility is brought to pass through the said coupling.
Another way to understand the effect would be to recall that in the body frame
of a free rotator the angular-momentum and angular-velocity vectors precess
relative to the principal axes. Therefore, the Barnett magnetisation
will, too, precess in the body frame. The resulting remagnetisation will be 
accompanied with energy dissipation (called paramagnetic relaxation). 

An immediate outcome from granule magnetisation is the subsequent coupling of 
the magnetic moment $\;\bf M\;$ (directed along the grain angular velocity)
with the interstellar magnetic field
$\;\bf B\;$: the magnetic
moment precesses about the magnetic line. What is important, is that 
this precession goes at an intermediate rate. On the one hand, it
is slower than the grain's spin about its instantaneous rotation axis. On the
other hand, the precession period is much shorter than the typical time 
scale at which the relative-to-$\bf B\;$ alignment gets established\footnote{
A more exact statement is that the period
of precession (about $\;\bf B\;$) of the magnetic moment $\;\bf M\;$ (and of
$\,Z\,$ aligned therewith) is much shorter than the mean time between two
sequent flip-overs of a spinning granule (Purcell 1979, Roberge et al 1993)}. 
The latter was proven by Dolginov \& Mytrofanov (1976), for magnetisation
resulting from the Barnett effect, and by Martin (1971), for magnetisation 
resulting from the grain's charge.

Paramagnetic relaxation of the body in the external interstellar magnetic
field had long been thought as the cause of the grain alignment. 
The ``Barnett equivalent magnetic field'', i.e. the equivalent external
magnetic field that would cause the same magnetization of the grain  
material, is $H_{BE}=5.6 \times10^{-3} \omega_{(5)}$~G, 
which is much larger than the interstellar magnetic 
field. Therefore the Barnett relaxation takes place on time scale 
$t_{Bar}\approx
4\times 10^7 \omega_{(5)}^{-2}$, 
i.e., essentially instantly compared to the time of paramagnetic alignment
$\sim 10^{11}$~s that establishes in respect to the external magnetic 
field\footnote{Radiative
torques (Draine \& Weingartner 1996, 1997) are currently favoured by
observations as the mechanism of grain alignment (see Lazarian 2002).
The alignment time for the radiative-torque mechanism is much larger
than the internal dissipation time.} 
Thus Purcell (1979) concluded that ${\bf \vec J}$ should be parallel to the
axis of the grain major inertia.  

From the discussion above it looks that the precise 
time of internal dissipation
is irrelevant, if the internal alignment takes place faster than the
alignment in respect to magnetic field. 
This is not true, however. Spitzer \& McGlynn (1979), henceforth
 SM79 observed that adsorption of heavy elements on a grain  should result
in the resurfacing phenomenon that, e.g.  should remove early sites
of H$_2$ formation and create new ones. As the result,
H$_2$ torques will occasionally change their direction and spin the grain
down. SM79  showed that in the absence of
random torques the spinning down grain will
 flip over preserving the direction of its original angular momentum.
However, in the presence of random torques
 this direction will be altered with the maximal deviation inflicted
over a short period of time just before and after the flip, i.e.
during the time when the value of grain angular momentum is minimal.
The actual value of angular momentum during this critical
period depends on the ability of ${\bf \vec J}$ to deviate from
the axis of maximal inertia. SM79 observed that as the internal
 relaxation 
couples ${\bf \vec J}$ with the axis of maximal inertia, it 
makes randomisation of grains during crossover nearly complete.
Due to finite times of the internal relaxation the coupling is
not complete and the grains preserve partially
their alignment during crossovers.
Calculations in SP79 showed that the Barnett relaxation is 
fast enough to make randomisation of grains nearly complete.

Lazarian \& Draine (1997), henthforth LD97, 
revisited the problem of crossovers and
found out that the result is different if the thermal fluctuations
associated with the relaxation mechanism are accounted for (see
Lazarian \& Roberge 1997). LD97 observed that the thermal
fluctuations partially decouple ${\bf \vec J}$ and the axis of maximal
inertia and therefore the value of angular momentum at the moment
of a flip is substantially larger than SM79 assumed. Thus the
randomisation during a crossover is  reduced and  LD97 obtained
a nearly
perfect alignment  for interstellar grains
rotating suprathermally, provided that
the grains were larger than a certain critical size $a_c$.  The latter 
size was found by
equating the time of the crossover and the time of the internal
dissipation $t_{dis}$. For $a<a_c$
Lazarian \& Draine (1999a) found new physical effects, which they termed
``thermal flipping'' and ``thermal trapping''. The thermal flipping
 takes place
as the time of the crossover becomes larger than $t_{dis}$.           
In this situation thermal fluctuations will enable flipovers. However,
being random, thermal fluctuations are likely to produce not a single
flipover, but multiple ones. As the grain flips back and forth, the
regular (e.g. H$_2$) torques average out and the
grain can spend a lot of time rotating with thermal velocity, i.e.
being ``thermally trapped''. The paramagnetic alignment of 
grains rotating with 
thermal velocities is small (see above), 
and therefore grains with $a<a_{c}$ are
expected to be aligned only marginally. The picture of preferential
alignment of large grains, as we know, corresponds to the Serkowski
law and therefore the real issue is to find the value of $a_c$.
In fact, Lazarian \& Draine (1997) showed that $a_c\sim 10^{-5}$~cm 
for the Barnett relaxation mechanism.

These finding stumulated more interest to the competing mechanism
of internal energy dissipation, namely, to the inelastic relaxation
of energy. Lazarian \& Efroimsky (1999) redid the analysis of
Purcell (1979) and obtained the expression for the relaxation time
\begin{eqnarray}
t_{i} \; 
&\approx & a^{5.5} \; [1+(c/a)^2]^4 \; 
\left(\frac{c}{a}\protect\right)^{3/2} \; 
\left( \beta \; kT_{\rm gas} \protect\right)^{-3/2} \; 
\mu \; Q \; \rho^{1/2} \; \frac{2^{11} \; 3^{-2.5}}{63(c/a)^4+20}
\label{6.23}
\end{eqnarray}
where the grain is modelled by a $\;2a\;\times\;2c\;\times\;2c\;$ 
prizm ($\;a\,<\,c\;$).
Unlike our earlier expression (\ref{5.21}), the kinetic energy of
a rotating body is expressed using the factor of suprathermality,
i.e. as $\beta kT_{\rm gas}$. For crossovers thermal rotational 
velocities are important and for those
Lazarian \& Efroimsky concluded that inelastic relaxation may be
dominant for grains larger than $10^{-5}$~cm, provided that
the grains are substantially flat (e.g. axes ratio is 4:1). 
The inelastic relaxation should also dominate atomic-size grains
for which paramagnetic response fails (see discussion of paramagnetic
relaxation within such grains in Lazarian \& Draine 2000). As far as the
critical-size value $a_c$ was concerned, the study by Lazarian
\& Efroimsky (1999) did not change its appreciably apart from the
case of extremely oblate grains.

However,  in a
recent paper, Lazarian \& Draine (1999b) reported a new solid state effect
that they termed ``nuclear relaxation''. This is an analog of Barnett
relaxation effect that deals with nuclei. Similarly to unpaired electrons
nuclei tend to get oriented in a rotating body. However the nuclear analog
of ``Barnett equivalent'' magnetic field is much larger and Lazarian \&
Draine (1999) concluded that the nuclear relaxation can be a million times
faster than the Barnett relaxation. If this is true $a_c$ becomes of the
order $10^{-4}$~cm, which means that the majority of interstellar grains
undergo constant flipping and rotate essentially thermally in spite of
the presence of
uncompensated Purcell torques. For particles rotating at a
thermal Brownian rate, nuclear relaxation dominates the inelastic
one for chunks up to 2~m. However, such large pieces rotate at a
nonthermal rate and therefore inelastic relaxation rate that scales
as $\omega^{3}$ compared with $\omega^{2}$ for the Barnett and
nuclear relaxations is dominant. 

The scaling of inelastic relaxation with angular velocity ensures that
this mechanism is dominant for fast rotating dust particles. Such 
a rotation is possible due to differential scattering of left
and right hand polarized photons (Dolginov \& Mytrophanov 1976,
Draine \& Weingartner 1996, 1997). This difference in scattering
cross sections arises naturally for irregular grains 
(Draine \& Weingartner 1996). 
Unlike the H$_2$ torques proposed by Purcell (1979), the radiative torques
are not fixed in the body coordinates. Therefore they do not average out
during fast thermal flipping of the grain. Hence, it is likely that they can 
provide
means for suprathermal rotation for grains that are larger than the
wavelength of the incoming radiation. The rate of internal relaxation 
for suprathermally rotating grains is
important for special circumstances when the alignment happens very
fast, e.g. in the stellar winds and comet atmospheres. 

Another interesting consequence of the coupling of rotational and
vibrational degrees of freedom, resulting from the inelastic dissipation,
is the expected relation between microwave and infrared polarisation
arising from ultrasmall ($a<10^{-7}$) grains. Dipole emission from those
grains was invoked by Draine \& Lazarian (1998a,b) to explain the anomalous
microwave emission of galactic origin within the range of 10-90~GHz.
This emission was shown to interfere with the intended measurements
of Cosmic Microwave Background (CMB), and therefore attracted a lot
of attention (see Lazarian \& Prunet 2002 for a review). In view
of the attempts to measure the CMB polarisation, it is important to know
whether or not the emission from small grains will be polarised. This problem 
has not had been fully resolved as yet.  

It is well known that the infrared emission from ultrasmall granules (the one 
of wavelength 12 $\mu$m) takes place as the dust particles absorb UV photons. 
Can we check the alignment of ultrasmall grains via infrared polarimetry? The 
answer to this question depends on the efficiency of internal relaxation. 
Indeed, UV photons raise the grain temperature and randomise grain axes in 
relation to its angular momentum vector (see Lazarian \& Roberge 1997) on the 
time scale of rotation-vibrational coupling. This is the time of internal
relaxation. Taking the necessary figures for Barnett relaxation from Lazarian 
\& Draine (1999), we get the randomisation time of a $10^{-7}$~cm-sized grain 
to be $2\times 10^{-6}$~s, which is less than the grain cooling time. As a
result, the emanating infrared radiation would be polarised very marginally.
If, however, the Barnett relaxation is suppressed due to small-size effects 
(Lazarian \& Draine 2000), the randomisation time is determined by inelastic 
relaxation and is $\sim 0.1$~s, which would entail a partial polarisation of
the infrared emission\footnote{~We would remind, that even though the inelastic
relaxation of a dust granule is several orders more effective that was presumed
in (Purcell 1979), its intensity (as compered to the Barnett relaxation) 
depends upon the granule size (Lazarian \& Efroimsky 1999). For the smallest 
granules, the Barnett dissipation still dominates (Lazarian \& Draine 2000).}. 
The uncertainty of this estimate arises from extrapolarion of dust properties 
and a classical treatment of an essentially quantum system. Further research in
this direction is necessary.

\section{\underline{CONCLUSIONS}}

{\it 1.} In the article this far we have described the present situation in the
   studies of the dynamics of an unsupported top, and some of its 
   applications to tumbling asteroids and comets and to the cosmic-dust 
   particles. We addressed relaxation of excited (out of principal state) 
   rotators through energy dissipation resulting from nutation-caused stresses.

{\it 2.} In many spin states of an unsupported rotator, dissipation at frequencies 
   different from the 
   nutation frequency makes a major input into the inelastic-relaxation 
   process. These frequencies are overtones of some "basic" frequency, that is 
   LOWER than the precession frequency. This is a very unusual example of 
   nonlinearity: the principal frequency (precession rate) gives birth not only
   to  higher frequencies but also to lower frequencies.
   
{\it 3.} In many spin states, the inelastic 
   relaxation far more effective than believed hitherto.

{\it 4.} However, if the rotation states that are close to the separatrix on Fig.2,  
   the lingering effect takes place: both precession and precession-damping 
   processes slow down. Such states (especially those close to the homoclinic 
   point) may mimic the principal rotation state.  

{\it 5.} A finite resolution of radar-generated images puts a limit on our ability of
   recognising whether an object is nutating or not. Nutation-caused changes of
   the precession-cone half-angle may be observed. Our estimates show that we 
   may be very close to observation of the relaxational dynamics of wobbling 
   small Solar System bodies, dynamics that may say a lot about their structure
   and composition and also about their recent histories of impacts and tidal 
   interactions. Monitoring of a wobbling comet during about a year after it 
   leaves the 3 AU zone will, most probably, enable us to register its 
   precession relaxation. 

{\it 6.} Measurements of the damping rate will provide us with valuable 
   information on attenuation in small bodies, as well as on
   their recent histories of impacts and tidal interactions

{\it 7.} Since inelastic relaxation is far more effective than presumed earlier, 
   the number of asteroids expected to wobble with a recognisable half-angle of
   the precession cone must be lower than expected. (We mean the predictions 
   suggested in (Harris 1994).) Besides, some of the small bodies may be in the
   near-separatrix states: due to the afore mentioned lingering effect, these 
   rotators may be ``pretending'' to be in a simple rotation state.

{\it 8.} Though the presently available theory predicts a much higher relaxation rate
   than believed previously, this high rate may still be not high enough to 
   match the experimentally available data. In the closemost vicinity of the 
   principal spin state the relaxation rate must decrease and the rotator must 
   demonstrate the "exponentially-slow finish". Asteroid 433 Eros is a 
   consolidated rotator whose $Q$-factor should not be too low. It is possible 
   that this asteroid was disturbed sometimes in its recent history by the 
   tidal forces. Nevertheless, it shows no visible residual precession. Hence, 
   there may be a possibility that we shall have to seek even more effective 
   mechanisms of relaxation. One such mechanism may be creep-caused deformation
   leading to a subsequent change of the position of the principal axes in the 
   body.

{\it 9.} Inelastic coupling of the cosmic-dust grain's rotational and vibrational  
   degrees of freedom influences randomisation of grain axes when an ultrasmall
   grain absorbs a UV photon. As a result, the microwave dipole emission 
   arising from ultrasmall grains may be polarised, while the near-infrared 
   emission arising from the same grains may be unpolarised.\\ 

~\\

{\Large{\underline {\bf Acknowledgements}}}\\
~\\
The authors wish to express their gratitude to the colleagues who 
    participated, through direct collaboration as well as through advice 
    and discussion, in the afore described research. ME would like to deeply 
    thank William Newman for numerous fruitful discussions and for offering 
    several highly illustrative examples that were included in the text. 
    AL has the pleasure to thank Bruce Draine, Roger Hildebrand and John Mathis
    for stimulating exchanges. 
    VS wants to acknowledge the contribution from Anatoly Neishtadt, Daniel 
    Scheeres and Alexey Vasiliev.
    The work of AL was supported by the NSF through grant AST-0125544.
    The work of VS was supported through the NASA JURRISS Grant NAG5-8715 
    and INTAS Grant 00-221.

\pagebreak


\begin{thebibliography}{99} 

\bibitem{Abramovitz} Abramovitz, M., \& I.A. Stegun 1965, {\emph{Handbook of 
           Mathematical
           Functions, Chapter 16}}, National Bureau of Standards 

\bibitem{Ahrens} Ahrens, T.J. 1995 Ed., {\emph{Mineral Physics \& Crystallography. 
           A Handbook of Physical Constants.}} American Geophysical Union, 
           Washington DC 

\bibitem{d15}
Arnold, V.I.: {\it Mathematical Methods of Classical Mechanics.} Springer,
New York, 1978.
\bibitem{Asphaug 1} Asphaug, E., S.J. Ostro, R. Hudson, D.J. Scheeres, \& W.
                    Benz. 1998,
                    {\it Nature}, {Vol. \bf{393}}, pp. 437

\bibitem{Asphaug 2} Asphaug, E., \& Scheeres, D.J. 1999, Deconstructing 
                    Castalia: Evaluating a Postimpact State. {\it Icarus}, 
                    Vol. {\bf 139}, pp. 383 - 386

\bibitem{Asphaug 3} Asphaug, E., \& W. Benz 1994. Density of comet 
                    Shoemaker-Levy 9 deduced by modelling breakup of the 
                    parent ''rubble pile''. {\it{Nature}}, Vol. {\bf 370}, 
                    pp. 120 - 124

\bibitem{Asphaug 4} Asphaug, E., \& W. Benz 1996. Size, density, and 
                    structure of comet Shoemaker-Levy 9 inferred from the 
                    physics of tidal breakup. {\it{Icarus}}, Vol. {\bf 121}, 
                    pp. 225 - 248                    

\bibitem{Belton} Belton, M.J.S., B.E.A.Mueller, W.H. Julian, \& A.J. 
                 Anderson. 1991. The spin state and homogeneity of Comet 
                 Halley's nucleus. {\it Icarus}, Vol. {\bf{93}}, pp. 183 - 
                 193

\bibitem{Black} Black, G.J, P.D.Nicholson, W.Bottke, J.Burns, \& Allan W.  
           Harris 1999. On a possible rotation state of (433) Eros. {\it 
           Icarus}, Vol. {\bf 140}, pp. 239 - 242

\bibitem{Bottke 1} Bottke Jr., W.F., \& Jay Melosh 1996a. Formation of 
                   asteroid satellites and doublet craters by planetary tidal
                   forces. {\emph{Nature}}, Vol. {\bf 381}, pp. 51 - 53     

\bibitem{Bottke 2} Bottke Jr., W.F., \& Jay Melosh 1996b. Binary Asteroids 
                   and the Formation of Doublet Craters. {\emph{Icarus}}, 
                   Vol.{\bf  124}, pp. 372 - 391                 

\bibitem{Bottke 3} Bottke Jr., W.F., D.C.Richardson, P.Michel, \& S.G.Love 
                   1999. ~1620 Geographos and 433 Eros: shaped by planetary 
                   tides? {\emph{The Astronomical Journal}}, Vol. {\bf 117}, 
                   pp. 1921 - 1928    

\bibitem{Bottke 4} Bottke Jr., W.F., D.C.Richardson, \& S.Love 1998. {\emph{
                   Planetary and Space Sciences}},
                   Vol. {\bf 46}, pp. 311 - 322

\bibitem{Bottke 5} Bottke Jr., W.F. 1998. Are Asteroids Rubble Piles? {\it{
                   Paper for the 23rd Meeting of the International Seminars 
                   on Planetary Emergencies. Erice, Italy, 10 September 
                   1998}}. See also
         {\it{~~http://astrosun.tn.cornell.edu/staff/bottke/rubble/rub.html}}

\bibitem{Brennan} Brennan, B.J. 1981, in: Anelasticity in the Earth (F.D.Stacey, 
           M.S.Paterson and A.Nicolas, Editors), Geodynamics Series
           4, AGU, Washington.

\bibitem{Burns Safronov} Burns, J. \& V.Safronov 1973, {Asteroid Nutation 
           Angles}. {\it MNRAS}, Vol. {\bf 165}, p. 403 - 411

\bibitem{Burns 1} Burns, J. 1975, The Angular Momenta of Solar System Bodies:
           Implications
           for Asteroid Strengths. {\it Icarus}, Vol. {\bf 25}, p. 545 - 554 
          
\bibitem{Burns 2} Burns, J. 1977, in: {\it Planetary Satellites} (J.Burns, 
           Ed.), Univ. of Arizona Press, Tuscon

\bibitem{Burns 3} Burns, J. 1986, in: {\it Satellites} (J.Burns, Ed.),
           Univ. of Arizona Press, Tuscon

\bibitem{Chernous'ko} Chernous'ko, F. 1968 {{Motion of a Rigid Body with 
           Cavities Containing Viscous Liquid}} (in Russian)

\bibitem{Clark} Clark V.A., B.R.Tittman \& T.W.Spencer 1980. 
               {\it J. Geophys. Res.}, Vol. {\bf 85}, p. 5190

\bibitem{d8}
Crifo, J.F., \& A.V.Rodionov: 1999, `Modelling the circumnuclear coma 
of comets: objectives, methods and recent results', 
{\it Planetary and Space Science}, Vol.~{\bf 47}, pp.~797--826.

\bibitem{Zappala} dell'Oro, A., P. Paolicchi, A. Cellino, V. Zappala, P. Tanga,
                  \& P. Michel. 2001. The Role of Families in Determining 
                  Collision Probability in the Asteroid Main Belt.
                  {\it{Icarus}}, Vol.~{\bf 153}, pp.~{52 - 60}.

\bibitem{Denisov Novikov} Denisov, G.G., \& Novikov, V.V. 1987. {Free motions
                  of an elastic ellipsoid.} {\it Izvestija AN SSSR. Mekhanika
            Tverdogo Tela}, Vol. {\bf 22}, No 6, pp. 69 - 74 

\bibitem{Dolginov} Dolginov, A. Z., \& I. G.Mitrofanov. 1976.
           Orientation of Cosmic-Dust Grains.
           {\it{Astrophysics and Space Science}}, Vol. {\bf{43}}, pp.~291 - 317

\bibitem{Draine Weingartner} Draine, B.T. \& J.C. Weingartner. 1996.
             Radiative Torques on Interstellar Grains. I. Suprathermal Spin-up.
             {\it The Astrophysical Journal}, Vol.~{\bf 470}, pp.~551-565

\bibitem{Weingartner Draine} Draine, B.T. \& J.C. Weingartner.  1997. 
                  Radiative Torques on Interstellar Grains. I. Grain Alignment.
                 {\it The Astrophysical Journal}, Vol.~{\bf 480}, pp.~633 - 646

\bibitem{Efroimsky 2002} Efroimsky, Michael 2002. Euler, Jacobi, and Missions
                         to Comets and Asteroids. {\it{Advances in Space 
                         Research}}, Vol. {\bf{29}}, pp.~725 - 734

\bibitem{Efroimsky 2001} Efroimsky, Michael 2001. Relaxation of Wobbling 
                         Asteroids and Comets. Theoretical Problems. 
                         Perspectives of Experimental Observation. 
                         {\it{Planetary \& Space Science}}, {Vol. \bf{49}}, 
                         pp.~937 - 955
 
\bibitem{Efroimsky Lazarian} Efroimsky, Michael, \& A.Lazarian 2000. 
                             Inelastic Dissipation in Wobbling Asteroids and 
                             Comets. {\it{Monthly Notices of the Royal 
                             Astronomical Society of London}}, Vol. 
                             {\bf 311}, 
                             pp.~269 - 278
 
\bibitem{Efroimsky 2000} Efroimsky, Michael 2000. {Precession of a Freely 
                         Rotating Rigid Body. Inelastic Relaxation in the 
                         Vicinity of Poles}. {\it{Journal of Mathematical 
                         Physics}}, Vol.{\bf  41}, p. 1854 

\bibitem{Euler 1765} Euler, Leonhard 1765. {\it{Theoria motus corporum 
                     solidorum seu rigidorum ex primus nostrae cognitionis 
                     principiis stabilita et ad omnes motus qui in huiusmodi 
                     corpora cadera possunt accomodata.}} (In Latin). Recent 
                     edition: {\it{Leonhard Euler. Series II. Opera mechanica
                     et astronomica.}} Vol. 3-4. Birkhauser Verlag AG, 
                     Switzerland 1999

\bibitem{Fowles Cassiday} Fowles, G. \& G. Cassiday 1986 {\emph{Analytical 
           Mechanics.}} Harcourt Brace \& Co, Orlando FL

\bibitem{Giblin Farinella} Giblin, Ian, \& Paolo Farinella 1997. Tumbling 
           Fragments from 
           Experiments Simulating Asteroidal Catastrophic Disruption. {\it 
           Icarus}, Vol. {\bf 127}, pp. 424 - 430

\bibitem{Giblin} Giblin, Ian, G.Martelli, P.Farinella, P.Paolicchi, \& M. Di 
           Martino 
           1998.  The Properties of Fragments from Catastrophic Disruption 
           Events  {\it Icarus}, Vol.{\bf  134}, pp. 77 - 112

\bibitem{Goldreich Soter} Goldreich, P., \& S. Soter 1965. Q in the Solar 
           System. {\it 
           Icarus}, Vol. {\bf 5}, pp. 375 - 454

\bibitem{Harris 1} Harris, Alan W. 1994. Tumbling Asteroids. {\it Icarus}, 
           Vol. {\bf 107}, pp.
           209 - 211

\bibitem{Harris 2} Harris, Alan W. 1996. The rotational states of very small 
           asteroids: evidence for 
           ''rubble-pile structure''. {\it{Lunar. Planet. Sci.}}, Vol. {\bf 
           XXVII}, pp. 493 - 494           

\bibitem{Harris 3} Harris, Alan W. 1998. Making and Braking Asteroids. {\it 
           Nature}, Vol. {Vol. \bf 393}, pp. 418 - 419

\bibitem{Houpis Gombosi} Houpis, Harry L. F., \& Gombosi, Tamas I. 1986. In: 
           {\it Proc. 20-th
           ESLab Symp. on the Exploration of Halley's Comet, Vol. II.} 
           European Space Agency, Paris.

\bibitem{Hudson Ostro} Hudson, R.S., \& S. J. Ostro 1995. Shape and 
           Non-Principal-Axis Spin State of 
           Asteroid 4179 Toutatis. {\it Science}, Vol. {\bf 270}, pp. 84 - 86

\bibitem{Hubert Schwehm} Hubert, M. C. E., \& G. Schwehm 1991. {\it {Space 
           Science Review}}, 
           {Vol. \bf{56}}, p. 109 

\bibitem{Jackson} Jackson, Ian 1986, in: {\emph{Mineral and Rock Deformation. 
           Laboratory Studies}}, American Geophysical Union, Washington DC

\bibitem{Jacobi 1} Jacobi, Karl Gustav Jacob 1829 {\it{Fundamenta 
           nova theoria 
           functionum ellipticarum}} Berlin

\bibitem{Jacobi 2} Jacobi, Karl Gustav Jacob 1849, {\emph {Journal fur 
           reine und 
           angevandte Mathematik}} (Berlin), {Vol. \bf{39}}, pp. 293 - 350 

\bibitem{Jacobi 3} Jacobi, Karl Gustav Jacob 1882 {\emph{Gessamelte Werke}}, 
           {Vol. \bf{2}}, pp. 427 - 510. Berlin

\bibitem{je}
Jewitt, D.: 1997, `Cometary rotation: an overview',
{\it Earth, Moon, and Planets}, Vol.~{\bf 79}, pp.~35 - 53.

\bibitem{Jones Spitzer} Jones, R.V. \& L. Spitzer. 1967. Magnetic Alignment of 
                        Interstellar Grains. {\it The Astrophysical Journal}, 
                        Vol.~{\bf{147}}, pp.~943 - 964

\bibitem{d1}
Jorda, L., \& J.Licandro. 2002. ({\it in press}), `Modeling the rotation of comets',
Proceedings of the IAU Colloquim 168, held in Nanjing, China.
Pub. Astron. Soc. Pacific.

\bibitem{Julian} Julian, W. 1990. The Comet Halley nucleus. Random jets. {\it 
           Icarus}, Vol. {\bf 88}, pp. 355 - 371

\bibitem{Karato} Karato, S.-i. 1998. {\it{Pure and Applied Geophysics}},
                 Vol.~{\bf 153}, pp.~239

\bibitem{Klinger} Klinger J., A.-C. Levasseur-Regourd, N. Bouziani, \& Enzian 
           A. 1996. 
           Towards a model of cometary nuclei for engineering studies for 
           future space missions to comets. {\it Planet. Space. Sci.}, Vol. 
           {\bf 44}, pp. 637 - 653

\bibitem{Knopoff} Knopoff, L. 1963. Q.  {\it Reviews of Geophysics}, Vol. 
                  {\bf 2}, pp. 625 - 629 

\bibitem{Lagrange} Lagrange, J. 1813 {\it{Mecanique analitique.}} 
                   Paris 

\bibitem{Lambeck 1} Lambeck, Kurt 1980 {\emph{The Earth's Variable Rotation: 
           Geophysical Causes and Consequencies}}, Cambridge University 
           Press, Cambridge, U.K.

\bibitem{Lambeck 2} Lambeck, Kurt 1988 {\emph{Geophysical Geodesy}}, Oxford 
                  University
           Press, Oxford \& NY

\bibitem{Lamy Burns} Lamy, Philippe L., \& Joseph A. Burns 1972. Geometrical 
           Approach to a 
           Torque-Free Motion of a Rigid Body Having Internal Energy 
           Dissipation. 
           {\it{Amer. J. Phys.}},  {{Vol. \bf 40}}, pp. 441 - 445

\bibitem{Landau Lifshitz} Landau,  L.D. \& Lifshitz, E.M. 1976 
           {\emph{Mechanics}}, Pergamon Press, NY 

\bibitem{Landau Lifshitz} Landau,  L.D. \& Lifshitz, E.M. 1970 {\emph{Theory 
           of Elasticity}}, Pergamon Press, NY

\bibitem{Lazarian 1} Lazarian, A.  1994. Gold-Type Mechanisms of Grain 
                     Alignment. {\it MNRAS},  {Vol. \bf 268}, pp. 713 - 723

\bibitem{Lazarian 2} Lazarian, A.  2000. Physics of Grain Alignment.
In: {\it Cosmic Evolution and Galaxy 
Formation}, ASP Vol. {\bf 215}, pp. 69 - 79. Eds. Jose Franco, Elena Terlevich,
Omar Lopez-Cruz. See also {\it astro-ph/0003314}

\bibitem{Lazarian Draine 1} Lazarian, A. \& Draine, B.T. 1997.  Disorientation 
           of Suprathermally Rotating Grains and the Grain Alignment Problem.
           {\it The Astrophysical Journal}, Vol. {\bf 487}, pp. 248 - 258 

\bibitem{Lazarian Draine 2} Lazarian, A. \& Draine, B.T. 1999a.
Thermal Flipping and Thermal Trapping: New Elements in Grain Dynamics. 
{\it The Astrophysical Journal}, Vol. {\bf 516}, pp.  L37 - L40

\bibitem{Lazarian Draine 3} Lazarian, A. \& Draine, B.T. 1999b.
                            Nuclear Spin Relaxation within Interstellar Grains.
                {\it The Astrophysical Journal}, Vol. {\bf 520}, pp. L67 - L70

\bibitem{Lazarian Draine 4} Lazarian, A. \& Draine, B.T. 2000.
Resonance Relaxation.
{\it The Astrophysical Journal}, Vol. {\bf 536}, pp. L15 - L18

\bibitem{Lazarian Efroimsky} Lazarian, A. \& Michael Efroimsky 1999. 
           Inelastic Dissipation
           in a Freely Rotating Body: Application to Cosmic Dust Alignment.  
           {\it MNRAS}, Vol. {\bf 303}, pp. 673 - 684 

\bibitem{Lazarian Prunet} Lazarian, A., \& S. Prunet. 2002. 
                          Polarised Microwave Emission from Dust. 
In: {\it{Proc. of the AIP Conference "Astrophysical Polarized Backgrouds,"}} 
Eds.: S. Cecchini, S. Cortiglioni, R. Sault, and C. Sbarra, pp.~32 - 44

\bibitem{Legendre} Legendre, Adrien-Marie 1837  {\emph{Traite des fonctions 
           elliptiques }}

\bibitem{Marsden} Marsden, B. \& S. Nakano 1993, IAU Circ. No 5800, 22 May 
           1993; 
           Marsden, B. \& A. Carusi 1993, IAU Circ. No 5801, 22 May 1993.

\bibitem{d9}
Marsden, B.G., Z.Sekanina, \& D.K.Yeomans. 1973.
`Comets and nongravitational forces.V', {\it Astron. J.},
Vol.~{\bf 78}, pp.~211--225.

\bibitem{Marsden 2000} Marsden, J. 2000. {\it{Lectures on Mechanics.}} 
           Cambridge University Press.

\bibitem{Martin} Martin, P.G. 1971. On interstellar grain alignment by a magnetic field. {\it Monthly Notices of the Royal
           Astronomical Society}, Vol.~{\bf 153}, p.~279

\bibitem{Meech} Meech, K. J., M. J. S. Belton, B. Mueller, M. Dicksion \& 
           H. Li 
           1993. Nucleus Properties of P/Schwassmann-Wachmann 1. {\it{Astron.
           J.}}, Vol. {\bf 106}, pp. 1222 - 1236.

\bibitem{Melosh} Melosh, H.J. \& P. Schenk 1993. Split comets and the origin 
           of crater 
           chains on Ganymede and Callisto. {\it{Nature}}, Vol. {\bf 365}, 
           pp. 731 - 733

\bibitem{Miller} Miller, J.K., P.J. Antreasian, R.W. Gaskell, J. Giorgini, 
           C.E. Helfrich, W.M Owen, B.G. Williams, \& D.K. Yeomans, 1999,
           {\it Determination of Eros Physical Parameters for Near Earth 
           Asteroid Rendezvous Orbit Phase Navigation.} Paper AAS 99-463, 
           presented at the AAS/AIAA Astrodynamics Specialist Conference, 
           Girdwood, Alaska, August 1999.

\bibitem{Mitchell Richardson} Mitchell, J. W., D. L. Richardson. 2001.
                 A Simplified Kinetic Element Formulation for the Rotation of
                 a Perturbed Mass-Asymmetric Rigid Body.
                 {\it{Celestial Mechanics and Dynamical Astronomy}},
                 Vol.~{\bf{81}}, pp.~13 - 25

\bibitem{Molina} Molina, A., F. Moreno, \& F. Mart{\'i}nez-L{\'o}pez. 2002. 
                   Communication at the {\it ACM2002} International Conference.
                   Berlin, 29 July - 2 August 2002. 

\bibitem{Muinonen} Muinonen, K., \& J.S.V. Lagerros. Inversion of shape 
                   statistics for small solar system bodies. {\it{Astron. \& 
                   Astrophys.}} Vol. {\bf 333}, pp. 753 - 761 (1998)

\bibitem{Neishtadt} Neishtadt, A. I. 1980. {\it{Mechanics 
           of Solids}}, {Vol. \bf{15}}, No 6, pp. 21 

\bibitem{f4}
Neishtadt, A.I., D.J.Scheeres, V.V.Sidorenko, \& A.A.Vasiliev: 2002,
`Evolution of comet nucleus rotation', {\it Icarus}, Vol.~{\bf 157},
pp.~205--218.

\bibitem{Nowick Berry} Nowick, A., \& Berry, D. 1972 {\emph{Anelastic 
                       Relaxation in 
           Crystalline Solids}}, Acad. Press  

\bibitem{Okubo} Okubo, S. 1982, Geophys.J., Vol. 71, p. 647

\bibitem{Ostro 1} Ostro, S.J., R.F.Jurgens, K.D.Rosema, R.Whinkler,
           D.Howard, R.Rose, D.K.Slade, D.K.Youmans, D.B.Cambell, 
           P.Perillat,  J.F.Chandler, I.I.Shapiro,  R.S.Hudson, P.Palmer, 
           \& DePater, I. 1993. {\it BAAS}, Vol.{\bf 25}, p. 1126

\bibitem{Ostro 2} Ostro, S.J., R.S.Hudson, R.F.Jurgens, K.D.Rosema, R.Winkler, 
           D.Howard, R.Rose, M.A.Slade, D.K.Yeomans, J.D.Giorgini, 
           D.B.Campbell, P.Perillat, J.F.Chandler, \& I.I.Shapiro, 1995. 
           Radar Images of Asteroid 4179 Toutatis. {\it Science}, Vol. 
           {\bf 270}, 
           pp. 80 - 83

\bibitem{Ostro 3} Ostro, S.J., R.S.Hudson, K.D.Rosema, J.D.Giorgini, 
           R.F.Jurgens, 
           D.Yeomans, P.W.Chodas, R.Winkler, R.Rose, D.Choate, R.A.Cormier, 
           D.Kelley, R.Littlefair, L.A.Benner, M.L.Thomas, \& M.A.Slade
           1999. Asteroid 4179 Toutatis: 1996 Radar Observations. {\it 
           Icarus},
           Vol. {\bf 137}, pp. 122-139

\bibitem{Ostro 4} Ostro, S.J., P.Pravec, L.A.Benner, R.S.Hudson, L.Sarounova,
           M.Hicks, D.L.Rabinovitz, J.V.Scotti, D.J.Tholen, M.Wolf, 
           R.F.Jurgens, M.L.Thomas, J.D.Giorgini, P.W.Chodas, D.Yeomans, 
           R.Rose, R.Frye, K.D.Rosema, R.Winkler, \& M.A.Slade 1999.
           Radar and Optical Observations of Asteroid 1998 KY26. {\it 
           Science},
           Vol. {\bf 285}, pp. 557 - 559

\bibitem{Peale 1} Peale, S. J. 1973. Rotation of Solid Bodies in the Solar 
           System. 
           {\it{Rev. Geophys. Space Phys.}}, Vol. {\bf 11}, pp. 767 - 793

\bibitem{Peale Lissauer} Peale, S.J., \& J.J.Lissauer, 1989. Rotation of 
           Halley's 
           Comet. {\it Icarus}, Vol. {\bf 79}, pp. 396-430

\bibitem{Peale 2} Peale, S.J. 1992. {{On LAMs and SAMs for the rotation of
           Halley's Comet.}} In: {\emph {Asteroids, Comets, Meteors 1991.}} 
           A.Harris and E. Bowles Eds. Lunar and Planetary Institute, Houston
           TX, pp. 459 - 467.

\bibitem{Peale 3} Peale, S.J., P.Cassen \& R.T.Reynolds 1979. Meltin of Io by
            tidal dissipation. 
           {it{Science}}, Vol. {\bf 203}, pp. 892 - 894

\bibitem{Poisson} Poisson 1813 {\emph {J. Ecol. Polyt.}}, Cah. 16, p. 274-264

\bibitem{Pravec Harris} Pravec, P., \& A. W. Harris 2000. Fast and Slow 
           Rotation of Asteroids.
           {\it{Icarus}}, Vol. {\bf 148}, pp. 12 - 20

\bibitem{Prialnik} Prialnik, D. 1999. Modelling gas and dust release from 
           comet Hale-Bopp. 
           {\emph{Earth, Moon and Planets}}, Vol. {\bf 77}, pp. 223 - 230

\bibitem{Prendergast} Prendergast, Kevin H. 1958. The Effects of Imperfect 
           Elasticity in Problems of Celestial Mechanics. {\it Astronomical 
           Journal}, Vol. {\bf 63}, pp. 412- 414

\bibitem{Prokof'eva 1} Prokof'eva, V.V.,  L.G.Karachkina, \&  V.P.Tarashchuk 
           1997. 
           Light Variations of Asteroid 1620 Geographos during its Encounter 
           with the Earth in 1994. {\it{Astronomy Letters}}, Vol. {\bf  23}, 
           pp. 758 - 767. 

\bibitem{Prokof'eva 2} Prokof'eva, V.V., V.P.Tarashchuk, \& L.G.Karachkina. 
           1996. 
           Precession of asteroid 1620 Geographos. {\it{Odessa Astronomical 
           Publications}}, Vol. {\bf 9}, p. 188.

\bibitem{Purcell} Purcell, E.M. 1979. Suprathermal Rotation of Interstellar 
           Grains. 
           {\it Astrophysical Journal}, Vol. {\bf 231}, p. 404 - 416

\bibitem{Richardson Mitchell} Richardson, D.L., \& J.W. Mitchell. 1999.
          A Simplified Variation of Parameters Approach to Euler's Equations.
          {\it{Journal of Applied Mechanics}}, {Vol. \bf 66}, pp.~273 - 276

\bibitem{Richardson} Richardson, D.C., W.F.Bottke, \& S.G.Love 1998. Tidal 
                     Distortion and Disruption of Earth-Crossing Asteroids. 
                     {\it Icarus}, Vol. {\bf 134}, pp. 47 - 76

\bibitem{Rickman Jorda} Rickman, H., \& L. Jorda 1998. Comet 46P/Wirtanen, 
               The Target of the Rosetta Mission, {\it{Adv. Space Res.}}, 
               Vol. {\bf 21}, pp. 1491 - 1504

\bibitem{Ryabova} Ryabova, G.O. 2002. Asteroid 1620 Geographos: Rotation.
                  {\it{Solar System Research}}. Vol.!{\bf 36}, pp.~168 - 174

\bibitem{Roberge} Roberge, W.G., T. A. DeGraff, \& J.E. Flaherty. 1993. The Langevin 
           Equation and Its Application to Grain Alignment in Molecular Clouds
           {\it Astrophys. J.}, {\bf 418}, p.~287

\bibitem{Ryan Blevins} Ryan, M.P., \& J.Y.K.Blevins. 1987 U.S. Geological Survey
           Bulletin, Vol.1764, p.1

\bibitem{Sagdeev} Sagdeev R.Z., Szego K., Smith B.A., Larson S., Merenyi E., 
           Kondor 
           A., \& Toth I. 1989, The rotation of P/Halley. {\it Astron.J.}, Vol. 
           {\bf 97}, pp. 546 - 551

\bibitem{Samarasinha 1} Samarasinha, N.H., \& M.F. A'Hearn 1991, 
           Observational and dynamical
           constraints on the rotation of Comet P/Halley. {\it Icarus}, Vol. 
           {\bf 93}, pp. 194 - 225

\bibitem{Samarasinha 2} Samarasinha, N.H., B.E.A. Mueller, \& M.J.S. Belton.
           1996, 
           Comments on the Rotational State and Non-gravitational Forces of 
           Comet 46P/Wirtanen. {\it Planetary and Space Science}, Vol. {\bf 
           44}, pp. 
           275 - 281

\bibitem{Samarasinha 3} Samarasinha, N.H., \& M.J.S. Belton 1995, Long-term 
           evolution of 
           rotational stress and nongravitational effects for Halley-like 
           cometary nuclei. {\it Icarus}, Vol. {\bf 116}, pp. 340 - 358

\bibitem{Samarasinha 4} Samarasinha, N.H., B. Mueller \& M. Belton 1999. {\it
            {Earth, Moon and Planets}}, {Vol. \bf{77}}, p. 189 

\bibitem{Scheeres} Scheeres, D.J., S.J.Ostro, R.S.Hudson, S.Suzuki, \& E. de 
           Jong 
           1998. Dynamics of Orbits Close to Asteroid 4179 Toutatis. {\it 
           Icarus}, Vol. {\bf 132}, pp. 53-79 

\bibitem{Sekanina} Sekanina Z. 1982, in: {\emph{Comets}} (L.Wilkening, Ed.) 
           U. of Arizona Press, Tuscon, pp. 251 - 287

\bibitem{Stacey} Stacey, Frank D. 1992. {\emph{Physics of the Earth}}, 
                 Brookfield Press, Brisbane

\bibitem{Stooke Abergel} Stooke, P. J., \& A. Abergel. 1991.
                         Morphology of the Nucleus of Comet P/Halley
                         {\it{Astronomy \& Astrophysics}}, Vol.~{\bf 248},
                         pp.~656 - 668

\bibitem{Spitzer McGlynn} Spitzer, L., Jr. \& T.A. McGlynn. 1979. 
                Disorientation of Interstellar Grains in Suprathermal Rotation.
                {\it The Astrophysical Journal.} Vol. {\bf 231}, pp. 417 - 424

\bibitem{Synge Griffith} Synge, J.L., \& B. Griffith 1959. {\emph{Principles 
                of Mechanics, Chapter 14}}, McGraw-Hill NY

\bibitem{Thomas} Thomas, N., and 41 colleagues, 1998. {\it{Advances in Space 
                 Research}}, {Vol. \bf{21}}, pp. 1505

\bibitem{Thomson} Thomson, W.T. 1961 {\it{Introduction to Space 
                  Dynamics}.} Wiley, NY

\bibitem{Tittman} Tittman, B., L.Ahlberg, \& J.Curnow 1976, in {\it Proc. 
                  7-th Lunar Sci. Conf.}, pp. 3123 - 3132

\bibitem{Tschoegl} Tschoegl, N.W. 1989. {\emph{The Phenomenological Theory of
                   Linear Viscoelastic Behaviour. An Introduction.}} 
                   Springer-Verlag, NY 

\bibitem{Veverka} Vasil'ev, V.G., \& V.M.Kovtunenko. 1969. 
                  On stability of revolution of a spacecraft with
                  hinge-joined rods. {\it{Kosmicheskie Issledovanija}}, Vol. 
                  {\bf 7}, No 5. ({\it in Russian})

\bibitem{Veverka} Veverka, J., P.Thomas, A.Harch, B.Clark, J.F.Bell III, 
                  B.Carcich, J.Joseph, C.Chapman, W.Merline, M.Robinson, 
                  M.Malin, L.A.McFadden, S.Murchie, S.E.Hawkins III, 
                  R.Farquhar, N.Izenberg, \& A.Cheng 1997. NEAR's Flyby of 
                  253 Mathilde. Images of the Asteroid. 
                  {\it{Science}}, Vol. {\bf 278}, pp. 2109 - 2114

\bibitem{Weidenschilling}  Weidenschilling, S. 1981. How fast can an asteroid 
                           spin? {\it{Icarus}}, Vol.~{\bf 46}, pp.~124 - 126

\bibitem{r18}     Whipple, F.L.: 1951, 
                  `A comet model.I. Acceleration of comet Encke' 
                  {\it Astrophys. J.}, Vol.~{\bf 113}, pp.~464--474. 

\bibitem{Wilhelm} Wilhelm, Klaus. 1987. Rotation and precession of comet 
                  Halley. {\it Nature}, Vol. {\bf 327}, pp. 27 - 30

\bibitem{Yoder} Yoder, C.F. 1982. Tidal Rigidity of Phobos. {\it Icarus}, 
                Vol. {\bf 49}, pp. 327 - 346

\bibitem{Yeomans 1} Yeomans, D.K., J.-P.Barriot, D.W.Dunham, R.W.Farquhar, 
                    J.D.Gorgini, C.E.Helfrich, A.S.Konopliv, J.V.McAdams, 
                    J.K.Miller, W.M.Owen Jr, D.J.Scheeres, S.P.Synnott, \& 
                    B.G.Williams. 1997. Estimating the Mass of Asteroid 253 
                    Mathilde from Tracking Data During the NEAR Flyby. 
                    {\it{Science}}, Vol. {\bf 278}, pp. 2106 - 2108  

\bibitem{Yeomans 2} Yeomans, D.K., P.G.Antreasian, J.-P.Barriot, S.R.Chesley,
                    D.W.Dunham, R.W.Farquhar, J.D.Giorgini, C.E.Helfrich, 
                    A.S.Konopliv, J.V.McAdams, J.K.Miller, W.M.Owen Jr., 
                    D.J.Scheeres, P.C.Thomas, J.Veverka, \& B.G.Williams. 
                    2000. Radio Science Results During the NEAR-Schoemaker
                    Spacecraft Rendezvous with Eros. {\it{Science}}, Vol. 
                    {\bf 289}, pp. 2085 - 2088         

\end{thebibliography}
\end{document}